\documentclass[twocolumn,superscriptaddress,floatfix,preprintnumbers,prd ,nofootinbib,hyperref]{revtex4-2}
\pdfoutput=1
\usepackage[colorlinks=true,breaklinks=true]{hyperref}
\usepackage[normalem]{ulem}
\usepackage[T1]{fontenc}
\usepackage[utf8]{inputenc}
\hypersetup{allcolors=[rgb]{0.0 0.0 0.6},linkcolor=[rgb]{0.75 0.05 0.05}}
\usepackage{amsmath,amssymb}
\usepackage{mathtools}
\usepackage{cancel}
\usepackage[dvipsnames]{xcolor}
\usepackage{float}
\usepackage{aas_macros}
\usepackage{letltxmacro}
\LetLtxMacro{\oldcite}{\cite}
\renewcommand{\cite}[1]{\mbox{\oldcite{#1}}}
\long\def\exclude#1{}

\newcommand{\beq}{\begin{equation}}
\newcommand{\eeq}{\end{equation}}

\def\ga{\,\,\raise0.14em\hbox{$>$}\kern-0.76em\lower0.28em\hbox
{$\sim$}\,\,}

\newcommand{\ie}{\emph{i.e.~}}
\newcommand{\eg}{\emph{e.g.~}}

\newcommand{\levnlm}{\left. |nlm\right>}
\newcommand{\levtwo}{\left. |211\right>}
\newcommand{\levthree}{\left. |322\right>}
\newcommand{\levfour}{\left. |411\right>}
\newcommand{\levfourtwo}{\left. |422\right>}
\newcommand{\levfourthree}{\left. |433\right>}
\long\def\exclude#1{}

\allowdisplaybreaks

\bibpunct{[}{]}{,}{n}{}{,}

\hypersetup{
    colorlinks=true,
    citecolor=[rgb]{.1, .7, .6},
    linkcolor=[rgb]{.2, .55, .95},
    filecolor=magenta,
    urlcolor=[rgb]{.1, .7, .6},
}

\begin{document}

\title{Stepping Up Superradiance Constraints on Axions}

\author{Samuel J. Witte}
\email[]{samuel.witte@physics.ox.ac.uk}

\author{Andrew Mummery}
\affiliation{Rudolf Peierls Centre for Theoretical Physics, University of Oxford, Parks Road, Oxford OX1 3PU, UK}


\begin{abstract}
Light feebly-coupled bosonic particles can efficiently extract the rotational energy of rapidly spinning black holes on sub-astrophysical timescales via a phenomenon known as black hole superradiance. In the case of light axions, the feeble self-interactions of these particles can lead to a non-linear coupled evolution of many superradiant quasi-bound states, dramatically altering the rate at which the black hole is spun down. In this work, we extend the study of axion superradiance to higher order states, solving for the first time the coupled evolution of all states with $n \leq 5$ in the fully relativistic limit (with $n$ being the principal quantum number). Using a Bayesian framework, we re-derive constraints on axions using the inferred spins of solar mass black holes, demonstrating that previously adopted limit-setting procedures have underestimated current sensitivity to the axion decay constant $f_a$ by around one order of magnitude, and that the inclusion to higher order states allows one to reasonably capture the evolution of typical high-spin black holes across a much wider range of parameter space, thereby allowing constraints to be extended to more massive axions.  We conclude with an extensive discussion on the systematics associated with spin inference from x-ray observations.
\end{abstract}

\maketitle
\tableofcontents

\section{Introduction}

It was shown more than half a century ago that low energy bosonic waves can be amplified when scattered off of a rotating dissipative body~\cite{zel1971generation}. This phenomena, known as rotational superradiance, occurs when
\begin{equation}\label{eq:sr}
    \omega < m \, \Omega \, ,
\end{equation}
where $\omega$ and $m$ are the energy and azimuthal number of the incident radiation, and $\Omega$ is the rotational frequency of the body. Applied to the context of black holes, this process can be interpreted as the wave analogue of the well-known Penrose process~\cite{penrose1971extraction}, whereby particles can extract the black hole's rotational energy. The remarkable aspect of this phenomenon, however, stems from the fact that the gravitational potential of a Kerr black hole contains quasi-bound states that serve to confine massive bosonic fields near the black hole itself~\cite{Damour:1976kh,Detweiler:1977gy,Cardoso:2004nk}. Consequently, confined low-energy bosons satisfying Eq.~\ref{eq:sr} will iteratively scatter off the black hole; at each scattering the initial wave will be amplified, leading to an instability that causes these bound states to grow at an exponential rate (see~\cite{Brito:2015oca} for a general review of black hole superradiance). 

The importance of black hole superradiance as a probe of new fundamental physics has been recognized for many years (see \eg~\cite{Arvanitaki:2009fg}). The general idea being that the mere existence of a light boson with a characteristic mass $\mu \sim \mathcal{O}(0.1) / (G M)$\footnote{We work in units with $c=\hbar=1$, and will frequently work with quantities which have been normalized via appropriate powers of $(GM)$.} is sufficient to cause all rapidly rotating black holes of mass $M$ to lose an order one fraction of their rotational energy on sub-astrophysical timescales (with the effective `spin-down timescale' depending on the mass and spin $a$ of the black hole, and mass $\mu$ and spin of the boson, see~\eg~\cite{Detweiler:1980uk,Dolan:2007mj,East:2017ovw,Baryakhtar:2017ngi,Baumann:2019eav,Brito:2020lup,East:2023nsk}); the observation of high-spin black holes (as inferred \eg using x-ray observations~\cite{Reynolds13, McClintock14, draghis2023systematic}, gravitational waves~\cite{abbott2019gwtc,abbott2019binary,zackay2019highly,venumadhav2020new,abbott2021population}, or tidal disruption events~\cite{Mummery20, Wen20, Wen21, Du:2022trq,Mummery:2023meb}) thus serves as strong evidence against the existence of such a boson~\cite{Arvanitaki:2009fg,Arvanitaki:2010sy,cardoso2018constraining}. The energy extracted from the black hole is used to grow a high-density boson cloud, which itself can give rise to a variety of observational signatures, including \eg narrow-bandwidth gravitational waves and modifications to binary inspirals~\cite{Arvanitaki:2010sy,Yoshino:2013ofa,Arvanitaki:2014wva,Arvanitaki:2016qwi,Ferreira:2017pth,Hannuksela:2018izj,Zhang:2019eid,Palomba:2019vxe,Siemonsen:2019ebd,Zhu:2020tht,Brito:2017wnc,Brito:2017zvb,Baumann:2018vus,Tsukada:2020lgt,Baumann:2019ztm,Baryakhtar:2020gao,Baumann:2021fkf,Baumann:2022pkl,Tong:2022bbl,Siemonsen:2022yyf,Fan:2023jjj,Boskovic:2024fga,Tomaselli:2024bdd,May:2024npn}.

Black hole superradiance is often discussed as a general phenomenon that can broadly be used to search for light bosons with masses $10^{-21} \, {\rm eV} \lesssim \mu \lesssim 10^{-11} \, {\rm eV}$ (where the upper and lower boundaries should be treated as ball-park estimates that are obtained by requiring $\alpha \equiv G M \mu \sim \mathcal{O}(0.1)$, with $M$ taken to be the mass of the heaviest and lightest known black holes)\footnote{The presence of a minimum mass scale comes from requiring that the growth of the instability takes place on astrophysical timescales. If the boson acquires mass corrections near the black hole, or if there exists an alternative confinement mechanism, then this characteristic mass range may be notably different.}. However, this statement is not entirely true; even the existence of small self-interactions or feeble couplings to ambient particles can dramatically alter the expected evolution of these systems, in some cases killing any hope of a potential measurement (see \eg~\cite{Arvanitaki:2009fg,Arvanitaki:2010sy,Fukuda:2019ewf,Rosa:2017ury,Ikeda:2018nhb,Mathur:2020aqv,Blas:2020kaa,Blas:2020nbs,Baryakhtar:2020gao,Caputo:2021efm,Siemonsen:2022ivj,Spieksma:2023vwl,Ferreira:2024ktd}). A prime example of this arises in the context of axions, where the quartic self-interactions $\mathcal{L}_{\rm self} \propto \lambda \, a^4$ can quench the growth of axion-bound states, thereby dramatically elongating the time required for the axion field to extract a sizable fraction of the black hole spin.

The role of axion self-interactions in black hole superradiance was first studied in~\cite{Arvanitaki:2010sy}; here, it was pointed out that these interactions could induce a number of distinct effects, including: $\emph{i})$ the emission of relativistic axions, $\emph{ii})$ an effective mixing, or energy transfer, between various quasi-bound states, and $\emph{iii})$ a bosenova (a rapid collapse of the ambient axion cloud driven by attractive axion self-interactions, leading to a potentially violent disruption of the system -- see \eg~\cite{Yoshino:2012kn,Yoshino:2015nsa,Omiya:2022mwv} for numerical studies of this process). It has often been assumed that the primary role of self-interactions in superradiant systems was to establish a maximal occupation number of the axion quasi-bound states, with the scale set by the density at which self-interactions are expected to induce a bosenova. Recent work~\cite{Gruzinov:2016hcq,Baryakhtar:2020gao,Omiya:2022gwu}, however, has argued that bosanovae are unlikely to occur across most of the axion parameter space -- this conclusion stemmed from the fact that self-interactions drive energy transfer between different superradiant levels (\ie, they mix superradiant levels), and this energy transfer tends to push the system to quasi-equilibrium configurations with occupation numbers below the bosenova threshold. Since the rate of spin extraction scales with the axion occupation number, the existence of such quasi-equilibrium configurations unavoidably weakens superradiance constraints. More specifically, Ref.~\cite{Baryakhtar:2020gao} studied how the coupled evolution of the two fastest growing superradiant states spin down black holes\footnote{Ref.~\cite{Baryakhtar:2020gao} also includes a discussion on how the two-state equilibrium can be disrupted at low masses by the next fastest growing superradiant level, alongside a discussion on the potential impact of higher-order states appearing on longer timescales; the direct impact of these states on the spin down, however, is not performed. }, and concluded that previously-derived limits on the axion decay constant $f_a$, which had been derived using x-ray observations of solar-mass-scale black holes, were overestimated by nearly $\sim 2$ orders of magnitude (see \eg~\cite{Stott:2018opm,Stott:2020gjj,Mehta:2020kwu,Mehta:2021pwf} for limits which were instead derived by applying a `bosenova' threshold on the relative occupation numbers of each bound state).

\begin{figure}
    \includegraphics[width=0.49\textwidth]{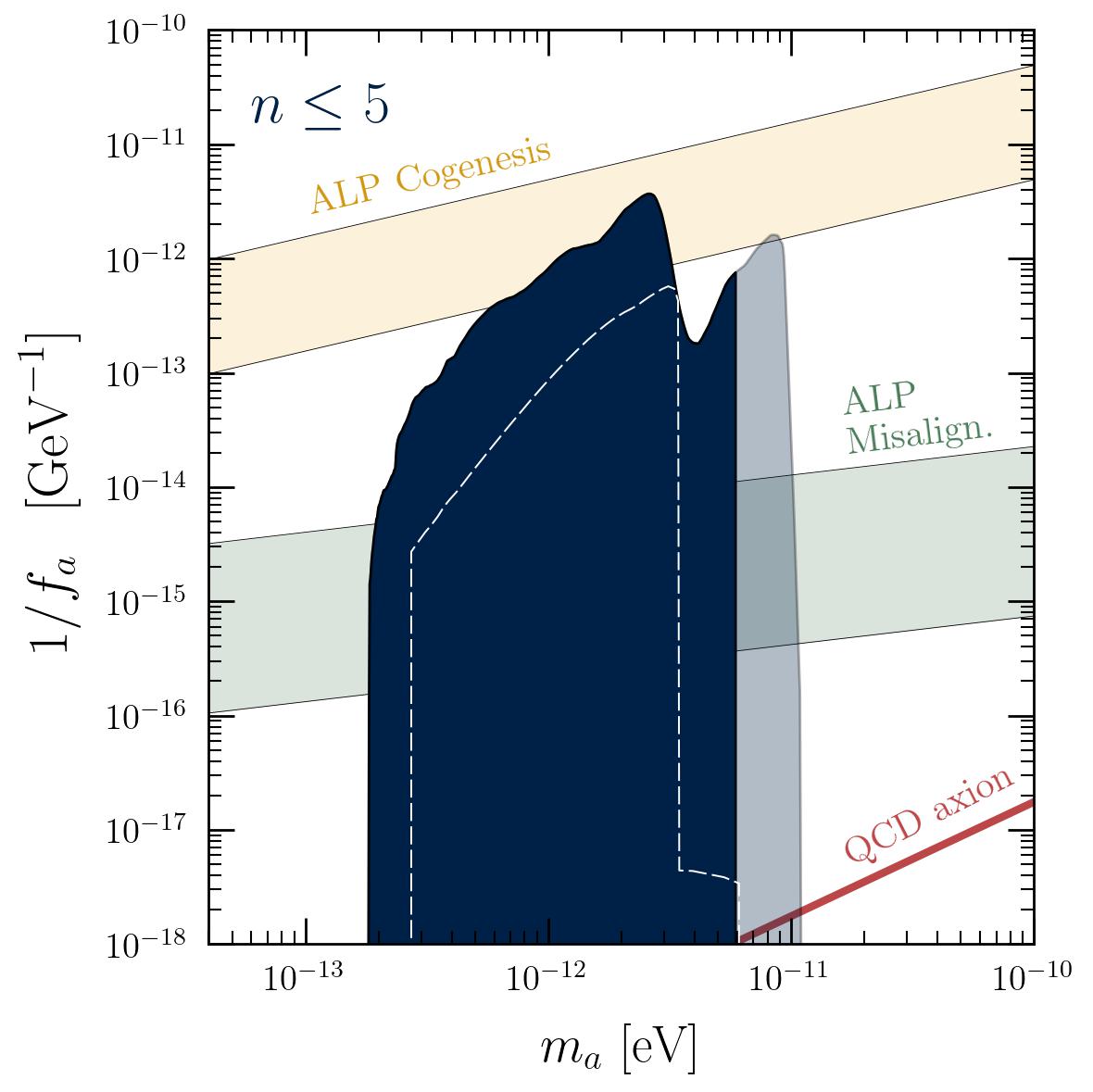}
    \caption{\label{fig:main} Superradiance limits, derived at the $2 \sigma$ level, using a combination of Cyg X-1 and GRS 1915+105 (see Table~\ref{tab:bhs}; in the case of Cyg X-1 we use the `conservative' inference of the spin). Our fiducial limit (dark blue) covers only the $m \leq 2$ spin-down region, as higher-order levels may alter the spin-down for $m > 2$ (shaded blue).
    Constraints are derived  using $n \leq 5$ superradiant levels, and are compared with various benchmarks, including: the QCD axion (red), the parameter space in which an axion-like particle can explain the origin of the baryon asymmetry (`ALP Cogenesis'; yellow)~\cite{Co:2020xlh}, and the axion parameter space in which the misalignment mechanism generates the dark matter relic abundance without invoking a fine-tuning of the initial field value. We also show for comparison previous constraints derived in~\cite{Baryakhtar:2020gao} (white dashed).  }
\end{figure}

Given the non-trivial evolution that can be induced by the presence of even small self interactions, it is natural to ask whether the precise signatures being induced by axion superradiance are, in fact, well understood (in the sense that one can make precision predictions about spin down, or the gravitational wave emission produced by these objects), and whether as a community we are prepared to interpret the forthcoming data that will allow for the detection, or exclusion, of light axions. We would argue that the answer to both questions is: no. Ignoring momentarily all of the potential issues associated with the fact that black holes are not isolated `clean' systems, let us emphasize that the description of the superradiant evolution in terms of a coupled two-level system is only adequate for sufficiently light axions ($\alpha \lesssim 0.1$) and at early times, and a complete description (characterized by the time evolution of all quasi-bound states, as well as the temporal evolution of all radiated energy, over the lifetime of the black hole) requires understanding the evolution of tens, hundreds, or potentially even thousands of coupled quasi-bound states. In this work, we make a step toward understanding the role and importance of higher order states in the superradiant evolution\footnote{It is worth highlighting that Ref.~\cite{Omiya:2024xlz} recently studied the gravitational wave signature produced from the coupled evolution of the $\levtwo$, $\levthree$, $\levfourthree$, and $\left| 544 \right>$ states, focusing primarily on the $\alpha \lesssim 0.5$ regime. This work had neglected higher order overtones based a conclusion drawn in~\cite{Omiya:2022mwv}. As we show below, these states cannot be neglected (something which had already been highlighted in Ref.~\cite{Baryakhtar:2020gao}), and can play a non-trivial role in both the evolution of spin down and the emission of gravitational waves.}, focusing in particular on the impact of self-interactions on the evolution of, and the spin down induced by, the $n \leq 5$ superradiant levels (with $n$ labeling the principal quantum number of each bound state); the impact of higher order states on the gravitational wave spectrum will be studied in future work.  One of the main results of this manuscript is a novel derivation of the superradiance limits on axions from the observation of highly spinning solar mass black holes. This limit is derived within a Bayesian framework, and includes fully relativistic scattering rates for all relevant processes up to $n=5$ (whereas previous work had focused primarily on $n \leq 3$) -- the resulting $2\sigma$ limit from the two most constraining black holes is provided in Fig.~\ref{fig:main} (dark and light blue), and is shown alongside various theory-motivated benchmarks, including the QCD axion (red), the parameter space in which axion-like particles can explain the baryon asymmetry of the Universe (`ALP Cogenesis', yellow)~\cite{Co:2020xlh}, and the parameter space in which the misalignment mechanism allows for axions to account for the entirety of dark matter without any fine-tuning of the initial conditions (`ALP Misalignment', green), see \eg~\cite{Irastorza:2018dyq}.  We argue in the following sections that the constraints highlighted in dark blue should be largely stable to the inclusion of states with $n = 6,7$. 
For the sake of completeness, we also include an extensive discussion on the potential systematics associated with inferring black hole spins from x-ray observations.

The rest of this manuscript is organized as follows. Sec.~\ref{sec:BHSR} provides a detailed overview of black hole superradiance, focusing on: the derivation of the spectrum and growth rates for a free particle in the hydrogen-like limit (Sec.~\ref{sec:hyrdg}), the derivation of the spectrum and the growth rates for a free particle in the relativistic limit (Sec.~\ref{sec:rel_cor}), the role of energy dissipation and energy transfer (at a conceptual and mathematical level) of $\phi^4$ self-interactions in the hydrogen-like and relativistic limits (Sec.~\ref{subsec:si}), and an analysis of how the inclusion of self-interaction induced scattering between different bound states influences the evolution of the bound state occupation numbers and black hole spin for a range of different parameters (Sec.~\ref{sec:example_evo}). Having developed the full machinery in Sec.~\ref{sec:BHSR} required to compute the superradiant evolution around an arbitrary black hole with the inclusion of self-interactions, we develop in Sec.~\ref{sec:analysis} the statistical analysis and limit setting procedure used to produce the limits shown in Fig.~\ref{fig:main}. We include a discussion on systematics and supermassive black holes in Secs.~\ref{sec:discussion} and \ref{sec:smbh}, and conclude in Sec.~\ref{sec:conc}.

\section{Black Hole Superradiance} \label{sec:BHSR}

The goal of this section is develop from general formalism required to compute the superradiant evolution of scalar fields around rotating black holes. In that light, we begin by reviewing black hole superradiance in the context of non-interacting spin-0 fields; in particular, we first focus on deriving the growth rate and eigenstates of the various quasi-bound states in the hydrogen-like limit (where the calculations are dramatically simplified, and intuition can easily be gained by mapping the problem into electromagnetic equivalent scenario), before generalizing the result to the relativistic limit. We then demonstrate how self-interactions modify this picture, once again both computing the results first in the hydrogen-like limit, and then with a fully relativistic treatment based on Greens functions. Having developed the formalism required to compute the growth and evolution of quasi-bound states (as well as radiative emission processes), we provide an extensive array of examples illustrating the role and impact on self-interactions on black hole spin down and occupation numbers of various states.

Let us begin with the case of a non-interacting scalar field. The growth rate of the superradiant instability for a generic scalar field $\Phi$ with mass $\mu$ can be derived by studying the quasi-bound state spectrum of the Klein-Gordon equation in a Kerr background (see  \eg~\cite{Zouros:1979iw,Detweiler:1980uk,Furuhashi:2004jk,Cardoso:2004nk,Dolan:2007mj,bao2022improved})\footnote{See also \eg~\cite{Rosa:2011my,Witek:2012tr,Pani:2012vp,Pani:2012bp,Endlich:2016jgc,Baryakhtar:2017ngi,East:2017ovw,East:2017mrj,East:2018glu,Dolan:2018dqv,Brito:2020lup,brito2020black,Brito:2015oca} for discussions of quasi-bound states arising for spin-1 and spin-2 fields.}. Let us start from the generalized Klein-Gordon equation in the presence of a source $T(t, r,\theta, \phi)$,
\begin{eqnarray}
    \left( \Box - \mu^2 \right) \Phi = T \, .
\end{eqnarray}
In Boyer-Lindquist coordinates this reduces to
\begin{multline}
 \partial_r (\Delta \partial_r)\Phi - \frac{a^2}{\Delta} \partial_\phi^2 \Phi - \frac{4 GM r a}{\Delta}\partial_\phi \partial_t \Phi - \frac{(r^2+a^2)^2}{\Delta}\partial_t^2 \Phi   \\  - \mu^2 r^2 \, \Phi + \frac{1}{\sin\theta}\partial_\theta(\sin\theta \, \partial_\theta \Phi) + \frac{1}{\sin^2\theta}\partial_\phi^2\Phi + a^2 \sin^2\theta \partial_t^2 \Phi  \\[10pt]  - \mu^2 a^2 \cos^2\theta \, \Phi = \rho^2 T \hspace{2cm} \label{eq:KG_exp}
\end{multline}
where $\Delta = r^2 - 2 GM r + a^2$, $\rho^2 = r^2 + a^2 \cos^2 \theta$, and $a$ is the black hole spin (for convenience, we often work with the dimensionless quantities normlized by appropriate factors of $GM$; e.g. the dimensionless spin is defined as $\tilde{a} \equiv a / (G M) $). Eq.~\ref{eq:KG_exp} has been written in a way such that the separability is immediately apparent; adopting a separable Ansatz ~\cite{Brill:1972xj,Carter:1968ks}
\begin{eqnarray}
    \Phi(\vec{r}, t) = R(r) \, S(\theta) \, e^{-i\omega t} \, e^{i m \phi} ,
\end{eqnarray}
and momentarily setting the source term to zero $T = 0$, one finds that Eq.~\ref{eq:KG_exp} reduces to the following differential equations for the radial and angular functions:
\begin{eqnarray}
    \frac{1}{\sin \theta}   \frac{d}{d \theta} \left(\sin \theta \frac{d S}{d \theta} \right) +   \left[ a^2 (\omega^2 - \mu^2) \cos^2\theta - \right.  \nonumber \\[5pt] \left.  \frac{m^2}{\sin^2 \theta} + \Lambda_{lm} \right] S(\theta)  = 0,   \, \hspace{.5cm} \label{eq:spheriodal} \\[15pt]
     \frac{d}{dr}  \left( \Delta \frac{d R}{d r} \right) + \left[ \frac{\omega^2 (r^2 + a^2)^2 - 4 G M a m \omega r + m^2 a^2 }{\Delta}  \right. \nonumber \\[5pt] \left. - \left( \omega^2 a^2 + \mu^2 r^2 + \Lambda_{lm} \right) \right] R_{lm}(r) = 0 \, . \, \hspace{.5cm}   \label{eq:radialR}
\end{eqnarray}
The angular solutions are the oblate spheroidal harmonics (reducing to the spherical harmonics in the limit $a \rightarrow 0$), with eigenvalues $\Lambda_{lm}$, while the eigenvalues of the radial solution $\omega$ characterize the energies of the quasi-bound states.

It is often convenient to redefine the radial wave function as $R(r) \equiv u(r) / \sqrt{r^2 + a^2}$, and pass to tortoise coordinates, 
\begin{eqnarray}
    r^* = r + \frac{2 }{r_+ - r_-} \left[r_+\log\left(\frac{r - r_+}{2} \right) - r_-\log\left(\frac{r - r_-}{2}  \right)  \right] \hspace{.2cm} \, ,
\end{eqnarray}
where $r_\pm = GM (1 + \sqrt{1 - \tilde{a}^2})$ are the inner and outer horizons. In this limit, Eq.~\ref{eq:radialR} reduces to
\begin{eqnarray}\label{eq:homoRad}
    \frac{d^2u}{dr^{* \, 2}} + \left(\omega^2 - V(\omega, r^*) \right) u = 0
\end{eqnarray}
with the potential
\begin{multline}
    V(\omega, r^*) = \frac{\Delta \mu^2}{r^2 + a^2} + \frac{\Delta (3 r^2 - 4 GM r + a^2)}{(r^2 + a^2)^3} \\[3pt] + \frac{4 GM r a m \omega - a^2 m^2 +  \Delta (\Lambda_{\ell m} + a^2 (\omega^2 - \mu^2))}{(r^2 + a^2)^2 } \\[3pt]
     - \frac{3 \Delta^2 r^2}{(r^2 + a^2)^4} \, ,
\end{multline}
where $r$ is understood to be a function of $r^*$. Below we discuss techniques for solving for the eigenvalues and wavefunctions of the system, and then return to the question of how a non-trivial source term (which in the case of axions is provided by quartic self-interactions) modifies the evolution of the system.

\subsection{The Hydrogen-like Limit}\label{sec:hyrdg}

In general, closed form solutions characterizing the quasi-bound states and their eigenvalues do not exist. Analytic expressions, however, can be derived perturbatively in the  limit $\alpha \ll 1$ and  $\tilde{a}  \ll 1$; below, we briefly review this calculation, but refer the reader to \eg~\cite{Detweiler:1980uk,Baumann:2019eav} for additional details. In the aforementioned limit,  the angular solutions of Eq.~\ref{eq:spheriodal} become spherical harmonics $Y_\ell^m(\theta, \phi)$ with $\Lambda_{\ell m} = \ell (\ell + 1)$, while the radial equation  (Eq.~\ref{eq:radialR}) reduces  to 
\begin{multline}\label{eq:radial_appr2}
    \left[-\frac{1}{r^2}\frac{d}{dr}\left(r^2 \frac{d}{dr}\right) - \frac{2 \alpha \mu}{r} \right. \\ \left. + \frac{\ell (\ell + 1)}{r^2} + (\mu^2 -\omega^2) \right] R_{\ell m} = 0 \, .
\end{multline}
This is equivalent to the time-independent Schr\"{o}dinger equation for the hydrogen atom (hence why this is often referred to as the `hydrogen-like' limit), with $\alpha$ playing the gravitationally equivalent role of the fine structure constant. 

Imposing boundary conditions such that the bound states vanish at large $r$, and are well-behaved at the origin\footnote{Recall that the Schr\"{o}dinger equation for the hydrogen atom also contains an un-physical solution which scales as $R \sim r^{-1}$ as $r \rightarrow 0$. This solution is discarded as it is divergent at the origin (and thus un-normalizable).  } discretizes the energy spectrum; the collection of discrete bound states can  be characterized by quantum numbers $\levnlm$ ($n$ being the principal quantum number, rather than the overtone number), with the quantum numbers obeying the conventional selection rules: $n \geq l+1$ and $|m| \leq l$. These states will have an energy given by $E_n = \mu (1 - \alpha^2 / (2n^2))$, and the radial wave function in this limit will be Bohr-like, given by
\begin{eqnarray}\label{eq:radial_bnd_NR}
    R_{n \ell}^{\rm far} = C_0^{\rm far}
    e^{-r/(n a_0)} \left(\frac{2r}{n a_0} \right)^l L_{n-l-1}^{2l + 1}\left( \frac{2r}{n a_0}\right)
\end{eqnarray}
where $a_0 \equiv (\alpha \mu)^{-1}$, $L_{n-l-1}^{2l+1}(x)$ is the generalized Laguerre polynomial, and $C_0^{\rm far}$ is a normalization coefficient. Throughout this work we will normalize the wavefunction such that $\int dV \, |\Phi|^2 = 1$, implying
\begin{eqnarray}
    C_0^{\rm far} = \left[\left(\frac{2}{n a_0} \right)^3 \, \frac{(n-l-1)!}{2 n (n + l)!} \right]^{1/2} \, .
\end{eqnarray}

The eigenvalues in the hydrogen-like limit are real, however it is known that the true eigenvalues of the system (\ie the eigenvalues of Eq.~\ref{eq:homoRad}) are, in general, complex $\omega_{nlm} = E_{nlm} + i \Gamma_{nlm}$. In order to solve for the complex eigenvalues of the system, one can independently solve for the radial solution at large $r$ (i.e., Eq.~\ref{eq:radial_appr2}) and at $r \rightarrow r_+$, with the latter being governed by~\cite{Baumann:2019eav} 
\begin{multline}\label{eq:nearRad}
\left[\frac{d^2}{dz^2} + \left(\frac{1}{z} + \frac{1}{z+1} \right)\frac{d}{dz} - \frac{\ell (\ell + 1)}{z (z+1)} \right. \\ \left. + \frac{\alpha^2 (\tilde{a} m \mu - 2 \tilde{r}_+ \omega)^2}{(\tilde{r}_- - \tilde{r}_+)^2 z^2 (1 + z)^2 \mu^2} \right] R^{\rm near}_{n\ell} = 0    
\end{multline}
where we have introduced the variable $z \equiv (r - \alpha \tilde{r}_+) / (\alpha (\tilde{r}_+ - \tilde{r}_-))$, and $\tilde{r}_\pm \equiv r_\pm / (G M)$ is the normalized outer/inner horizon radius. The solution to Eq.~\ref{eq:nearRad} is given by~\cite{Baumann:2019eav}
\small
\begin{eqnarray}
    R_{n \ell}^{\rm near} = C_0^{\rm near} \left(\frac{z}{z+1} \right)^{i \zeta} {}_2F_1\left(-\ell, \ell + 1, 1-2i\zeta, 1+z\right) \, \hspace{.3cm}
\end{eqnarray}
\normalsize
with $C_0^{\rm near}$ the normalization, ${}_2 F_1$ the hypergeometric function, and $\zeta \equiv (\mu \tilde{a} m - 2 \alpha \tilde{r}_+ \omega) / (\mu (\tilde{r}_+ - \tilde{r}_-))$.

The energy spectrum of a given quasi-bound state can be directly obtained by perturbatively matching the near- and far-field solutions in an intermediate regime (see \eg ~\cite{Rosa:2009ei,Baumann:2018vus,Baumann:2019eav}); keeping corrections beyond the hydrogen-like limit, the real part of the energy is given by
\begin{multline}
    E_{nlm} \simeq \mu \left(1 - \frac{\alpha^2}{2 n^2} - \left[\frac{1}{8 n} + \frac{6}{2 l + 1} - \frac{2}{n} \right] \frac{\alpha^4}{n^3} \right. \\ \left. + \cdots \right) ,
\end{multline}
and the imaginary part by
\begin{eqnarray}\label{eq:gammaDet}
    \Gamma_{nlm} \simeq 2 \tilde{r}_+ C_{nl} \, g_{lm} \, (m \Omega_H - \omega_{nlm}) \, \alpha^{4 l + 5} \, + \cdots 
\end{eqnarray}
where we have introduced the following factors: 
\begin{eqnarray}
    C_{nl} \equiv \frac{2^{4 l + 1} (n+l)!}{n^{2l+4} (n-l-1)!} \left[\frac{l!}{(2l)! (2l+1)!} \right]^2 , \\
    g_{lm} \equiv \prod_{k=1}^l \left(k^2 (1-\tilde{a}^2) + (\tilde{a} m - 2 \tilde{r}_+ \alpha)^2 \right) \, ,
\end{eqnarray}
 with $\Omega_H = \tilde{a} / (2 r_+)$.  Note that the pre-factor of Eq.~\ref{eq:gammaDet} directly contains the condition for superradiance highlighted in the introduction (see Eq.~\ref{eq:sr}) -- namely, the $\Gamma_{nlm}$ is only positive, corresponding to growing states, when $m \Omega_H > \omega_{nlm}$.

\subsection{The Relativistic Solution }\label{sec:rel_cor}

The  approximation outlined above provides a good estimate of the energies and bound state wavefunctions at small $\alpha$, but can have non-negligible deviations from the true solution at $\alpha \gtrsim 0.1$ (see \eg Fig. 9 of~\cite{Baumann:2019eav}). In order to improve upon these approximations we adopt the continued fraction method, which was originally developed in~\cite{leaver1985analytic}, and later refined in~\cite{konoplya2006stability,cardoso2005superradiant,Dolan:2007mj}. Here, one begins by expanding the radial equation in a well-chosen basis that respects the requirement of purely in-going waves at the horizon,~\cite{Dolan:2007mj}
\begin{eqnarray}\label{eq:radial_expand}
    R(r) = \left(r - r_+ \right)^{-i\sigma} \left(r - r_- \right)^{i\sigma + \chi - 1}  \times \nonumber \\ e^{qr} \, \sum_{n=0}^{\infty} a_n \left(\frac{r-r_+}{r-r_-} \right)^n \, ,
\end{eqnarray}
where 
\begin{eqnarray}
    \sigma &=& \frac{2 r_+ GM (\omega - m \Omega)}{r_+ - r_-} \\ 
    q  &=& \pm \sqrt{\mu^2 - \omega^2} \\
    \chi  &=& GM \frac{\mu^2 - 2 \omega^2}{q} \, .
\end{eqnarray}
The sign of $q$ determines the asymptotic behavior at large $r$, with Real$(q)>0$ yielding a divergent wavefunction (corresponding to the quasi-normal mode spectrum) and Real$(q)<0$ yielding a vanishing wavefunction at large $r$ (corresponding to quasi-bound state spectrum). Here, we focus on quasi-bound states (as these correspond to the growing superradiant modes), and defer a discussion of quasi-normal modes to the following section.

\begin{figure}
    \includegraphics[width=0.49\textwidth]{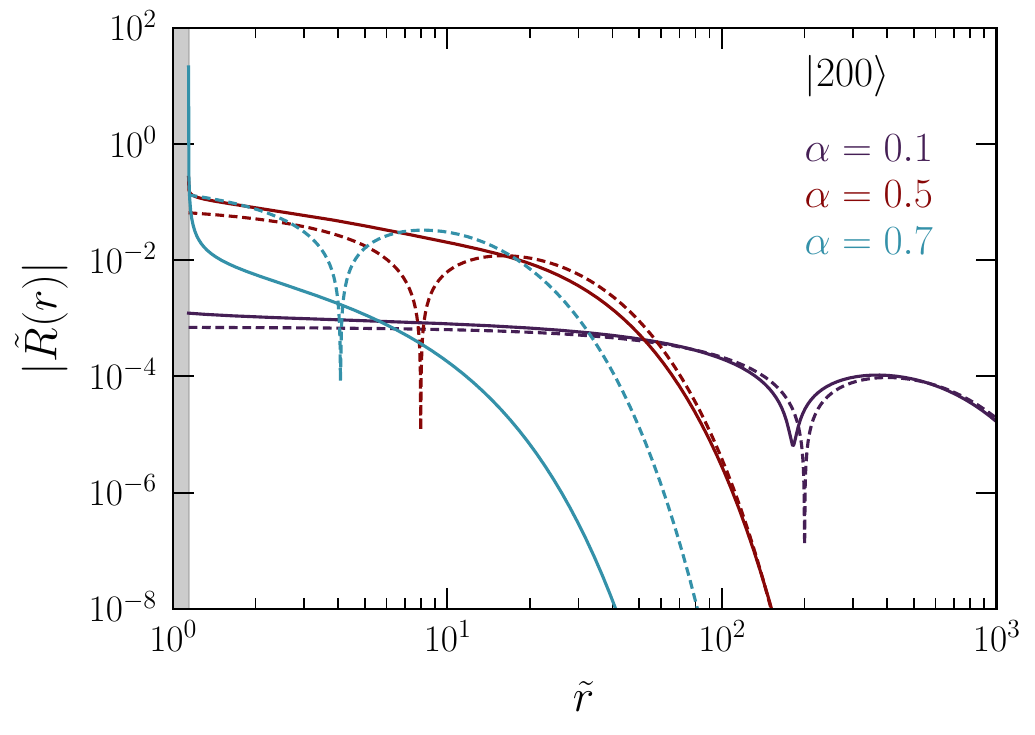}
    \includegraphics[width=0.49\textwidth]{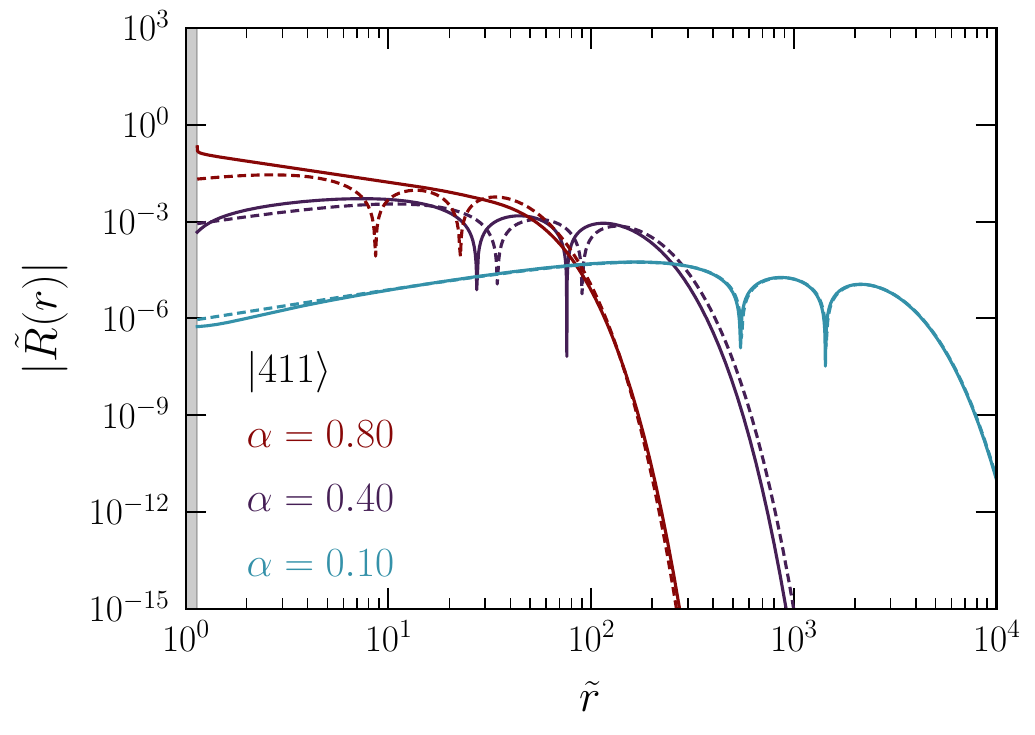}
    \includegraphics[width=0.49\textwidth]{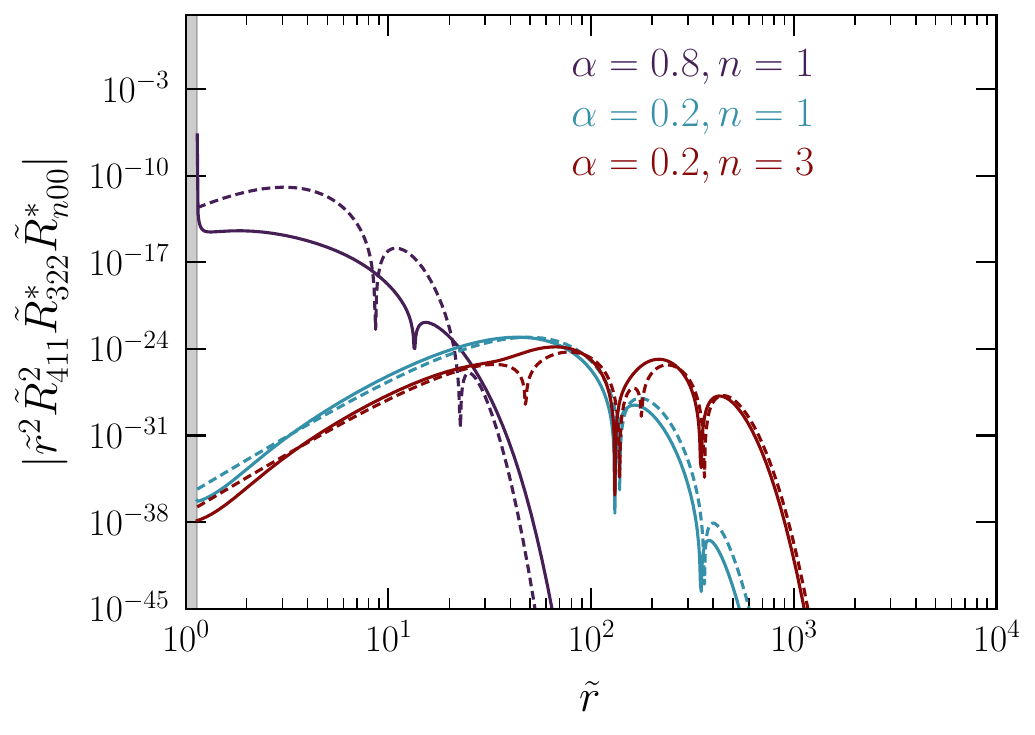}
    \caption{Top and Center: Relativistic (solid) and non-relativistic (dashed) radial wave functions for various bound states and values of $\alpha$, with the tilde notation denoting normalized quantities. Bottom: Integrand entering the projection of the source term associated to the $\left. |411 \right> \times \left. |411 \right> \rightarrow \left. |322 \right> \times \left. |n00 \right>$ rate for various choices of $n$ and $\alpha$ (see Eq.~\ref{eq:cnlm}). Results are shown using the relativistic (solid) and non-relativistic (dashed) wave functions. All results are computed using $\tilde{a} = 0.99$.  }
    \label{fig:WF}
\end{figure}

Plugging Eq.~\ref{eq:radial_expand} into Eq.~\ref{eq:radialR}, one can derive a three-term recurrence relation for the coefficients $a_n$,~\cite{Dolan:2007mj}
\begin{eqnarray}
    \alpha_0 a_1 + \beta_0 a_0 &=& 0 \\
    \alpha_n a_{n+1} + \beta_n a_n + \gamma_n a_{n-1} &=& 0 \, \hspace{.5cm} n > 0 \, ,
\end{eqnarray}
where
\begin{eqnarray}
    \alpha_n = n^2 + (c_0 + 1)n + c_0 \\
    \beta_n = -2 n^2 + (c_1 + 2) n + c_3 \\
    \gamma_n = n^2 + (c_2 - 3)n + c_4 \, .
\end{eqnarray}
Here, the constants $c_i$ are given by
\begin{eqnarray}
    c_0 &=& 1 - 2 i \tilde{\omega} - \frac{2i}{b}\left(\tilde{\omega} - \frac{\tilde{a} m}{2}\right) \\[11pt]
    c_1 &=& -4 + 4i (\tilde{\omega} - i \tilde{q} (1 + b)) \nonumber \\[8pt]  &
    +&\frac{4i}{b} \left( \tilde{\omega} - \frac{\tilde{a} m}{2}\right) - \frac{2 (\tilde{\omega}^2 + q^2)}{q} \\[11pt]
    c_2 &=& 3 - 2i\tilde{\omega} - 2\frac{q^2 - \tilde{\omega}^2}{q} - \frac{2i}{b}\left(\tilde{\omega} - \frac{\tilde{a} m}{2}\right) \\[11pt]
    c_3 &=& \frac{2 i (\tilde{\omega} - i q)^3}{q} + 2 (\tilde{\omega} - i q)^2 b + q^2 \tilde{a}^2  \nonumber \\[8pt] &+&  2 i q \tilde{a} m - \Lambda_{\ell m} - 1 - \frac{(\tilde{\omega}-iq)^2}{q} + 2 q b \nonumber \\[8pt] &+& \frac{2i}{b}\left(\frac{(\tilde{\omega} - i q)^2}{q} + 1 \right)\left(\tilde{\omega} - \frac{\tilde{a} m}{2}\right) \\[11pt]
    c_4 &=& \frac{(\tilde{\omega}-iq)^4}{q^2} + \frac{2 i \tilde{\omega}(\tilde{\omega} - i q)^2}{q} \nonumber \\[8pt] &-& \frac{2i}{b}\frac{(\tilde{\omega} - i q)^2}{q}\left(\tilde{\omega} - \frac{\tilde{a} m}{2}\right)
\end{eqnarray}
where $b=\sqrt{1-\tilde{a}^2}$ (and, as before, the ``$\tilde{X}$" implies $X$ has been normalized by the appropriate factors of $GM$).

The complex eigenstates of the system are then obtained by solving 
\begin{eqnarray}
    \frac{\beta_0}{\alpha_0} - \frac{\gamma_1}{\beta_1 - \alpha_1 \left[ \frac{\gamma_2}{\beta_2 - \alpha_2 \left[\frac{\gamma_3}{\beta_3 - \cdots} \right]}  \right]} = 0 \, ,
\end{eqnarray}
where the recursion is run to some large integer $N$. In practice, obtaining convergence requires one to evaluate at a sufficiently large value of $N$ (for the calculations here, a value of $N \sim 10^3-10^4$ is typically sufficient), and one needs to ensure high floating point precision in order to avoid the accumulation of rounding errors\footnote{In practice, we use the default `BigFloat' type in JuliaLang, which uses 256 bits of precision (approximately corresponding to around 77 decimal digits).}. Using this approach,  our determination of the energy eigenvalues are in good agreement with those of~\cite{Dolan:2007mj,Pani:2012bp,Baumann:2019eav}.

Having computed the complex energies of each quasi-bound state, one can return the recursion relations above, and use these energies to directly solve for each coefficient $a_n$, which then allows for the computation of the full relativistic wave function as expressed in Eq.~\ref{eq:radial_expand}. An alternative approach is to directly plug the eigenvalues into the radial differential equation and integrate from large $r$ to $r \rightarrow r_+$, adopting initial conditions at large $r$ consistent with $R \rightarrow 0$. We have verified that both approaches yield identical wave functions.
Examples illustrating the behavior of the magnitude of the non-relativistic radial wave function (Eq.~\ref{eq:radial_bnd_NR}) and the relativistic wave function are shown in the middle and top panels of Fig.~\ref{fig:WF} for various values of $\alpha$. Here, one can see the convergence of the two in the limit $\alpha \rightarrow 0$, whereas deviations spanning many orders of magnitude appear for $\alpha \gtrsim 0.2$.

\subsection{The Effect of Self-Interactions}\label{subsec:si}

The discussion above has been framed in the context that  axions can be treated as free non-interacting particles. Axions, however, generically have self-interactions, which can drive scattering processes that significantly alter the evolution of superradiant states\footnote{One may ask whether interactions between axions and Standard Model particles could also lead to a suppression of the superradiant growth. While this may be possible in certain contexts, the role of self-interactions are generally expected to provide the dominant effect. This conclusion simply follows from the fact that the axion occupation numbers $N_a$ of a growing superradiant bound state are sufficiently large to compensate for the suppression in the interaction strength.  }. The effect of self interactions on the growth of superradiant clouds has been studied in a variety of contexts (see e.g.~\cite{Arvanitaki:2010sy,Yoshino:2012kn,Yoshino:2015nsa,Omiya:2022mwv,Gruzinov:2016hcq,Baryakhtar:2020gao,Omiya:2024xlz}\footnote{See also~\cite{Branco:2023frw} for a recent study on the role of axion self-interactions around astroid scale primordial black holes.}), with one of the most recent works~\cite{Baryakhtar:2020gao} concluding that the dominant role of self interactions (at least for $\alpha \lesssim 0.3$) is to dissipate energy from the growing superradiant levels,  eventually driving the system toward quasi-stable equilibria with bound state occupation numbers well below what one would predict for the case of a free particle -- the net result being a suppression in the rate of spin extraction, and thus a weakening of superradiant limits. Below, we begin by reviewing the state-of-the-art treatment of self interactions, which has often focused on deriving the rate of energy exchange/loss in the hydrogen-like limit\footnote{It is worth mentioning that self-interactions also play an important role in modifying the expected gravitational wave signal that emerges from these systems, see\eg~\cite{Collaviti:2024mvh,Omiya:2024xlz} for a recent analysis.}. We then use a Green's function approach to derive the fully relativistic wavefunction induced by leading order self-interactions for all relevant scattering processes at $n \leq 5$. We study the evolution of superradiant systems with these states, and demonstrate how self-interactions between these states alter the spin down evolution of solar mass black holes.

We will focus here on a generic quartic axion self interaction, $\mathcal{L}\supset -\frac{\lambda}{4!} \Phi^4$. In the case of the QCD axion, this arises as the leading order contribution in the $\Phi/f_a \ll 1$ limit from the potential $V(\Phi) = \mu^2 \, f_a^2 \, [1 - \cos(\Phi / f_a) ]$, where $f_a$ is the axion decay constant. In what follows, we will use the definition $\lambda \equiv \mu^2 / f_a^2$ to make connection with the QCD axion, but this can be generically re-scaled for more general potentials. In general, cubic self-interactions could also be present, however these are not expected in most cases to be contribute significantly~\cite{Baryakhtar:2020gao}. Higher order terms may also contribute as  $\Phi/f_a \rightarrow \mathcal{O}(1)$; this corresponds to the highly non-linear regime (where previous works have predicted the appearance of a bosenova -- an explosive event expelling an order one fraction of the bound axions); while this is an interesting direction of research, it remains unclear whether these systems can reach the critical occupation numbers for these events to occur~\cite{Baryakhtar:2020gao} (and, in fact, the results of this work are in agreement that the critical occupation numbers where a bosenova could occur are likely not realized), and studying the dynamics in this regime requires dedicated numerical simulations in order to make qualitative predictions.

In the presence of quartic self interactions, the equation of motion becomes
\begin{eqnarray}\label{eq:eom_si}
    (\Box - \mu^2) \Phi = -\frac{\lambda}{6} \, \Phi^3 \, .
\end{eqnarray}
In the limit where self-interactions can be thought of as perturbative, it is natural to expand the axion field as $\Phi = \sum \lambda^i \, \Phi^{(i)}$. The zeroth order wave function is naturally that of the free particle (the solution having been discussed the proceeding sections), $ (\Box - \mu^2) \Phi^{(0)} = 0$, with the first order correction being given by $ (\Box - \mu^2) \Phi^{(1)} = - \lambda  (\Phi^{(0)})^3 / 6$. 
Since $\Phi^{(0)}$ is the free particle solution, one can decompose this into the relevant contribution of each of the bound superradiant states:
\begin{eqnarray}
    \Phi^{(0)} = \sum_{\levnlm} \sqrt{\frac{N_{n\ell m}}{2 \mu}} \, \psi_{n\ell m} \,  e^{-i \omega_{n\ell m} t} + {\rm c.c.}  \, ,
\end{eqnarray}
where $N_{n\ell m}$ are the occupation numbers, $\psi_{n\ell m}$ the complex wave functions of each state $\levnlm$, and ${\rm c.c.}$ denotes the complex conjugate. Plugging this solution in the source term, one finds terms proportional to
\begin{multline}\label{eq:rhs_expand}
     -\frac{\lambda}{6} ({\Phi^{(0)}})^3  = -\frac{\lambda}{6}\sum_{n\ell m} \left(\frac{N_{n\ell m}}{2 \mu} \right)^{3/2} \left[\psi_{n\ell m}^3 e^{-3 i \omega_{n\ell m} t} + \right.  \\[5pt] \left. 3 \psi_{n\ell m}^2 \psi_{n\ell m}^* e^{- i \omega_{n\ell m} t} + {\rm c.c.} \right] +  \\[5pt]
     -\frac{\lambda}{6}\sum_{n\ell m}\sum_{(n\ell m)^\prime} \left(\frac{N_{n\ell m} \sqrt{N_{(n\ell m)^\prime}}}{(2 \mu)^{3/2}} \right)  \times   \\[5pt] \left[ 3 \psi_{n\ell m}^2 \psi_{(n\ell m)^\prime} e^{-i (2\omega_{n\ell m} + \omega_{(n\ell m)^\prime}) t}   \right. \\[5pt] \left. + 6 \psi_{n\ell m} \psi_{n\ell m}^* \psi_{(n\ell m)^\prime} e^{-i \omega_{(n\ell m)^\prime} t} + \right.  \\[5pt] \left. 3  \psi_{n\ell m}^2 \psi_{(n\ell m)^\prime}^* e^{-i (2\omega_{n\ell m} - \omega_{(n\ell m)^\prime}) t}  + {\rm c.c.} \right] +  \\[5pt] 
     -\frac{\lambda}{6}\sum_{n\ell m}\sum_{(n\ell m)^\prime}\sum_{(n\ell m)^{\prime\prime}} \left(\frac{\sqrt{N_{n\ell m} N_{(n\ell m)^\prime} N_{(n\ell m)^{\prime\prime}}}}{(2 \mu)^{3/2}} \right)   \times \\[5pt]
     \left[6  \psi_{n\ell m}  \psi_{(n\ell m)^\prime}  \psi_{(n\ell m)^{\prime\prime}} e^{-i (\omega_{n\ell m} + \omega_{(n\ell m)^\prime}+ \omega_{(n\ell m)^{\prime\prime}}) t} + \right. \\[5pt] \left. 6  \psi_{n\ell m}  \psi_{(n\ell m)^\prime}  \psi_{(n\ell m)^{\prime\prime}}^* e^{-i (\omega_{n\ell m} + \omega_{(n\ell m)^\prime}- \omega_{(n\ell m)^{\prime\prime}}) t} + {\rm c.c. }\right]
\end{multline}
where we have explicitly expanded the summations such that $n\ell m \neq (n\ell m)^\prime \neq (n \ell m)^{\prime \prime}$. One can directly identify the role of each term in the above summation by analyzing the induced energy of the harmonic: 
\begin{itemize}
    \item $\psi_{n\ell m}^3 e^{-3 i \omega_{n\ell m} t}$: Production of a relativistic  axion with energy $\omega = 3\omega_{n\ell m} \simeq 3 \mu$ from a $3 \rightarrow 1$ scattering process, with the three axions in the $n\ell m$ state. 
    \item $\psi_{n\ell m}^2 \psi_{(n\ell m)^\prime} e^{-i (2\omega_{n\ell m} + \omega_{(n\ell m)^\prime}) t}$: Production of a relativistic axion with energy $\omega \simeq 3 \mu$ from a $3 \rightarrow 1$ scattering process, with the two axions in the $n\ell m$ state and one in the $(n\ell m)^\prime$. 
    \item $ \psi_{n\ell m}  \psi_{(n\ell m)^\prime}  \psi_{(n\ell m)^{\prime\prime}} e^{-i (\omega_{n\ell m} + \omega_{(n\ell m)^\prime}+ \omega_{(n\ell m)^{\prime\prime}}) t} $: Production of a relativistic axion with energy $\omega \simeq 3 \mu$ from a $3 \rightarrow 1$ scattering process, with one axion in the $n\ell m$ state, one in the $(n\ell m)^\prime$, and one in the $(n\ell m)^{\prime\prime}$. 
    \item $ \psi_{n\ell m}^2 \psi_{n\ell m}^* e^{- i \omega_{n\ell m} t}$: Self-energy correction to the $n\ell m$ state.
    \item $\psi_{n\ell m} \psi_{n\ell m}^* \psi_{(n\ell m)^\prime} e^{-i \omega_{(n\ell m)^\prime} t}$: Correction to the energy of state $(n\ell m)^\prime$ from mixing with the state $n\ell m$.
    \item $\psi_{n\ell m}^2 \psi_{(n\ell m)^\prime}^* e^{-i (2\omega_{n\ell m} - \omega_{(n\ell m)^\prime}) t}$: Two-to-two scattering process, with the two initial states in the $n\ell m$ quasi-bound state, and $(n\ell m)^\prime$ in the final state. The other final leg corresponds either to a non-relativistic bound axion which dissipates its energy through the horizon ($2\omega_{n\ell m} - \omega_{(n\ell m)^\prime} < \mu$), or to a out-going non-relativistic axions ($2\omega_{n\ell m} - \omega_{(n\ell m)^\prime} \geq \mu$).
    \item $\psi_{n\ell m}  \psi_{(n\ell m)^\prime}  \psi_{(n\ell m)^{\prime\prime}}^* e^{-i (\omega_{n\ell m} + \omega_{(n\ell m)^\prime}- \omega_{(n\ell m)^{\prime\prime}}) t}$: Two-to-two scattering process, with initial states $n\ell m$ and $(n\ell m)^\prime$, one final state in the $(n\ell m)^{\prime \prime}$ quasi-bound level and the other final leg corresponds either to a non-relativistic bound axion which dissipates its energy through the horizon ($\omega_{n\ell m} + \omega_{(n\ell m)^\prime} - \omega_{(n\ell m)^{\prime\prime}} < \mu$), or to out-going non-relativistic axions ($2\omega_{n\ell m} + \omega_{(n\ell m)^\prime} - \omega_{(n\ell m)^{\prime\prime}} \geq \mu$).
\end{itemize}
Note that three-to-one scattering processes are heavily suppressed (at least so long as the density remains small relative to the bosenova threshold), and can typically be neglected in the evolution of the superradiant system (see \eg~\cite{Baryakhtar:2020gao}) -- as such, we focus on the role of energy corrections and energy dissipation via two-to-two processes.

{\renewcommand{\arraystretch}{1.6}
\begin{table}
    \centering
    \begin{tabular}{|c|c|}
    \hline
      Process & Rate  \\ \hline
      $\gamma_{211 \times 211}^{322 \times {\rm BH}}$   & $4\times10^{-7} \, \alpha^{11} \, (M_{p}/f_a)^4 \, \tilde{r}_p \, \mu$ \\ \hline
      $\gamma_{211 \times 411}^{322 \times {\rm BH}}$ & $2.5\times10^{-8} \, \alpha^{11} \, (M_p / f_a)^4 \, \tilde{r}_p \, \mu $ \\ \hline
      $\gamma_{411 \times 411}^{322 \times {\rm BH}}$ & $1.6\times10^{-11} \, \alpha^{11} \, (M_p / f_a)^4 \, \tilde{r}_p \, \mu $
      \\ \hline
      $\gamma_{211 \times 211}^{422 \times {\rm BH}}$ & $1.5\times10^{-7} \, \alpha^{11} \, (M_p / f_a)^4 \, 
      \tilde{r}_p \, \mu $ \\ \hline
      $\gamma_{411 \times 411}^{422 \times {\rm BH}}$ & $2.3\times10^{-7} \, \alpha^{7} \, (M_p / f_a)^4 \, \tilde{r}_p \, \mu $ \\ \hline
  
     $\gamma_{211 \times 422}^{433 \times {\rm BH}}$ & $7.8\times10^{-11} \, \alpha^{7} \, (M_p / f_a)^4 \, \tilde{r}_p \, \mu $ 
      \\ \hline
      $\gamma_{211 \times 322}^{433 \times {\rm BH}}$ & $9.1\times10^{-8} \, \alpha^{11} \, (M_p / f_a)^4 \, \mu $ \\ \hline
      $\gamma_{322 \times 411}^{433 \times {\rm BH}}$   & $3.8 \times 10^{-11} \, \alpha^7 \, (M_p / f_a)^4\, \tilde{r}_p \, \mu $  \\ \hline
       $\gamma_{211 \times 411}^{422 \times {\rm BH}}$   & $3.2 \times 10^{-11} \, \alpha^7 \, (M_p / f_a)^4\, \tilde{r}_p \, \mu $  \\ \hline
       $\gamma_{411 \times 422}^{433 \times {\rm BH}}$   & $2.3 \times 10^{-11} \, \alpha^7 \, (M_p / f_a)^4\, \tilde{r}_p \, \mu $  \\ \hline
      $\gamma_{211 \times 311}^{322 \times {\rm BH}}$   & $3.1 \times 10^{-10} \, \alpha^7 \, (M_p / f_a)^4\, \tilde{r}_p \, \mu $  \\ \hline
       $\gamma_{311 \times 311}^{322 \times {\rm BH}}$   & $1.6 \times 10^{-10} \, \alpha^7 \, (M_p / f_a)^4\, \tilde{r}_p \, \mu $  \\ \hline
       $\gamma_{211 \times 311}^{422 \times {\rm BH}}$   & $2.7 \times 10^{-7} \, \alpha^{11} \, (M_p / f_a)^4\, \tilde{r}_p \, \mu $  \\ \hline
       $\gamma_{311 \times 311}^{422 \times {\rm BH}}$   & $1.7 \times 10^{-11} \, \alpha^{11} \, (M_p / f_a)^4\, \tilde{r}_p \, \mu $  \\ \hline
       $\gamma_{311 \times 322}^{433 \times {\rm BH}}$   & $7.0 \times 10^{-8} \, \alpha^{11} \, (M_p / f_a)^4\, \tilde{r}_p \, \mu $  \\ \hline
       $\gamma_{311 \times 411}^{322 \times {\rm BH}}$   & $1.9 \times 10^{-10} \, \alpha^7 \, (M_p / f_a)^4\, \tilde{r}_p \, \mu $  \\ \hline
       $\gamma_{311 \times 411}^{422 \times {\rm BH}}$   & $3.8 \times 10^{-13} \, \alpha^7 \, (M_p / f_a)^4\, \tilde{r}_p \, \mu $  \\ \hline
       $\gamma_{311 \times 422}^{433 \times {\rm BH}}$   & $7.7 \times 10^{-12} \, \alpha^7 \, (M_p / f_a)^4\, \tilde{r}_p \, \mu $  \\ \hline
      $\gamma_{322 \times 322}^{211 \times \infty}$   & $10^{-8} \, \alpha^8 \, (M_p / f_a)^4 \, \mu $ \\ \hline
      $\gamma_{322 \times 322}^{\rm GW}$ & $3\times10^{-8} \, \alpha^{18} \, \mu $ \\ \hline  
      $\gamma_{411 \times 411}^{211 \times \infty}$ & $1.7\times10^{-9} \, \alpha^{8} \, (M_p / f_a)^4 \, \mu $ \\ \hline
      $\gamma_{411 \times 433}^{211 \times \infty}$ & $1.1\times10^{-10} \, \alpha^{8} \, (M_p / f_a)^4 \, \mu $ \\ \hline
      $\gamma_{422 \times 422}^{211 \times \infty}$ & $1.6\times10^{-9} \, \alpha^{8} \, (M_p / f_a)^4 \, \mu $ \\ \hline
      $\gamma_{422 \times 433}^{211 \times \infty}$ & $6.1\times10^{-10} \, \alpha^{8} \, (M_p / f_a)^4 \, \mu $ \\ \hline               
      $\gamma_{322 \times 411}^{211 \times \infty}$ & $3.8\times10^{-9} \, \alpha^{8} \, (M_p / f_a)^4  \, \mu $ \\ \hline 
      $\gamma_{411 \times 422}^{211 \times \infty}$ & $2.2\times 10^{-9} \, \alpha^{8} \, (M_p / f_a)^4 \, \mu $ \\ \hline
      $\gamma_{433 \times 433}^{211 \times \infty}$ & $9.2\times10^{-11} \, \alpha^{8} \, (M_p / f_a)^4 \, \mu $ \\ \hline
      $\gamma_{322 \times 433}^{211 \times \infty}$ & $2.6\times10^{-9} \, \alpha^{8} \, (M_p / f_a)^4 \, \mu $ \\ \hline
        $\gamma_{422 \times 322}^{211 \times \infty}$ & $1.6\times10^{-8} \, \alpha^{8} \, (M_p / f_a)^4 \, \mu $ \\ \hline
      $\gamma_{311 \times 311}^{211 \times \infty}$ & $5.1\times10^{-8} \, \alpha^{8} \, (M_p / f_a)^4 \, \mu $ \\ \hline
      $\gamma_{311 \times 322}^{211 \times \infty}$ & $1.2\times10^{-8} \, \alpha^{8} \, (M_p / f_a)^4 \, \mu $ \\ \hline
      $\gamma_{311 \times 411}^{211 \times \infty}$ & $1.9\times10^{-8} \, \alpha^{8} \, (M_p / f_a)^4 \, \mu $ \\ \hline
      $\gamma_{311 \times 422}^{211 \times \infty}$ & $7.0\times10^{-9} \, \alpha^{8} \, (M_p / f_a)^4 \, \mu $ \\ \hline
      $\gamma_{311 \times 433}^{211 \times \infty}$ & $2.2\times10^{-10} \, \alpha^{8} \, (M_p / f_a)^4 \, \mu $ \\ \hline
         $\gamma_{211^3}^{\infty}$ & $1.5 \times 10^{-8} \, \alpha^{21} \, (M_p / f_a)^4 \, \mu$ \\ \hline
         $\gamma_{211 \times 211}^{\rm GW}$ & $10^{-2} \,  \alpha^{14} \, \mu $ \\ \hline
      $\gamma_{322 \rightarrow 211}^{\rm GW}$ &  $5 \times 10^{-6} \, \alpha^{10} \, \mu $\\ \hline
    \end{tabular}
    \caption{\label{tab:rates} Non-relativistic limit of the scattering rates driving the growth and evolution of the superradiant system (up to $n=4$). Relativistic corrections to some of these rates as a function of $\alpha$ are shown in Fig.~\ref{fig:rel_rates}. }
\end{table}

{\renewcommand{\arraystretch}{1.6}
\begin{table}
    \centering
    \begin{tabular}{|c|c|}
    \hline
      Process & Rate  \\ \hline
      $\gamma_{322 \times 322}^{544 \times {\rm BH}}$   & $1.9 \times 10^{-9} \, \alpha^{11} \, (M_p / f_a)^4 \, \tilde{r}_p \mu $ \\ \hline
     $\gamma_{322 \times 411}^{533 \times {\rm BH}}$   & $2 \times 10^{-8} \, \alpha^{11} \, (M_p / f_a)^4 \,  \tilde{r}_p \, \mu $ \\ \hline
     $\gamma_{322 \times 422}^{544 \times {\rm BH}}$   & $1.3 \times 10^{-11} \, \alpha^{11} \, (M_p / f_a)^4 \,  \tilde{r}_p \, \mu $\\ \hline
      $\gamma_{411 \times 411}^{522 \times {\rm BH}}$   & $9.0 \times 10^{-11} \, \alpha^{11} \, (M_p / f_a)^4 \,  \tilde{r}_p \, \mu$ \\ \hline
      $\gamma_{322 \times 522}^{544 \times {\rm BH}}$   &  $3.4 \times 10^{-12} \, \alpha^{7} \, (M_p / f_a)^4 \,  \tilde{r}_p \, \mu$   \\ \hline
      $\gamma_{422 \times 422}^{544 \times {\rm BH}}$   &  $2.3 \times 10^{-9} \, \alpha^{11} \, (M_p / f_a)^4 \,  \tilde{r}_p \, \mu$ \\ \hline
      $\gamma_{422 \times 522}^{544 \times {\rm BH}}$   & $3.7 \times 10^{-14} \, \alpha^{7} \, (M_p / f_a)^4 \,  \tilde{r}_p \, \mu$   \\ \hline
      $\gamma_{522 \times 522}^{544 \times {\rm BH}}$   &  $2.7 \times 10^{-13} \, \alpha^{7} \, (M_p / f_a)^4 \,  \tilde{r}_p \, \mu$ \\ \hline
    $\gamma_{211 \times 211}^{522 \times {\rm BH}}$   &  $7.5 \times 10^{-8} \, \alpha^{11} \, (M_p / f_a)^4 \,  \tilde{r}_p \, \mu$ \\ \hline
    $\gamma_{411 \times 411}^{522 \times {\rm BH}}$   &  $9.0 \times 10^{-11} \, \alpha^{11} \, (M_p / f_a)^4 \,  \tilde{r}_p \, \mu$ \\ \hline
    $\gamma_{211 \times 311}^{522 \times {\rm BH}}$   &  $1.0 \times 10^{-7} \, \alpha^{11} \, (M_p / f_a)^4 \,  \tilde{r}_p \, \mu$ \\ \hline
    $\gamma_{211 \times 411}^{522 \times {\rm BH}}$   &  $9.9 \times 10^{-8} \, \alpha^{7} \, (M_p / f_a)^4 \,  \tilde{r}_p \, \mu$ \\ \hline
    $\gamma_{211 \times 511}^{522 \times {\rm BH}}$   &  $6.6 \times 10^{-12} \, \alpha^{7} \, (M_p / f_a)^4 \,  \tilde{r}_p \, \mu$ \\ \hline
    $\gamma_{211 \times 522}^{533 \times {\rm BH}}$   &  $2.6 \times 10^{-11} \, \alpha^{7} \, (M_p / f_a)^4 \,  \tilde{r}_p \, \mu$ \\ \hline
    $\gamma_{211 \times 533}^{544 \times {\rm BH}}$   &  $4.6 \times 10^{-13} \, \alpha^{7} \, (M_p / f_a)^4 \,  \tilde{r}_p \, \mu$ \\ \hline      
      $\gamma_{422 \times 544}^{322 \times \infty}$   & $2.5\times10^{-11} \, \alpha^{8} \, (M_p / f_a)^4 \, \mu $ \\ \hline
      $\gamma_{433 \times 544}^{322 \times \infty}$   &  $7.8\times10^{-10} \, \alpha^{8} \, (M_p / f_a)^4 \, \mu $ \\ \hline
      $\gamma_{522 \times 544}^{322 \times  \infty}$   &  $2.2\times10^{-11} \, \alpha^{8} \, (M_p / f_a)^4 \, \mu $ \\ \hline
      $\gamma_{533 \times 544}^{322 \times \infty}$   &   $1.8\times10^{-10} \, \alpha^{8} \, (M_p / f_a)^4 \, \mu $ \\ \hline
      $\gamma_{544 \times 544}^{322 \times \infty}$   &   $4.3\times10^{-11} \, \alpha^{8} \, (M_p / f_a)^4 \, \mu $ \\ \hline
      $\gamma_{422 \times 533}^{322 \times \infty}$   &  $1.2\times10^{-9} \, \alpha^{8} \, (M_p / f_a)^4 \, \mu $ \\ \hline
      $\gamma_{433 \times 533}^{322 \times \infty}$   &   $2.8\times 10^{-9} \, \alpha^{8} \, (M_p / f_a)^4 \, \mu $ \\ \hline
      $\gamma_{522 \times 533}^{322 \times \infty}$   &  $4.4\times 10^{-10} \, \alpha^{8} \, (M_p / f_a)^4 \, \mu $ \\ \hline
      $\gamma_{533 \times 533}^{322 \times \infty}$   &  $1.3\times 10^{-9} \, \alpha^{8} \, (M_p / f_a)^4 \, \mu $ \\ \hline
      $\gamma_{433 \times 522}^{322 \times \infty}$   &  $6.3\times 10^{-10} \, \alpha^{8} \, (M_p / f_a)^4 \, \mu $ \\ \hline
      $\gamma_{422 \times 522}^{322 \times \infty}$   &  $1.6\times 10^{-9} \, \alpha^{8} \, (M_p / f_a)^4 \, \mu $ \\ \hline
      $\gamma_{522 \times 522}^{322 \times \infty}$   &  $1.6\times 10^{-10} \, \alpha^{8} \, (M_p / f_a)^4 \, \mu $ \\ \hline
    $\gamma_{322 \times 511}^{211 \times \infty}$   &  $1.7\times 10^{-9} \, \alpha^{8} \, (M_p / f_a)^4 \, \mu $ \\ \hline
    $\gamma_{422 \times 522}^{211 \times \infty}$   &  $3.7\times 10^{-9} \, \alpha^{8} \, (M_p / f_a)^4 \, \mu $ \\ \hline

    \end{tabular}
    \caption{\label{tab:rates_n5}  Leading order behavior of a subset of scattering rates arising at $n=5$, computed in the hydrogen-like limit. Relativistic corrections to some of these rates as a function of $\alpha$ are shown in Fig.~\ref{fig:rates_n5}. }
\end{table}

\subsubsection{Hydrogen-like Limit}

We now turn toward solving for the first order correction, $\Phi^{(1)}$, focusing specifically on the semi-analytic derivation in the hydrogen-like limit; this subsection is largely a review of what  has previously been outlined in~\cite{Baryakhtar:2020gao}, but which we include both for the sake of completeness, and to provide the reader with intuition which may serve as a useful guide as we later move into the fully relativistic limit.

Let us begin by noting that the discrete frequency spectrum of the source term simplifies the structure of the first order correction, such that $\Phi^{(1)}$ can generically be written as
\begin{multline}\label{eq:phi1_expd}
\Phi^{(1)} = \sum_{bd} \psi_{bd} e^{- i \omega_{bd} t} + \sum_{{ nre}} \psi_{ nre} e^{- i \omega_{nre} t} \\[5pt] + \sum_{ re} \psi_{ re} e^{- i \omega_{re} t} +  \sum_{\levnlm}  \psi^{(1)}_{n\ell m} e^{- i \omega_{n\ell m} t}\, + {\rm cc} \, ,
\end{multline}
where the sums run over the bound states which dissipate their energy through the horizon ($bd$; with $\omega_{bd} < \mu$), the emission of non-relativistic states (nre; with $\omega_{nre} \simeq \mu + \epsilon$ with $0 \ll \epsilon \ll \mu$), the emission of relativistic states (re; with $\omega_{re} \simeq 3 \mu$), and the first order corrections to the bound superradiant states $\levnlm$. Note that the definition above is such that each of the  contributions in Eq.~\ref{eq:phi1_expd} is inherently orthogonal, and thus any contribution can be isolated by projecting onto the energy mode of interest. We will begin by analyzing the self-energy corrections included in $\psi^{(1)}_{n\ell m}$.

For energy eigenstates of the system, $\psi_{n\ell m}$, the leading order correction $\psi_{n\ell m}^{(1)}$ from the source  will serve to modify the energy of the state itself. This can be seen by adopting an Ansatz for a given state $\psi_{n\ell m}^{(1)} = \sqrt{\frac{N_{n\ell m}}{2 \mu}} \psi_{n\ell m} e^{-i \lambda \delta \omega_{n\ell m}} $, and inserting this into the equation of motion -- in the hydrogen-like limit this yields
\begin{multline}
    \left[-\frac{1}{r^2}\frac{d}{dr}\left(r^2 \frac{d}{dr}\right) - \frac{2 \alpha \mu}{r} + \frac{\ell (\ell + 1)}{r^2} + (\mu^2 -\omega^2) \right. \\ \left. - 2 \lambda \omega \delta\omega \right]  \psi_{n\ell m}  = -\lambda \,  \psi_{n\ell m} \sum_{(n\ell m)^\prime}\frac{N_{(n\ell m)^\prime}}{2 \mu} |\psi_{(n\ell m)^\prime}|^2  \, ,
\end{multline}
where we have only kept the leading order correction. The standard operator acting on the basis state vanishes, i.e. $\mathcal{O}\psi_{n\ell m} = 0$ with $\mathcal{O}\equiv -\nabla^2 - 2\alpha \mu / r + \ell (\ell + 1)/r^2 + (\mu^2-\alpha^2)$, allowing one to isolate the energy shift by projecting onto $\psi_{n\ell m}^*$, \ie 
\begin{multline}
    - 2 \omega \delta\omega \int dV \psi_{n\ell m} \psi_{n\ell m}^*  = - 2 \omega \delta\omega = \\ -\frac{\xi}{2 \mu}\sum_{\levnlm^\prime} N_{(n\ell m)^\prime}  \int dV \,  |\psi_{n\ell m}|^2 |\psi_{(n\ell m)^\prime}|^2  \, ,
\end{multline}
where we have assumed a normalization such that $\int dV |\psi_{n\ell m}|^2 = 1$, and $\xi = \frac{1}{2} / 1$ for identical/different initial states.

Below, we provide the leading order corrections to the energies of a few of the relevant quasi-bound states:
\small
\begin{eqnarray}
    \lambda (\delta \omega_{211}) &\simeq& -\frac{\lambda }{4 \omega_{211} \mu} \alpha^3 \mu^3 \left[ 4.6 \times 10^{-4} \, N_{211} + 1.4 \times 10^{-4} \, N_{322} \nonumber \right. \\ &+& \left. \cdots \right] \nonumber \\
    &\simeq & -\alpha^5 \mu \left(\frac{M_{\rm pl}}{f_a} \right)^2 \left[1.2 \times 10^{-4} \epsilon_{211}  + 3.5 \times 10^{-5} \epsilon_{322} \nonumber \right.  \\  &+& \left. \cdots \right] \\ 
     \lambda (\delta \omega_{322}) &\simeq&  -\frac{\lambda }{4 \omega_{322} \mu} \alpha^3 \mu^3 \left[1.4 \times 10^{-4} N_{211} + 5.7 \times 10^{-5} N_{322}  \nonumber \right. \\ &+& \left.   3.1 \times 10^{-5} N_{433} +  3.9 \times 10^{-6} N_{544} + \cdots  \right] \nonumber \\
     &\simeq& -\alpha^5 \mu \left(\frac{M_{\rm pl}}{f_a} \right)^2 \left[3.5 \times 10^{-5} \epsilon_{211}  \nonumber \right. \\ &+& \left. 1.4 \times 10^{-5} \epsilon_{322} + 7.7 \times 10^{-6}  \epsilon_{433} \nonumber \right. \\ &+& \left. 9.7 \times 10^{-7} \epsilon_{544} + \cdots \right] \\ 
     \lambda (\delta \omega_{433}) &\simeq& -\frac{\lambda }{4 \omega_{433} \mu} \alpha^3 \mu^3 \left[ 9.0 \times 10^{-6} \, N_{211} + 3.1 \times 10^{-5} \, N_{322} \nonumber \right. \\ &+& \left. 1.3 \times 10^{-5} \, N_{433} + 9.6 \times 10^{-6} \, N_{544} + \cdots \right] \nonumber \\
    &\simeq & -\alpha^5 \mu \left(\frac{M_{\rm pl}}{f_a} \right)^2 \left[2.2 \times 10^{-6} \epsilon_{211}  \nonumber \right.  \\  &+& \left. 7.7 \times 10^{-6} \epsilon_{322}  +3.3 \times 10^{-6} \epsilon_{433} \nonumber \right.  \\  &+& \left. 2.4 \times 10^{-6} \epsilon_{544}  \cdots \right]  \\
     \lambda (\delta \omega_{544}) &\simeq& -\frac{\lambda }{4 \omega_{544} \mu} \alpha^3 \mu^3 \left[ 3.1 \times 10^{-7} \, N_{211} + 3.9 \times 10^{-6} \, N_{322} \nonumber \right. \\ &+& \left. 9.6 \times 10^{-6} \, N_{433} + 4.3 \times 10^{-6} \, N_{544} + \cdots \right] \nonumber \\
    &\simeq & -\alpha^5 \mu \left(\frac{M_{\rm pl}}{f_a} \right)^2 \left[7.9 \times 10^{-8} \epsilon_{211}  \nonumber \right.  \\  &+& \left. 9.7 \times 10^{-7} \epsilon_{322}  +2.4 \times 10^{-6} \epsilon_{433} \nonumber \right.  \\  &+& \left. 1.1 \times 10^{-6} \epsilon_{533}  \cdots \right] \\
    &\cdots& \nonumber
\end{eqnarray}
\normalsize
where we have restricted our attention to a few of the relevant quasi-bound states, and defined normalized occupation numbers $\epsilon_X \equiv N / (GM^2)$. One can generalize this approach in a straightforward manner to compute the relativistic corrections valid at larger $\alpha$ -- we find that these corrections to the leading coefficients of each term are typically at the $\mathcal{O}(1)$ level for the $\alpha$ values of interest.

In order to determine if the frequency shifts play an important role in the spin down one can compute the ratio of the energy shift induced by self interactions to the leading order energy shift induced by gravity $(\delta \omega)_{\rm grav} \simeq - \alpha^2 \mu / (2 n^2)$ (note that small perturbations at this level to the real component of the energy eigenstates will not significantly alter the superradiant growth, but, as discussed below, such corrections due play an important role in setting the strength of the self-interaction scattering processes -- it is for this reason that we use this as a reasonable, and conservative, threshold for comparison). The maximum shift obtainable can be computed using the characteristic occupation number required for a bosenova (where higher order self-interactions would necessarily disrupt the growth of the cloud)~\cite{Baryakhtar:2020gao}
 \begin{eqnarray}\label{eq:bnova}
  N_{bn} \sim    \, G M^2 \, \left(\frac{n}{2} \right)^4 \, \frac{1024 \pi f_a^2}{9 \alpha^3 \, M_{\rm Pl}^2} \, ,
\end{eqnarray} 
which in normalized units is given by
\begin{eqnarray}
    \epsilon_{bn} \simeq \frac{357}{\alpha^3}  \,  \left(\frac{n}{2} \right)^4 \left(\frac{f_a}{M_{\rm Pl}}\right)^2  \, .
\end{eqnarray}
Comparing these corrections, we find that self-interaction induced corrections (even including an order one relativistic correction) could be as large as $\sim \mathcal{O}(10\%) (\delta \omega)_{\rm grav}$, but only for occupation numbers saturated to the bosenova threshold. As we show below, the occupation numbers associated to a bosenova are in general not expected to be realized for the range of parameter space studied here, and thus we conclude that the energy shifts are not expected to significantly alter the evolution of spin down (however they can play an important role in the search for narrow frequency gravitational wave signatures emitted from the clouds, see~\cite{Collaviti:2024mvh,Omiya:2024xlz}). 

Determining the rate at which self interactions dissipate energy, either into bound oscillations (which lose energy through the horizon) or to infinity, requires directly solving for the value of $\Psi^{(1)}$ at the horizon or at large radial distances, and in turn computing
\begin{eqnarray}\label{eq:e_horiz}
   \frac{dE}{dt} = \int_{\partial \Sigma} T^\mu_0 n_\mu d
   \Omega
\end{eqnarray}
where the surface $\partial \Sigma$ is taken to be either the outer horizon or a two-sphere at large $r$, $d\Omega = \rho^2 \sin\theta d\theta d\phi$, the normal vector $n_\mu = -\delta^1_\mu$, and $T$ the stress energy tensor. 

Energy losses through the horizon are dominated by $\psi_{bd}$; in the hydrogen-like limit, this function can be determined by decomposing the wavefunction into a complete set of basis states of the operator $\mathcal{O}^\prime \equiv -\nabla^2 - 2\alpha \mu / r + \ell (\ell + 1)/r^2$, i.e.  
\begin{eqnarray}
    \Phi_1 = \sum_{n\ell m} c_{n\ell m} \psi_{n\ell m} + \sum_{\ell m} \int dk \,  c(k) \, \psi_{k\ell m} \, ,
\end{eqnarray}
where the sum runs over the bound states $\psi_{n\ell m}$ (whose wavefunctions are provided in Eq.~\ref{eq:radial_bnd_NR}), and the integration is performed over the continuum states $\psi_{k\ell m} = Y_\ell^m R_{k\ell}$, with the radial function defined by
\begin{multline}
    R_{k\ell} = \frac{2 k e^{\pi / (2 k a_0)} |\Gamma(\ell + 1 - i / (k a_0))|}{(2\ell +1)!} e^{-i k r} \\ \times (2 k r)^\ell {}_1F_1(i/(k a_0) + \ell +1, 2 \ell + 2, 2 i k r) \, .
\end{multline}
Here, $a_0 \equiv (\alpha \mu)^{-1}$, $\Gamma$ is the Gamma function,  and ${}_1F_1$ is the confluent hypergeometric function of the first kind. Using the fact that $\mathcal{O}\psi_{n\ell m} = -\left(\alpha^2 \mu^2 / n^2 \right) \psi_{n\ell m}$ and $\mathcal{O}^\prime\psi_{k\ell m} = k^2 \psi_{k\ell m}$, one can transform the equation of motion for $\Psi^{(1)}$ into
\begin{multline}
    \sum_{n\ell m}\left(\mu^2 - \omega^2 - \frac{\alpha^2 \mu^2}{n^2} \right) c_{n\ell m} \, \psi_{n\ell m} \, - \\[8pt] \sum_{\ell m}  \int dk \, \left(\mu^2 - \omega^2 + k^2 \right) \,  c_{\ell m}(k) \, \psi_{k\ell m} = -\frac{\lambda}{6} \Phi^{(0)^3} \, .
\end{multline}
The orthonormality of the basis states
\begin{eqnarray}
    \int d^3 r \, \psi_{n\ell m}^* \psi_{n^\prime \ell^\prime m^\prime} &=& \delta_{n,n^\prime} \delta_{\ell,\ell^\prime} \delta_{m,m^\prime} \\
     \int d^3 r \, \psi_{k\ell m}^*  \psi_{k\ell m} &=& 2\pi \delta(k^\prime - k) \, \delta_{\ell,\ell^\prime} \delta_{m,m^\prime} \\
     \int d^3 r \, \psi_{k\ell m}^* \psi_{n^\prime \ell^\prime m^\prime} &=& 0 \, 
\end{eqnarray}
allows one to project onto $\psi_{n^\prime\ell^\prime m^\prime}^*$ or $ \psi_{k^\prime\ell^\prime m^\prime}^*$ and extract the coefficients, yielding
\begin{eqnarray} 
    c_{n\ell m} &=& \frac{\xi}{\left(\mu^2 - \omega^2 - \frac{\alpha^2 \mu^2}{n^2} \right)}  \,  \int \, d^3r \, \psi_{1} \psi_{2} \psi_{3}^* \psi_{n\ell m}^*  \, \label{eq:cnlm} \\
    c_{\ell m}(k) &=& \frac{\xi}{2\pi \left(\mu^2 - \omega^2 + k^2 \right)}  \, \int \, d^3 r \, \psi_{1} \psi_{2} \psi_{3}^* \psi_{k\ell m}^*
\end{eqnarray}
with $\xi = \frac{1}{2} / 1$ for identical/different initial states, and $\psi_i$ ($i=1,2,3$) being the superradiant states entering the source term (\ie those states for which $\omega \equiv \omega_1 + \omega_2 - \omega_3$). Note that for bound oscillations, $\mu > \omega$, so $\left(\mu^2 - \omega^2 - \frac{\alpha^2 \mu^2}{n^2} \right)$ can vanish, violating the perturbative expansion. In general, one can imagine expanding the original operator $\mathcal{O}$ to include the leading order relativistic correction -- this induces corrections to momentum eigenvalues, thereby `de-tuning' this resonance and restoring the validity of the expansion. The same issue does not occur for $c_{\ell m}(k)$ in this case, as $(\mu^2 - \omega^2), k^2 > 0$.

\begin{figure*}
    \includegraphics[width=0.49\textwidth]{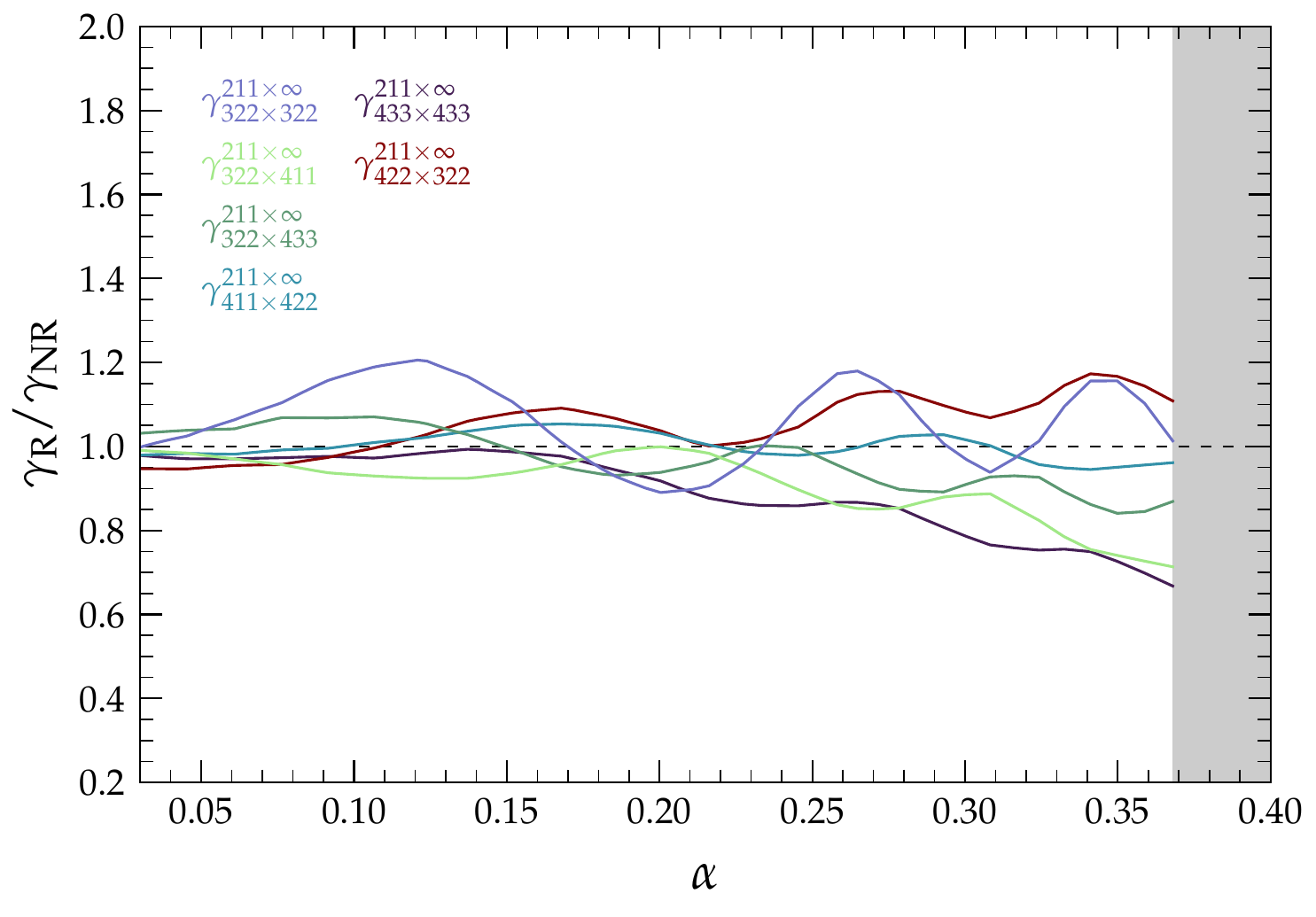}
    \includegraphics[width=0.49\textwidth]{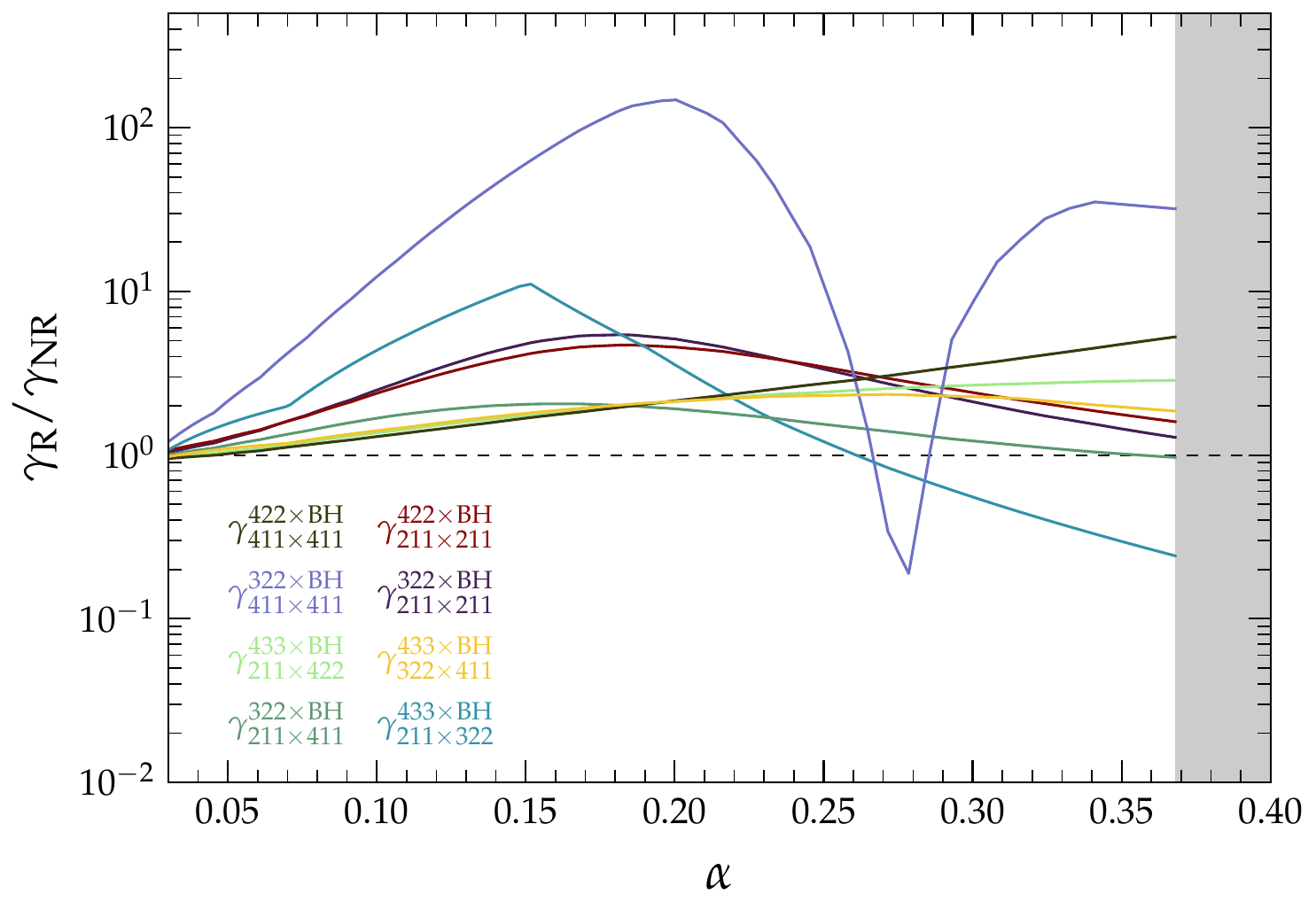}
    \caption{Ratio of the relativistic scattering rates $\gamma_{\rm R}$, computed using the Green's function method, to the non-relativistic limit $\gamma_{\rm NR}$ shown in Table~\ref{tab:rates} for the various rates at $n \leq 4$ (for the sake of clarity we only show a subset of the scattering rates). The values of $\alpha$ for which all states are not superradiant is highlighted in gray. Results are computing using $\tilde{a} = 0.95$.}
    \label{fig:rel_rates}
\end{figure*}

In the limit $r \rightarrow r_+$, $\Psi^{(1)}$ will be dominated by the $\ell=m=0$ components of $\psi_{bd}$ (since the continuum spectrum and the higher-$\ell$ components are suppressed in the limit $r \rightarrow r_+$). Neglecting all other terms, one can compute the energy losses via bound oscillations using Eq.~\ref{eq:e_horiz}, which yields
\begin{eqnarray}\label{eq:dedt_bnd}
   \frac{dE}{dt} &=& \mathcal{A}_H |\omega|^2 |\lambda R_1(r_+)|^2 \nonumber \\ &\simeq& 4 \alpha^2 (1 + \sqrt{1-\tilde{a}^2}) \lambda^2 |R^{(1)}(r_+)|^2 \, .
\end{eqnarray}
where the horizon area $\mathcal{A}_H = 8 \pi r_g r_+$, and $R^{(1)}$ is the radial wavefunction (i.e. without the spherical harmonic). Note that the existence of a horizon is not present in the hydrogen-like limit, and thus it is natural to evaluate $R^{(1)}$ in the limit that $r \rightarrow 0$ (in the small $\alpha$ limit, this apparent inconsistency introduces a negligible correction).

Examples illustrating the radial integrand entering Eq.~\ref{eq:cnlm} are shown in the bottom panel of Fig.~\ref{fig:WF}, using both the non-relativistic and relativistic calculations outlined in the proceeding section, as a function of the normalized radial coordinate.  Here, one can see that large energy losses could be due either to large overlap of the wave functions entering the source term, or due to an enhanced rate of energy dissipation through the horizon, which occurs for large $\alpha$.

Radiative energy losses at infinity can be treated analogously. These will be dominated by the $\psi_{nre}$ (recall that we neglect $\psi_{re}$, as $3\rightarrow 1$ processes are less efficient\footnote{This can be understood by noting that the final state momentum in $3\rightarrow 1$ processes is of order the axion mass, while for non-relativistic emission it is suppressed by a factor of $\alpha$. Axion production is largely controlled by the overlap of the free and bound state wave functions near $r \sim 1/k$, and the bound state wave functions are largely suppressed in the small $r$ limit.}), which scales at large $r$ as $\psi_{nre} \sim e^{i k r} / r$. In this case,  $\ell$ and $m$ are fixed by the angular structure of the source term (e.g., the $\levthree \times \levthree \rightarrow \levtwo \times \infty$ requires a final state with $m = 3$, and $\ell = 3$ and or $\ell = 5$, with the $\ell=5$ case being suppressed). The bound states vanish at large $r$, and $c(k)$ is apparently singular at $k = \sqrt{\omega^2 - \mu^2}$; as with the bound states, however, relativistic corrections to $\omega$ will in general contain a small imaginary component -- inserting a perturbation of this form allows one to compute the $\int dk$ using the residue theorem, yielding
\small
\begin{multline}
    \psi_{nre}(r\rightarrow \infty) \sim \frac{i \xi}{2 k} \frac{e^{i k r}}{r} \left[\int dr^\prime \psi_1(r^\prime) \psi_2(r^\prime) \psi_3^*(r^\prime) \psi_{k \ell m}(r^\prime) \right]
\end{multline}
\normalsize
where we have only kept the leading order $\ell$ mode. Using Eq.~\ref{eq:e_horiz}, this yields a rate of energy loss given by
 \begin{eqnarray}\label{eq:dedt_inf}
 \frac{dE}{dt} \simeq  \frac{\omega \, \lambda^2 \, \xi^2}{2 k} \left|\int dr^\prime \psi_1(r^\prime) \psi_2(r^\prime) \psi_3^*(r^\prime) \psi_{k \ell m}(r^\prime) \right|^2  \, .
 \end{eqnarray}
 Note that this expression  is equivalent to that derived in ~\cite{Baryakhtar:2020gao}.

Using the rate of energy loss defined in Eq.~\ref{eq:dedt_bnd} and Eq.~\ref{eq:dedt_inf}, one can determine the rate of change of the occupation number of a superradiant state $\levnlm$; for the process  $\levnlm \times \left.|n^\prime \ell^\prime m^\prime\right> \rightarrow \left.|n^{\prime\prime} \ell^{\prime\prime} m^{\prime\prime} \right> \times {\rm BH}$, the rate of change is given by
\begin{eqnarray}
    \dot{N}_{n\ell m} &\simeq& - \frac{1}{\mu}\frac{dE}{dt} \nonumber \\ &\simeq& \frac{4 \alpha^2 (1 + \sqrt{1- \tilde{a}^2})\lambda^2}{\mu} \left( \frac{|R_1(r_+)|^2}{N_{n\ell m}N_{n^\prime \ell^\prime m^\prime}N_{n^{\prime\prime}\ell^{\prime\prime} m^{\prime\prime}}} \right) \nonumber \\ &\times& N_{n\ell m} N_{n^\prime \ell^\prime m^\prime} N_{n^{\prime\prime}\ell^{\prime\prime} m^{\prime\prime}} \nonumber \\[12pt] &\equiv& \Gamma_{n\ell m \times n^\prime \ell^\prime m^\prime}^{n^{\prime\prime}\ell^{\prime\prime} m^{\prime\prime} \times {\rm BH}} \times N_{n\ell m} N_{n^\prime \ell^\prime m^\prime} N_{n^{\prime\prime}\ell^{\prime\prime} m^{\prime\prime}} \, , 
\end{eqnarray}
where in the last line we have introduced the scattering rate $\Gamma$, following the notation of~\cite{Baryakhtar:2020gao} (with analogous expressions for the energy loss at infinity being derived in the same manner). As we will discuss below, we will work with normalized occupation numbers $\epsilon_i \equiv N_i / (G M^2)$; in this case, the rate of change can be expressed as $\dot{\epsilon}_i \propto \gamma_{i \times j}^{k \times {\rm BH}} {\epsilon}_i {\epsilon}_j {\epsilon}_k$ (where we have momentarily compressed each state to a single identifying index), with the relationship between $\gamma$ and $\Gamma$ being obtained via the substitution $\lambda^2 \rightarrow \alpha^4 (M_{p} / f_a)^4$. We have compiled a list of the relevant rate coefficients, which have been computed in the hydrogen-like limit using the procedure outlined above, for the leading processes involving the relevant states at $n \leq 4$ (see Table~\ref{tab:rates}) and $n = 5$ (see Table~\ref{tab:rates_n5}). We find good agreement with the results of Ref.~\cite{Baryakhtar:2020gao} when comparisons are possible, except for two processes -- the origin of this discrepancy is unclear, however we do obtain identical results using the hydrogen-like calculation and computing energy losses using Green's functions in the small-$\alpha$ limit. It is worth noting that despite the non-existence of continuum states near the Kerr black hole, the small $\alpha$ limit of the relativistic calculations do not show any notable differences with respect to the hydrogen-like limit -- we believe this is likely a consequence of the fact that the source term peaks at an increasingly large distance from the black hole in this limit, implying the produced axions are increasingly insensitive to the existence of the horizon (which in the end is what is responsible for generating the discretized spectrum).

\begin{figure*}
    \includegraphics[width=0.46\textwidth]{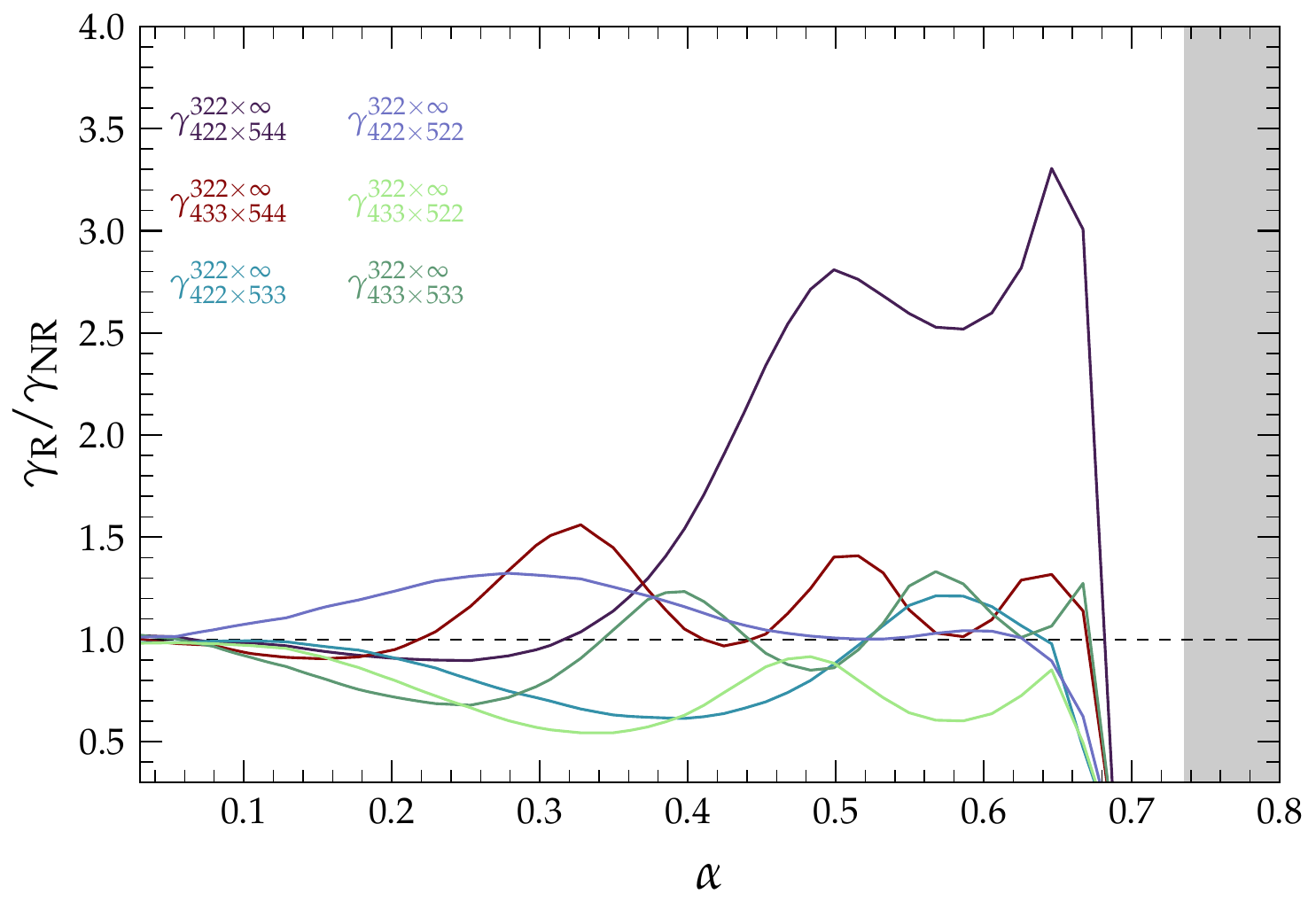}
    \includegraphics[width=0.46\textwidth]{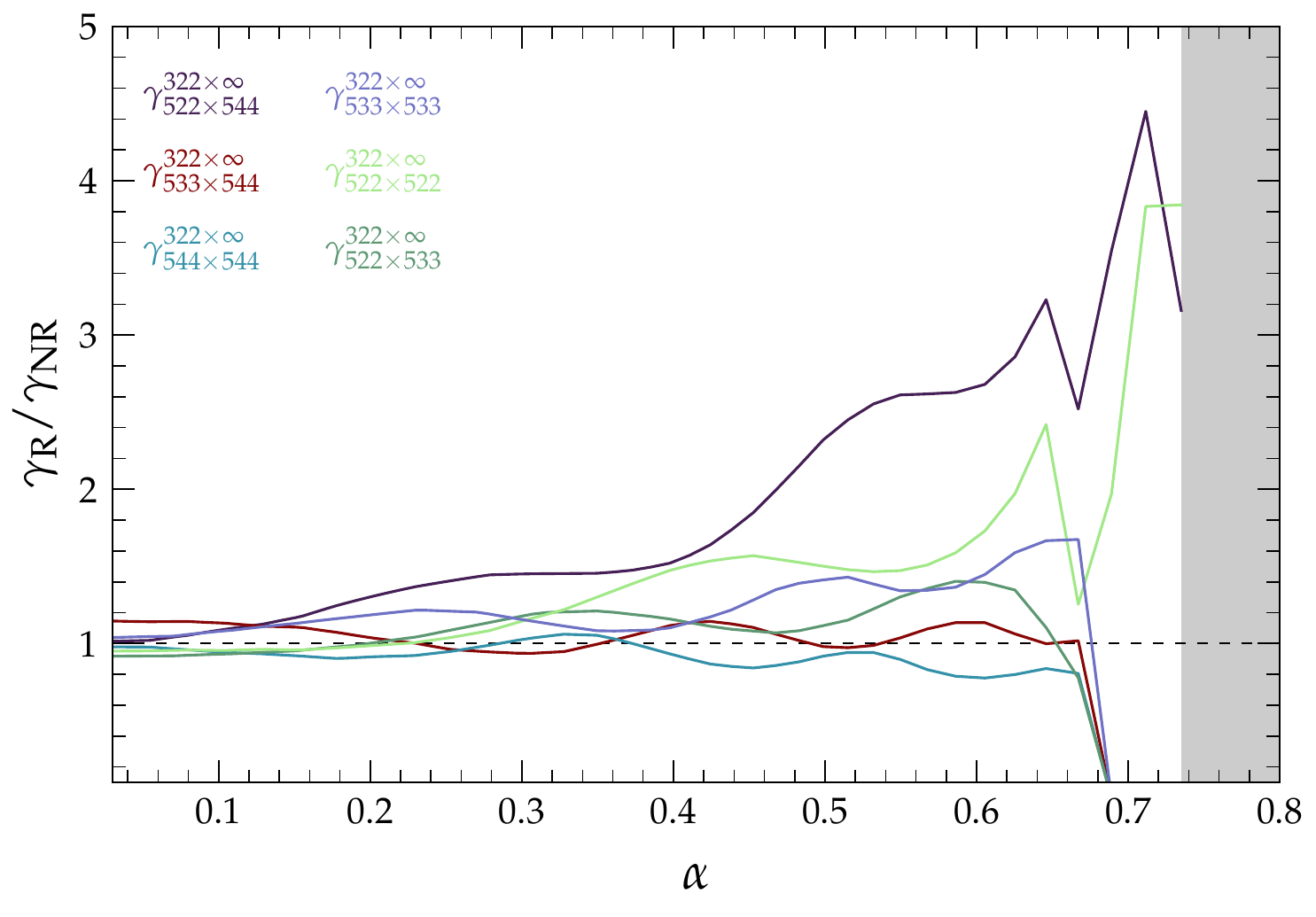}
     \includegraphics[width=0.46\textwidth]{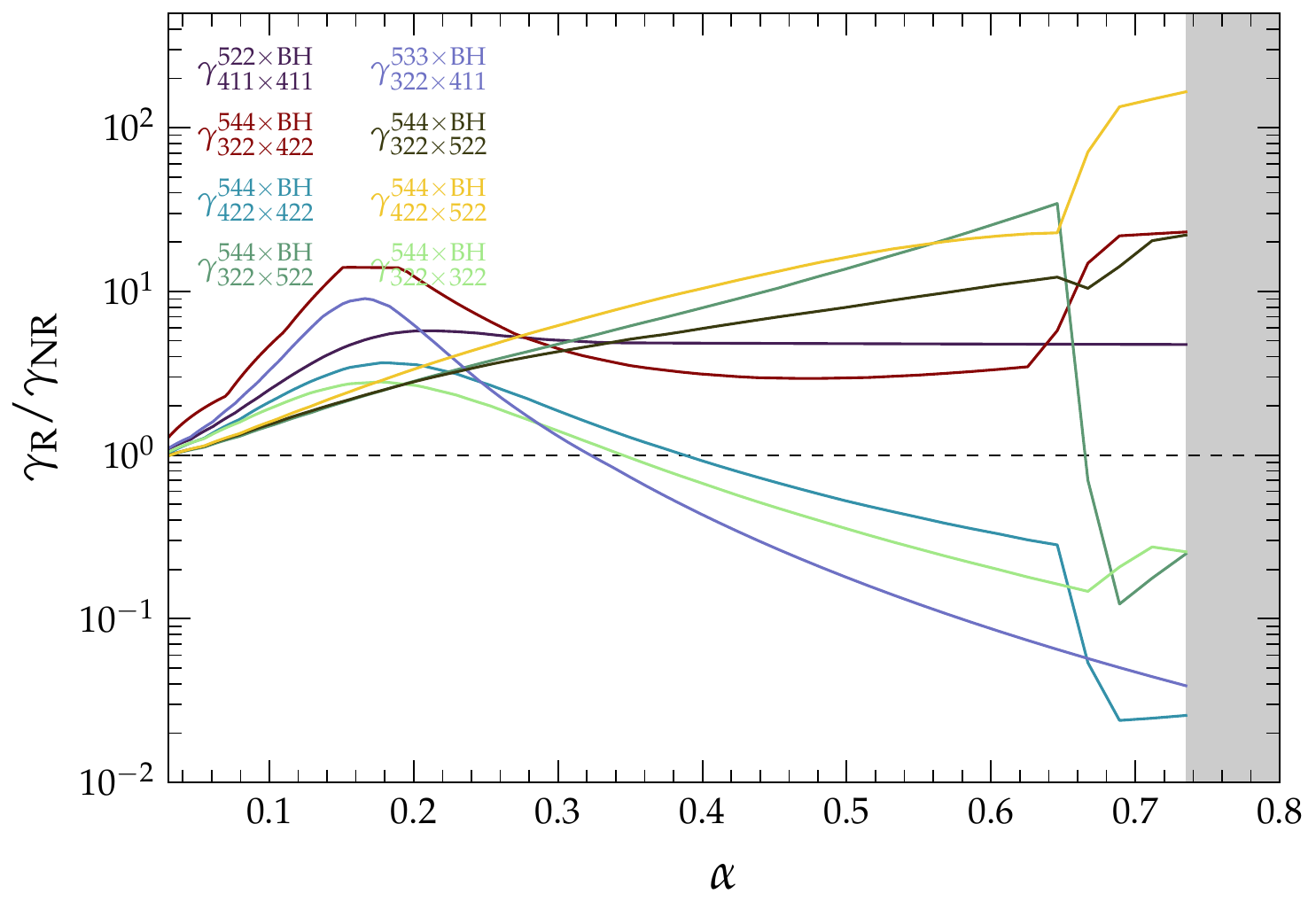}
    \caption{ Same as Fig.~\ref{fig:rel_rates}, but for a selection of relevant scattering processes arising at $n=5$ (see Table~\ref{tab:rates_n5}). For clarity, we divide the rates  across three plots. Note that the sharp downturn seen at large $\alpha$ for the radiative energy losses arises when relativistic corrections to the energy of the bound states cause the final state to transition from a free particle to a bound oscillation -- this tends to lead to an  exponential suppression the amplitude of the wavefunction at large $r$, except in two cases where the bound oscillation is sufficiently strong to efficiently dissipate energy through the horizon.  }
    \label{fig:rates_n5}
\end{figure*}

Let us briefly summarize the result outlined above. In the hydrogen-like limit, dissipative scattering processes induced by the presence of self-interactions are computed by expanding the perturbed wavefunction in a basis of the discrete bound states and the continuum free spectrum; the basis coefficients can be computed by projecting the source term onto the basis functions, and the rate of energy dissipation can in turn be determined by integrating the perturbed wavefunction over the horizon or over the sky in the $r\rightarrow\infty $ limit. This process determines the rate of depletion/growth of any superradiant state due to the quartic self interactions. We now turn our attention to the issue of generalizing the above calculation to obtain the energy dissipation rates in the relativistic limit.

\subsubsection{Relativistic Treatment}

Generalizing the procedure outlined above to the fully relativistic case carries a number of additional challenges. First, note that light scalar fields in a Kerr background do not contain a continuum component in the energy spectrum; rather, these continuum states are comprised of a combination of discrete `quasi-normal modes', a prompt signal, and a slowly decaying  mode. This already presents a rather significant departure (at least conceptually) from the classical limit outlined above, and suggests that one is not necessarily guaranteed to recover the hydrogen-like result as one takes the $\alpha \rightarrow 0$ limit (although for bound transitions the dominant contribution to the energy loss is often through the bound, rather than the continuum states). Now the presence of a fully discrete spectrum does not inherently present an impediment to the aforementioned approach. Complications, however, begin to arise when one realizes that the radial wavefunctions of quasi-normal modes are  divergent at the horizon and spatial infinity, implying the conventional definition of an inner product (which is used, for example, to project the source term onto the basis states and determine the coefficients $c_{n\ell m}$) is not sufficient to e.g. establish orthogonality, to normalize wavefunctions, or to perform a spectral decomposition. Recent progress has been made~\cite{Green:2022htq,Cannizzaro:2023jle}  in establishing suitable definitions of the relativistic inner product, which explicitly demonstrates the quasi-normal modes and quasi-bound states are indeed orthogonal, and in general can be used to derive well-defined spectral coefficients for these basis functions (see also a counter-term subtraction method developed for the Schwarzschild limit~\cite{Sberna:2021eui}). A more formidable obstacle which inhibits the spectral decomposition, however, stems from the fact that the set of eigenfunctions of the quasi-bound and quasi-normal states do not form a complete basis (note that this differs from the hydrogen-limit, where the Hermitian nature of the operator guarantees completeness); in general, the complete basis also includes a prompt (high-energy) component, and a slowly decaying low-energy tail~\cite{ching1996wave}. In that sense, the quasi-bound and quasi-normal modes only offer an approximate description of the system at intermediate times (after the prompt early-time died off, and before the quasi-normal modes have decayed away).

\begin{figure*}
    \includegraphics[width=0.49\textwidth]{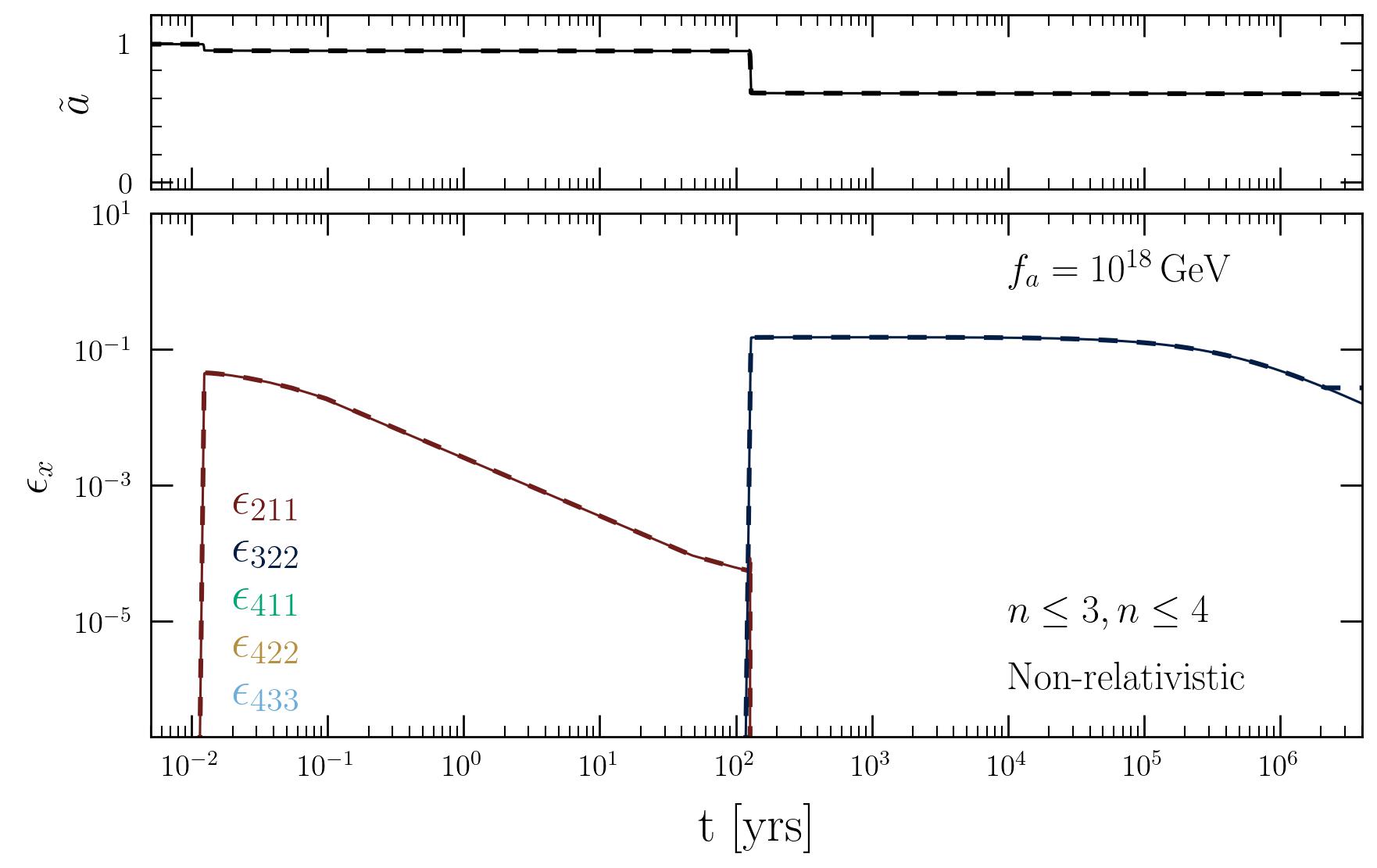}
    \includegraphics[width=0.49\textwidth]{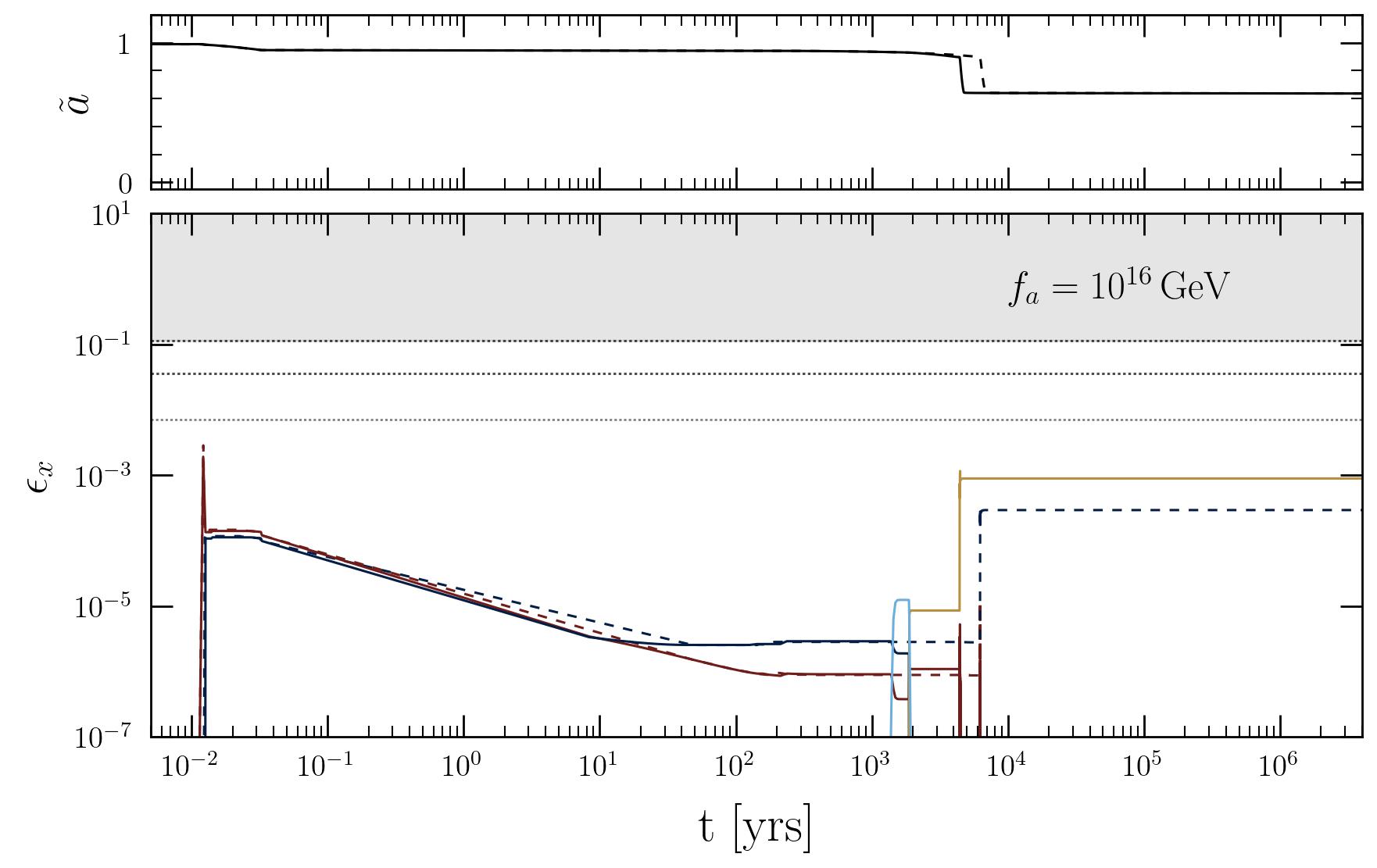}
    \includegraphics[width=0.49\textwidth]{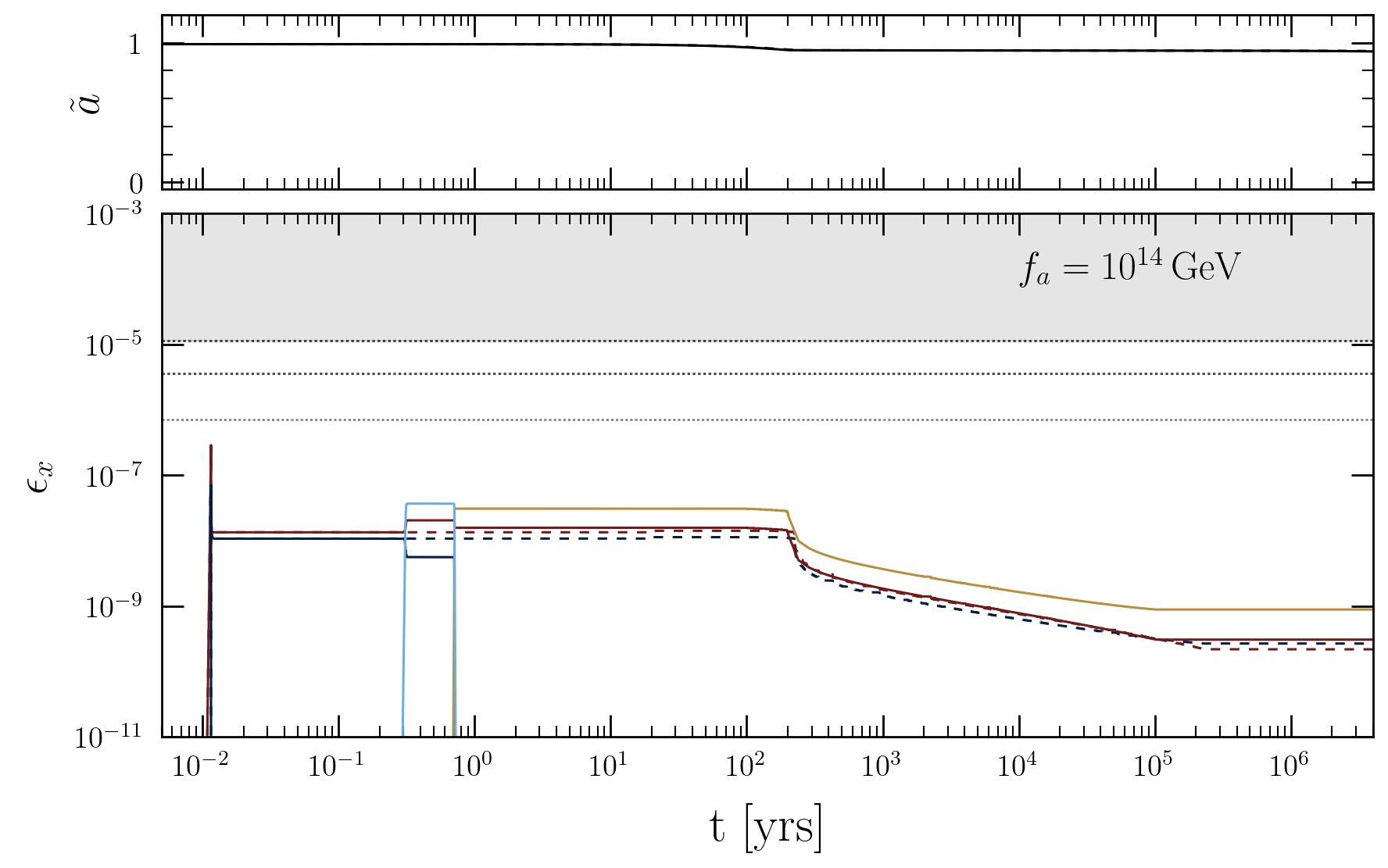}
    \includegraphics[width=0.49\textwidth]{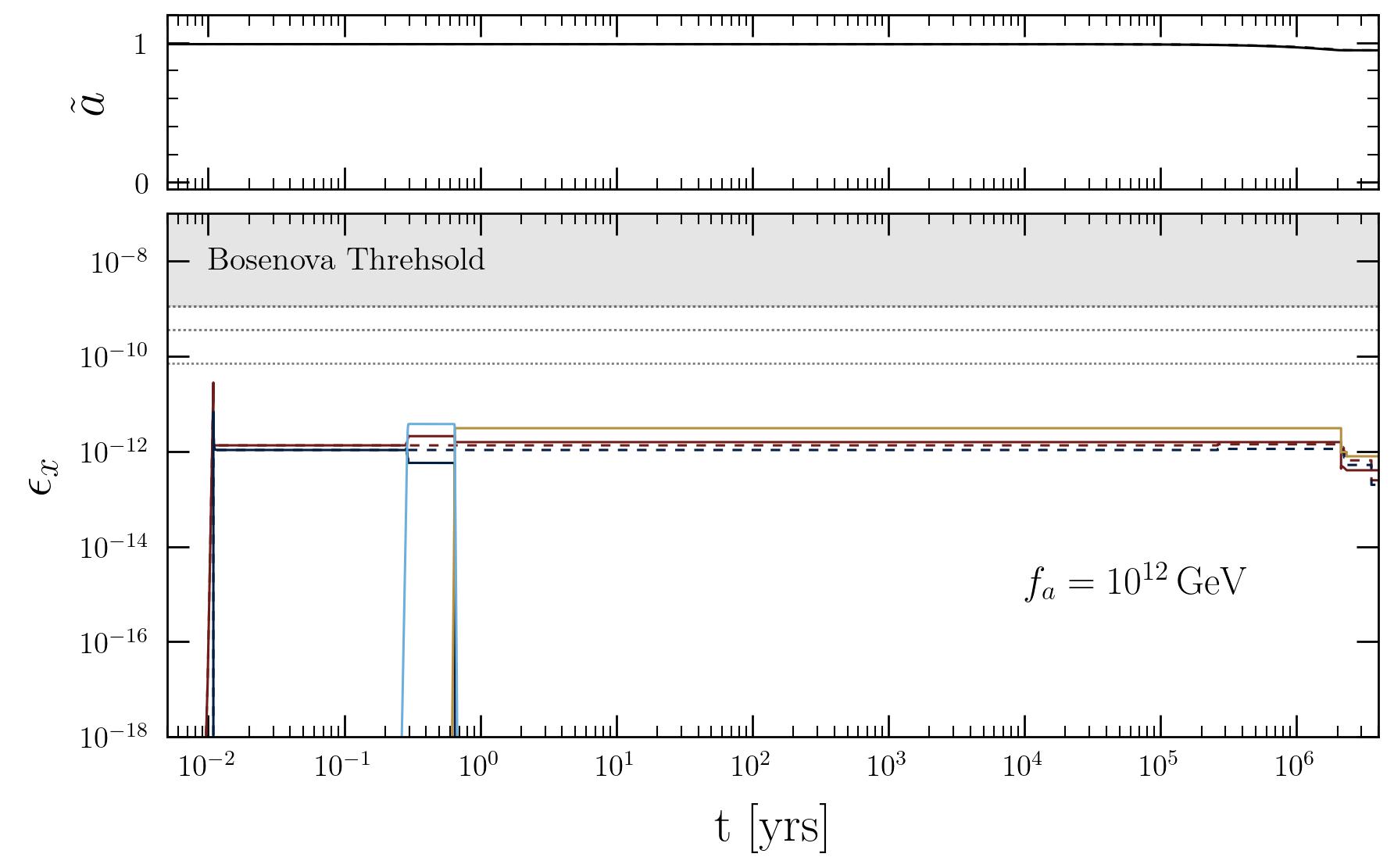}
    \caption{\label{fig:evolve} Evolution of the $\levtwo$ (red), $\levthree$ (blue), $\levfour$ (green), $\levfourtwo$  (gold), and $\levfourthree$ (light blue)  levels, and the dimensionless spin (top panel of each figure) for various axion decay constants ranging from $f_a = 10^{18}$ GeV (top left) to $f_a = 10^{12}$ GeV (bottom right). Dashed and solid lines show the evolution in the two-level system (evolving only the $\levtwo$ and $\levthree$ states) and the five-level system (evolving relevant states with $n \leq 4$), respectively. Calculations are shown using the non-relativistic limit of the scattering rates provided in Table~\ref{tab:rates} (examples of the evolution of the full system, namely $n \leq 5$ with relativistic corrections, are shown below). All panels display results for a black hole of mass $M = 22.2 \, M_\odot$, spin $\tilde{a} = 0.99$, and an axion mass $\mu = 2 \times 10^{-12}$ eV (corresponding to $\alpha \sim 0.33$). The maximum occupation number achievable before self-interactions induce a bosenova is shown using black horizontal dotted lines (one line for each value of $n$), with the shaded region denoting values of $\epsilon_x$ exceeding the $n=4$ bosenova threshold (no such lines are seen in the top left panel, as the occupation numbers occur at larger $\epsilon_x$, and are thus irrelevant for the evolution of the system). Note that growth is assumed to saturate at these thresholds.  }
\end{figure*}

An alternative approach which avoids the spectral decomposition outlined above (thereby naturally including contributions not only from the quasi-bound states and quasi-normal modes, but also from the prompt and slowly decaying components) is to solve the radial equation using Green's functions. This approach has been applied in a variety of contexts, including e.g. to solve for gravitational wave emission sourced by an in-falling point particle~\cite{Mino:1997bx,Kokkotas:1999bd,Piovano:2020zin} (which mimics the physics, for example, of an extreme mass ratio inspiral), and has been applied in a similar context to compute scattering rates and self energy corrections of axions~\cite{Omiya:2020vji,Omiya:2022mwv}. The general approach in this case is as follows. One begins by writing down the two solutions, call them $R^{\rm in}_{\ell m \omega}$ and $R^{\rm up}_{\ell m \omega}$, of the homogeneous equation (i.e. Eq.~\ref{eq:radialR}), where asymptotically these solutions scale as
\begin{eqnarray}
    R^{\rm in} \sim \frac{e^{-i (\omega - m \Omega_H) r_*}}{\Delta} \hspace{.4cm} r_* \rightarrow -\infty \\
    R^{\rm up} \sim \frac{e^{-i k r_*}}{r} \hspace{.4cm} r_* \rightarrow \infty \, .
\end{eqnarray}
These solutions can be obtained directly by integrating Eq.~\ref{eq:homoRad}  inward (outward) from infinity (the outer horizon). The solution to the inhomogeneous radial equation
\begin{multline}
     \frac{d}{dr}  \left( \Delta \frac{d R}{d r} \right) + \left[ \frac{\omega^2 (r^2 + a^2)^2 - 4 G M a m \omega r + m^2 a^2 }{\Delta}  \right.  \\[5pt] \left. - \left( \omega^2 a^2 + \mu^2 r^2 + \Lambda_{lm} \right) \right] R_{lm}(r) = \mathcal{T}_{\ell m \omega}(r) \, 
\end{multline}
which is purely in-going at the horizon and out-going at infinity is given by
\begin{multline}
    R = \frac{1}{W_{\ell m \omega}} \left[R^{\rm up}_{\ell m \omega}(r) \int_{r_+}^{r} dr^\prime  R^{\rm in}_{\ell m \omega}(r^\prime) \, \mathcal{T}_{\ell m \omega}(r^\prime)  \right. \\ \left. + R^{\rm in}_{\ell m \omega}(r) \int_{r}^\infty dr^\prime R^{\rm up}_{\ell m \omega}(r^\prime) \mathcal{T}_{\ell m \omega}(r^\prime) \right] \, .
\end{multline}
Here, $W_{\ell m \omega}$ is the $\Delta-$scaled Wronskian
\begin{eqnarray}
    W_{\ell m \omega} = \Delta \left[R^{\rm up}_{\ell m \omega} \frac{d R^{\rm in}_{\ell m \omega}}{d r} - R^{\rm in}_{\ell m \omega} \frac{d R^{\rm up}_{\ell m \omega}}{dr} \right] \, ,
\end{eqnarray}
which is independent of $r$. For the case of self-interactions, the source term of the $\ell m \omega$ mode is simply given by
\small
\begin{eqnarray}
    \mathcal{T}_{\ell m \omega}(r) = -\frac{\lambda}{6} \,  \int d\phi \, d(\cos\theta) \, \rho^2 (\Phi^{(0)}_\omega)^3 \, S_{\ell m}^*  \, ,
\end{eqnarray}
\normalsize
where $\Phi^{(0)}_\omega$ is the term in $\Phi^{(0)} \propto e^{- i \omega t}$, and the integrals serve to project the initial source term onto the appropriate value of $\ell$ and $m$. With the radial solution in hand, one can compute energy losses using Eq.~\ref{eq:e_horiz}, and thereby rate coefficients (i.e. $\Gamma/\gamma$), in an analogous manner to what has been outlined above. It should be mentioned that at large $\alpha$, the relativistic corrections from the energy can cause a change in the final state (e.g. from non-relativistic emission to bound absorption at the horizon) -- we comment on this briefly below, mentioning here only that such a shift is trivially computed in this formalism as one can compute the full radial wavefunction for any set of $\omega\ell m$.  This is the approach adopted throughout much of this work, and which allows one to extend the results of the hydrogen-like limit to the relativistic regime. The ratio of the relativistic to hydrogen-like rates as a function of $\alpha$ are shown in Figs.~\ref{fig:rel_rates} and ~\ref{fig:rates_n5}.

Before continuing, let us note that we only compute the relativistic rates for values of $\alpha$ small enough such that all three quasi-bound states entering the source term are superradiant. At larger values of $\alpha$ the imaginary component of the energy of at least one state flips sign, causing the state to undergo exponential decay (rather than exponential growth) -- this decay is typically sufficiently strong to drive the occupation number to zero, rendering the scattering process of interest irrelevant.

\subsection{Evolution of Superradiant Systems} \label{sec:example_evo}
The sections above have served to provide a general outline that allows one to compute the growth rates, evolution, and spin down of rapidly rotating black holes in the presence of a free scalar field, or a scalar field with quartic self-interactions. Having developed the generalized formalism, we now turn our attention toward demonstrating the potential impact on realistic systems. We begin by studying the coupled evolution with $n\leq 3$, and incrementally include the $n=4$ and $n=5$ states. For completeness, we also demonstrate the relative importance of including relativistic corrections to the self-interaction induced scattering rates. 

\subsubsection{Evolution at $n \leq 3$}
We now turn our attention to studying the evolution of the superradiant system with self-interactions. We begin by focusing our attention on states with $n \leq 3$, which includes the $\levtwo$, $\left. |311\right> $, $\left. |321\right> $ , and the $\levthree$ states. The $\levtwo$ and $\levthree$ are both leading order superradiant states, and thus must be included. The superradiance growth rate of the $\left. |321\right>$ state is extremely slow, and since selection rules prevent the growth of this state via the other states, this state can be neglected. In principle, $\left. |311\right> $ state can grow via superradiance on timescales which are only slightly longer than the $\levtwo$ state, but there has been some disagreement in the literature as to whether self-interactions are sufficient strong to prohibit the growth of the the $\left. |311\right> $ state; in particular, Ref.~\cite{Baryakhtar:2020gao} has argued that the $\left. |311\right> \times \levtwo \rightarrow \levthree \times \left. |200\right>$ scattering process always suppresses the growth, while Ref.~\cite{Omiya:2022mwv} instead finds non-negligible occupation numbers of the $\left. |311\right> $ state. In order to remain open to both possibilities, we include this state in the evolution. We are thus left with the following coupled differential equations
\begin{eqnarray}
 \dot{\epsilon}_{211} &=& \gamma_{211}^{\rm SR} \epsilon_{211} -2\gamma_{211\times 211}^{322 \times {\rm BH}} \epsilon_{211}^2 \epsilon_{322} \nonumber \\[5pt] & -& \gamma_{211\times 311}^{322 \times {\rm BH}} \epsilon_{211} \epsilon_{311} \epsilon_{322} + \gamma_{311\times 311}^{211 \times \infty} \epsilon_{311} \epsilon_{311} \epsilon_{211} \nonumber \\[5pt] & +& \gamma_{311\times 322}^{211 \times \infty} \epsilon_{311} \epsilon_{322} \epsilon_{211} + \gamma_{322\times 322}^{211 \times \infty} \epsilon_{322} \epsilon_{322} \epsilon_{211} \nonumber \\[5pt] & -& 2 \gamma^{\rm GW}_{211\times211} \epsilon_{211}^2 - 3  \gamma^{\infty}_{211^3} \epsilon_{211}^3 \nonumber \\[5pt] & +& \gamma^{\rm GW}_{322\rightarrow 211}  \epsilon_{211}  \epsilon_{322} \label{eq:dE_1}
 \\[15pt]
\dot{\epsilon}_{311} &=& \gamma_{311}^{\rm SR} \epsilon_{311} - \gamma_{211\times 311}^{322 \times {\rm BH}} \epsilon_{211} \epsilon_{311} \epsilon_{322}\nonumber \\[5pt] &-& 2\gamma_{311\times 311}^{211 \times \infty} \epsilon_{311}^2 \epsilon_{211} - \gamma_{311\times 322}^{211 \times \infty} \epsilon_{311} \epsilon_{322} \epsilon_{211} \nonumber \\[5pt] &-& 2\gamma_{311\times 311}^{322 \times {\rm BH}} \epsilon_{311}^2 \epsilon_{322}
\\[15pt]
\dot{\epsilon}_{322} &=& \gamma_{322}^{\rm SR} \epsilon_{322} + \gamma_{211\times 211}^{322 \times {\rm BH}} \epsilon_{211} \epsilon_{211} \epsilon_{322}\nonumber \\[5pt] &+& \gamma_{211\times 311}^{322 \times {\rm BH}} \epsilon_{211} \epsilon_{311} \epsilon_{322} - \gamma_{311\times 322}^{211 \times \infty} \epsilon_{311} \epsilon_{322} \epsilon_{211} \nonumber \\[5pt] &-& 2\gamma_{322\times 322}^{211 \times \infty} \epsilon_{322}^2 \epsilon_{211} + \gamma_{311\times 311}^{322 \times {\rm BH}} \epsilon_{311} \epsilon_{311} \epsilon_{322} \nonumber \\[5pt] &-& 2 \gamma^{\rm GW}_{322\times322} \epsilon_{322}^2 - \gamma^{\rm GW}_{322\rightarrow 211}  \epsilon_{211}  \epsilon_{322}
\\[15pt]
 \dot{\tilde{a}} &=& -\gamma_{211}^{\rm SR} \epsilon_{211} -\gamma_{311}^{\rm SR} \epsilon_{311} - 2 \, \gamma_{322}^{\rm SR} \epsilon_{322}    \\[15pt]
 \frac{\dot{M}}{\mu G M^2} &=&  -\gamma_{211}^{\rm SR} \epsilon_{211} -\gamma_{322}^{\rm SR} \epsilon_{322} \nonumber \\[5pt] &+& \gamma_{211\times211}^{322\times{\rm BH}} \epsilon_{211}^2 \epsilon_{322}  + \gamma_{311\times 311}^{322 \times {\rm BH}} \epsilon_{311}^2 \epsilon_{322} \nonumber \\[5pt] &+& \gamma_{211\times 311}^{322 \times {\rm BH}} \epsilon_{211} \epsilon_{311} \epsilon_{322} \label{eq:dE_1_end}
 \end{eqnarray}
 where we have defined the normalized occupation number of each state $x$ as $\epsilon_{x} \equiv N_x / (G M^2)$, with $N_x$ is the un-normalized occupation number. In the analysis that follows, we do not find that the $\left. |311\right> $ level reaches non-negligible occupation numbers (in some cases, moderate growth is seen, but is then quickly quenched), and thus we find that this level can in fact be safely neglected.  This implies the $n\leq 3$ evolution is effectively well described by a two-level system, defined only by the $\levtwo$ and $\levthree$ states.

We illustrate in Fig.~\ref{fig:evolve} (using dashed lines) the evolution of the two-level system for various values of $f_a$, taking $M = 22.2 M_\odot$, $\tilde{a} = 0.99$, and $\alpha \sim 0.33$. Here, we adopt the non-relativistic limit of the scattering rates $\gamma$ provided in Table~\ref{tab:rates}, and return to the question of relativistic corrections below. The top panels of each figure show the evolution of the dimensionless spin (the spin evolution of the two-level system is shown with a dashed line, which is, on occasion, overlapping with the spin evolution of the higher order model, shown in solid). The bottom panels display the evolution of the normalized occupation number of each state, with the black horizontal dashed lines denoting the bosenova threshold of the $n=2$, 3, and 4 levels (with the gray shaded region covering the occupation numbers for the $n=4$ bosenova)\footnote{For $f_a = 10^{18} \, {\rm GeV}$, the bosenova threshold is above the maximal occupation number shown on the figure, and is never realizable (since spin down is gauranteed to occur prior to reaching such large occupation numbers).}. One can see that for large values of $f_a$ each superradiant level grows rapidly and independently, spinning down the black hole on their respective timescales. Here, the exponential decay of the $\levtwo$ state occurs because the imaginary component of the energy becomes negative when the $\levthree$ spin down begins. For larger values of $f_a$, the $\levtwo$ forces the $\levthree$ to grow at a comparable rate, driving the system to a temporarily equilibrium which is sustained until spin can be extracted by the black hole. Note the occupation numbers for the $n\leq 3$ system always remain well below the bosenova threshold.

In the limit that $f_a \rightarrow M_{\rm pl}$, self-interactions are negligible, and one can supplement Eqns.~\ref{eq:dE_1}-\ref{eq:dE_1_end} by analogous equations governing the growth of higher order states. Despite the fact that  the $\left. |n 11\right>$ and $\left. |n 22 \right>$ states may have large growth rates, these states are negligible in this limit since the $\levtwo$ and $\levthree$ states operate on shorter timescales, and will spin down the black hole before the higher $n$ states have had sufficient time to grow. This same statement does not apply when self-interactions are included, since the $\levtwo-\levthree$ equilibrium slows the spin-down process, in some cases allowing for higher-order states to play a non-trivial role in the evolution of the system.

For values of $\alpha \lesssim 0.22$, it has been argued that the two-level system accurately describes the early stages of spin down of black holes with self-interactions~\cite{Baryakhtar:2020gao} -- this statement stems from analyzing the relative ratio of growth and energy dissipation of higher-order states, and showing that at small $\alpha$ dissipation prevents most states from achieving large occupation numbers. This is not necessarily true for larger values of $\alpha$, and has thus far prohibited robust constraints from being derived across broad regions of axion parameter space. In the large $\alpha$ regime, higher order level mixing and relativistic corrections to the scattering rates play a crucial role. We now turn our attention towards these aspects of the problem.

\begin{figure*}
     \includegraphics[width=0.47\textwidth, trim={0cm 0cm 0cm 0cm}, clip]{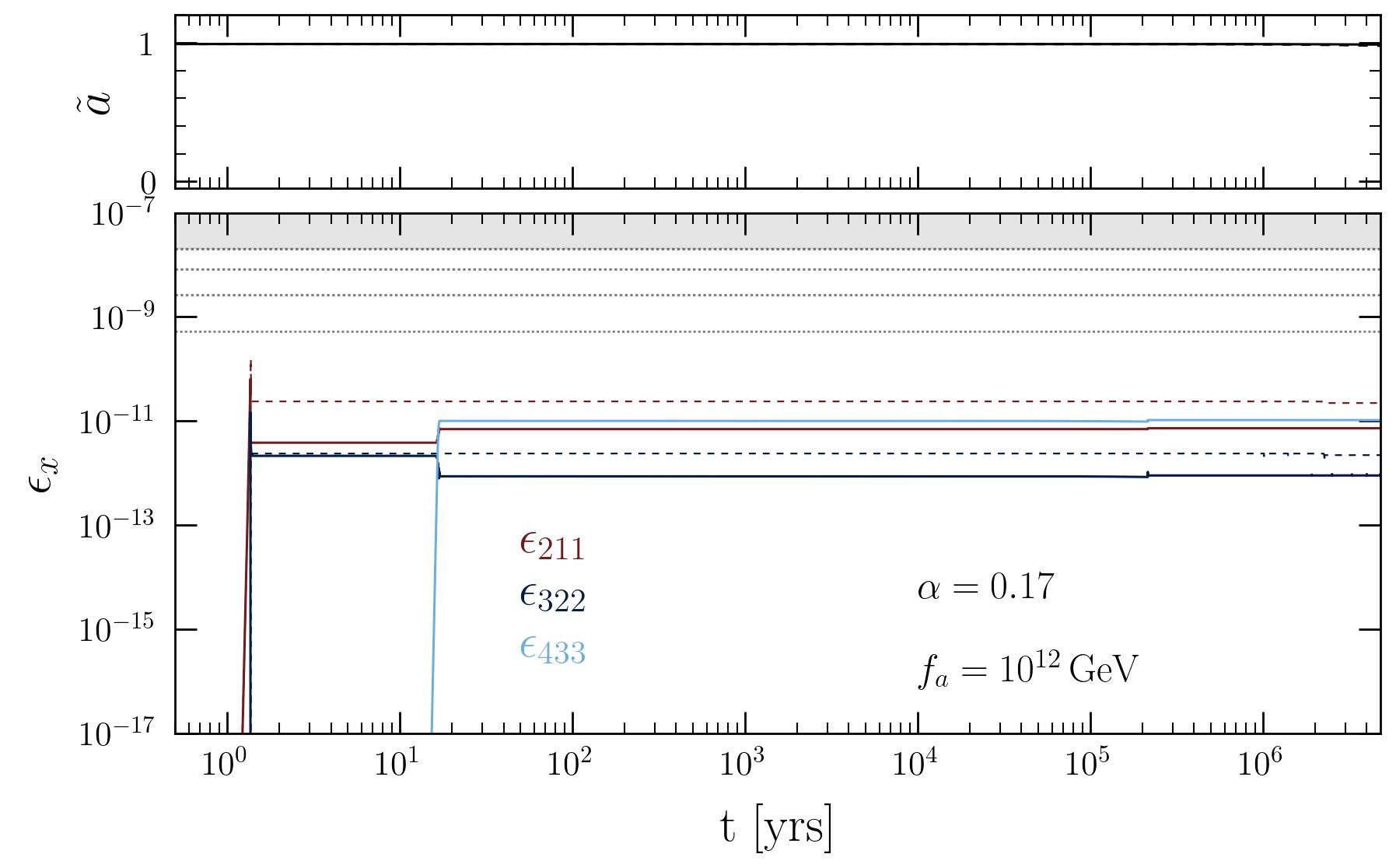}
     \includegraphics[width=0.47\textwidth, trim={0cm 0cm 0cm 0cm}, clip]{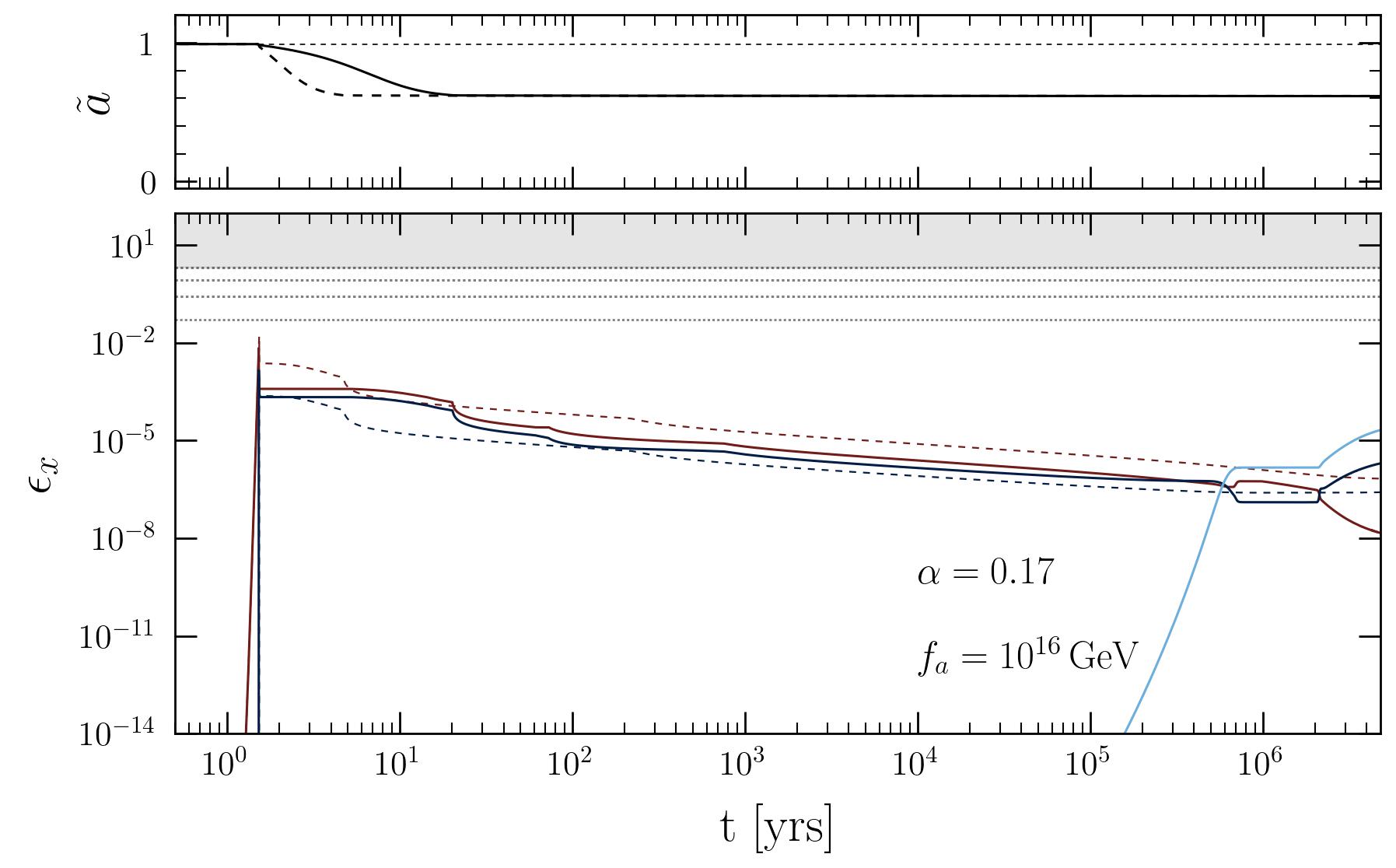}
     \caption{ Comparison between evolution of $n \leq 4$ superradiant states for $\alpha \simeq 0.17$, with (solid) and without (dashed) relativistic corrections. Results are shown for and $f_a = 10^{12}$ GeV (left) and $f_a = 10^{16}$ GeV (right). As can be seen, the growth of the $n=4$ states are sufficient to disrupt the $\levtwo-\levthree$ equilibrium. Note that the $\levfourthree$ does not grow in the non-relativistic analysis, and thus only the solid lines for this state appear (see text for further details). \label{fig:f12_lowmass} }
\end{figure*}

\subsubsection{Evolution at $n \leq 4$}
Let us now extend beyond the $n\leq 3$ system and include the next set of states arising at  $n=4$.

At $n=4$, one introduces the following six states:
\begin{itemize}
    \item $\left. |411 \right>$: This state can potentially grow via superradiance on short timescales, and it has been argued in Ref.~\cite{Baryakhtar:2020gao} (based on the rates derived in the hydrogen-like limit) that it can contribute to spin down for $\alpha \gtrsim 0.2$. We therefore include this level in our analysis.
    \item $\left. |421 \right>$: The superradiant growth rate is highly suppressed, and cannot drive the state to large occupation numbers on the relevant timescales. One may ask whether rates of the form $\left. |n\ell m \right> \times \left. |n^\prime \ell^\prime m^\prime \right> \rightarrow \left. |421 \right> \times {\rm BH}$ or $\left. |n\ell m \right> \times \left. |n^\prime \ell^\prime m^\prime \right> \rightarrow \left. |421 \right> \times \infty$ could lead to non-negligible occupations numbers; however, the former processes are absent due to selection rules, and the latter processes would require initial states with $n \gg 4$ in order to ensure the axion emission is not bound. Therefore we neglect this state.
    \item $\left. |431 \right>$: Similarly to the $\left. |421 \right>$ state, the superradiance timescale is too large to be relevant and this state cannot be grown via scattering processes; therefore we neglect this state in the evolution of the system.
    \item $\left. |422 \right>$: The superradiance timescale can be sufficiently short (relative to the black hole lifetime, or to the Salpeter timescale, see Sec.~\ref{sec:analysis}) such that the $\left. |422 \right>$ level becomes sizable. In addition, this level can potentially be grown at intermediate values of $\alpha$ via $\levtwo \times \levtwo$ scattering processes. As such, we do include this state in the evolution of the $n \leq 4$ system\footnote{We note that Ref.~\cite{Omiya:2022mwv} concluded that this level should not grow; we disagree with this conclusion. One can directly compute the ratio of the growth rate to the energy dissipation rate in the hydrogen like limit and show that this state can grow once the occupation number of the $\levfour$ state has become sufficiently large. This conclusion is in agreement with what was previously found in~\cite{Baryakhtar:2020gao}.}. 
    \item $\left. |432 \right>$: The superradiant growth rate for this state is heavily suppressed, and as with the $\left. |431 \right>$ and $\left. |421 \right>$ states, the $\left. |432 \right>$ state cannot be grown via $n \leq 4$ states. We neglect this state in the evolution.
    \item $\left. |433 \right>$: The superradiant growth rate is sufficiently large to spin down the black hole on relevant timescales. This state is also the first $m=3$ state to appear, and furthermore can be grown via various scattering processes (see Table~\ref{tab:rates}). We include this in the evolution. It is worth mentioning that Refs.~\cite{Baryakhtar:2020gao,Collaviti:2024mvh} have concluded that the $\levfourthree$ state cannot grow for $\alpha \lesssim 0.3$, which is in part used to justify the claim that the two-level system (comprised purely of the $\levtwo$ and $\levthree$ states) is sufficient to describe the early stages of spin down and the gravitational wave spectrum in this region of parameter space.  This conclusion is based on the fact that the rate of change of the occupation number of the $\left| n33 \right>$ state (assuming growth is driven by self-interactions, rather than superradiance) is proportional to $\dot{\epsilon}_{n33} \propto (\gamma_{322 \times n33}^{211 \times \infty} - \gamma_{211 \times 322}^{n33 \times {\rm BH}})$, which in the hydrogen-like limit is always negative for $\alpha \lesssim 0.3$. Let us point out, however, that the relativistic corrections to $ \gamma_{211 \times 322}^{433 \times {\rm BH}}$ are much larger than those to $\gamma_{322 \times 433}^{211 \times \infty}$, and re-adjust this threshold to lower values of $\alpha$ (implying the two-level system is only accurate for values of $\alpha \lesssim 0.15$, in agreement with the conclusion of~\cite{Omiya:2024xlz}). This is demonstrated explicitly in Fig.~\ref{fig:f12_lowmass}, where the solid (dashed) lines denote the evolution of the occupation numbers with (without) relativistic corrections. Similar conclusions were also recently obtained in~\cite{Omiya:2024xlz}, which studied the evolution of the four coupled $\ell=m \leq 4$ states (with the focus of studying the impact of self-interactions on the gravitational wave emission).
\end{itemize}
In the end, studying the evolution of states at $n\leq 4$ involves introducing three new states, namely: $\levfour$, $\levfourtwo$, and $\levfourthree$. We now turn our attention to evolution of this five-level system.

In order to move from the two-level to the five-level system, Eqns.~\ref{eq:dE_1}-\ref{eq:dE_1_end} must be augmented by the following:
\begin{eqnarray}
\dot{\epsilon}_{211} &=& \gamma_{211}^{\rm SR} \epsilon_{211} -2\gamma_{211\times 211}^{322 \times {\rm BH}} \epsilon_{211}^2 \epsilon_{322} \nonumber  \\[5pt] &-& 2\gamma_{211\times 211}^{422 \times {\rm BH}} \epsilon_{211}^2 \epsilon_{422} - \gamma_{211\times 322}^{433 \times {\rm BH}} \epsilon_{211} \epsilon_{322} \epsilon_{433} \nonumber  \\[5pt] &-& \gamma_{211\times 411}^{322 \times {\rm BH}} \epsilon_{211} \epsilon_{411} \epsilon_{322} - \gamma_{211\times 411}^{422 \times {\rm BH}} \epsilon_{211} \epsilon_{411} \epsilon_{422} \nonumber  \\[5pt] &-& \gamma_{211\times 422}^{433 \times {\rm BH}} \epsilon_{211} \epsilon_{422} \epsilon_{433} + \gamma_{322\times 322}^{211 \times \infty} \epsilon_{322} \epsilon_{322} \epsilon_{211} \nonumber \\[5pt] &+& \gamma_{322\times 411}^{211 \times \infty} \epsilon_{322} \epsilon_{411} \epsilon_{211} + \gamma_{322\times 422}^{211 \times \infty} \epsilon_{322} \epsilon_{422} \epsilon_{211} \nonumber  \\[5pt] &+& \gamma_{322\times 433}^{211 \times \infty} \epsilon_{322} \epsilon_{433} \epsilon_{211} + \gamma_{411\times 411}^{211 \times \infty} \epsilon_{411} \epsilon_{411} \epsilon_{211} \nonumber  \\[5pt] &+& \gamma_{411\times 422}^{211 \times \infty} \epsilon_{411} \epsilon_{422} \epsilon_{211} + \gamma_{411\times 433}^{211 \times \infty} \epsilon_{411} \epsilon_{433} \epsilon_{211}\nonumber   \\[5pt] &+& \gamma_{422\times 422}^{211 \times \infty} \epsilon_{422} \epsilon_{422} \epsilon_{211} + \gamma_{422\times 433}^{211 \times \infty} \epsilon_{422} \epsilon_{433} \epsilon_{211} \nonumber  \\[5pt] &+& \gamma_{433\times 433}^{211 \times \infty} \epsilon_{433} \epsilon_{433} \epsilon_{211} - 2 \gamma^{\rm GW}_{211\times211} \epsilon_{211}^2 \nonumber  \\[5pt]  &-& 3  \gamma^{\infty}_{211^3} \epsilon_{211}^3  + \gamma^{\rm GW}_{322\rightarrow 211}  \epsilon_{211}  \epsilon_{322}
\\[15pt]
\dot{\epsilon}_{322} &=& \gamma_{322}^{\rm SR} \epsilon_{322} + \gamma_{211\times 211}^{322 \times {\rm BH}} \epsilon_{211} \epsilon_{211} \epsilon_{322} \nonumber \\[5pt] &-& \gamma_{211\times 322}^{433 \times {\rm BH}} \epsilon_{211} \epsilon_{322} \epsilon_{433} + \gamma_{211\times 411}^{322 \times {\rm BH}} \epsilon_{211} \epsilon_{411} \epsilon_{322} \nonumber  \\[5pt] &-&2\gamma_{322\times 322}^{211 \times \infty} \epsilon_{322}^2 \epsilon_{211} - \gamma_{322\times 411}^{211 \times \infty} \epsilon_{322} \epsilon_{411} \epsilon_{211} \nonumber  \\[5pt] &-& \gamma_{322\times 422}^{211 \times \infty} \epsilon_{322} \epsilon_{422} \epsilon_{211} - \gamma_{322\times 433}^{211 \times \infty} \epsilon_{322} \epsilon_{433} \epsilon_{211} \nonumber  \\[5pt] &-& \gamma_{322\times 411}^{433 \times {\rm BH}} \epsilon_{322} \epsilon_{411} \epsilon_{433} + \gamma_{411\times 411}^{322 \times {\rm BH}} \epsilon_{411} \epsilon_{411} \epsilon_{322}  \nonumber  \\[5pt] &-& 2 \gamma^{\rm GW}_{322\times322} \epsilon_{322}^2 \nonumber - \gamma^{\rm GW}_{322\rightarrow 211}  \epsilon_{211}  \epsilon_{322}
\\[15pt]
\dot{\epsilon}_{411} &=& \gamma_{411}^{\rm SR} \epsilon_{411} - \gamma_{211\times 411}^{322 \times {\rm BH}} \epsilon_{211} \epsilon_{411} \epsilon_{322} \nonumber 
 \\[5pt] &-& \gamma_{211\times 411}^{422 \times {\rm BH}} \epsilon_{211} \epsilon_{411} \epsilon_{422} - \gamma_{322\times 411}^{211 \times \infty} \epsilon_{322} \epsilon_{411} \epsilon_{211} \nonumber  \\[5pt] &-& \gamma_{322\times 411}^{433 \times {\rm BH}} \epsilon_{322} \epsilon_{411} \epsilon_{433} -2\gamma_{411\times 411}^{211 \times \infty} \epsilon_{411}^2 \epsilon_{211} \nonumber  \\[5pt] &-& \gamma_{411\times 422}^{211 \times \infty} \epsilon_{411} \epsilon_{422} \epsilon_{211} - \gamma_{411\times 433}^{211 \times \infty} \epsilon_{411} \epsilon_{433} \epsilon_{211} \nonumber  \\[5pt] &-&2\gamma_{411\times 411}^{322 \times {\rm BH}} \epsilon_{411}^2 \epsilon_{322} -2\gamma_{411\times 411}^{422 \times {\rm BH}} \epsilon_{411}^2 \epsilon_{422} \nonumber  \\[5pt] &-& \gamma_{411\times 422}^{433 \times {\rm BH}} \epsilon_{411} \epsilon_{422} \epsilon_{433}
\\[15pt]
 \dot{\epsilon}_{422} &=& \gamma_{422}^{\rm SR} \epsilon_{422} + \gamma_{211\times 211}^{422 \times {\rm BH}} \epsilon_{211} \epsilon_{211} \epsilon_{422} \nonumber  \\[5pt]&+& \gamma_{211\times 411}^{422 \times {\rm BH}} \epsilon_{211} \epsilon_{411} \epsilon_{422} - \gamma_{211\times 422}^{433 \times {\rm BH}} \epsilon_{211} \epsilon_{422} \epsilon_{433} \nonumber  \\[5pt] &-& \gamma_{322\times 422}^{211 \times \infty} \epsilon_{322} \epsilon_{422} \epsilon_{211} - \gamma_{411\times 422}^{211 \times \infty} \epsilon_{411} \epsilon_{422} \epsilon_{211} \nonumber  \\[5pt] &-& 2\gamma_{422\times 422}^{211 \times \infty} \epsilon_{422}^2 \epsilon_{211} - \gamma_{422\times 433}^{211 \times \infty} \epsilon_{422} \epsilon_{433} \epsilon_{211}\nonumber  \\[5pt] &+& \gamma_{411\times 411}^{422 \times {\rm BH}} \epsilon_{411} \epsilon_{411} \epsilon_{422} - \gamma_{411\times 422}^{433 \times {\rm BH}} \epsilon_{411} \epsilon_{422} \epsilon_{433}
 \\[15pt]
 \dot{\epsilon}_{433} &=& \gamma_{433}^{\rm SR} \epsilon_{433} + \gamma_{211\times 322}^{433 \times {\rm BH}} \epsilon_{211} \epsilon_{322} \epsilon_{433} \nonumber  \\[5pt] &+& \gamma_{211\times 422}^{433 \times {\rm BH}} \epsilon_{211} \epsilon_{422} \epsilon_{433} - \gamma_{322\times 433}^{211 \times \infty} \epsilon_{322} \epsilon_{433} \epsilon_{211} \nonumber  \\[5pt] &+& \gamma_{322\times 411}^{433 \times {\rm BH}} \epsilon_{322} \epsilon_{411} \epsilon_{433} - \gamma_{411\times 433}^{211 \times \infty} \epsilon_{411} \epsilon_{433} \epsilon_{211}\nonumber  \\[5pt] &-& \gamma_{422\times 433}^{211 \times \infty} \epsilon_{422} \epsilon_{433} \epsilon_{211} -2\gamma_{433\times 433}^{211 \times \infty} \epsilon_{433}^2 \epsilon_{211} \nonumber  \\[5pt] &+& \gamma_{411\times 422}^{433 \times {\rm BH}} \epsilon_{411} \epsilon_{422} \epsilon_{433} 
 \\[15pt]
 \dot{\tilde{a}} &=& - \sum_{n\ell m} m \gamma_{n\ell m}^{\rm SR} \epsilon_{n\ell m}
 \\[15pt]
 \frac{\dot{M}}{\mu G M^2} &=& -\sum_{n\ell m} \gamma_{n\ell m}^{\rm SR} \epsilon_{n\ell m}  + \sum_{ijk} \gamma_{i \times j}^{k \times {\rm BH}} \epsilon_{i} \epsilon_j \epsilon_k
 \end{eqnarray}

Fig.~\ref{fig:evolve} illustrates the evolution of the black hole spin and quasi-bound state occupation numbers $\epsilon_x$, aside the two-level system as described above. As before, scattering rates are taken to their non-relativistic values listed in Table~\ref{tab:rates}. Here, one can see that the evolution of the $\levtwo$ and $\levthree$ states largely follow those of the two-level system at early times, however a shift of the occupation numbers when the $\levfourtwo$ and $\levfourthree$  states begin to become sizable. Before continuing, let us point out that Fig.~\ref{fig:evolve} appears to suggest that the inclusion of the $n = 4$ states has a minimal role of the rate of spin extraction -- this is somewhat misleading, and is merely a coincidental feature arising at this point in parameter space.

\begin{figure*}
     \includegraphics[width=0.49\textwidth, trim={0cm 0cm 0cm 0cm}, clip]{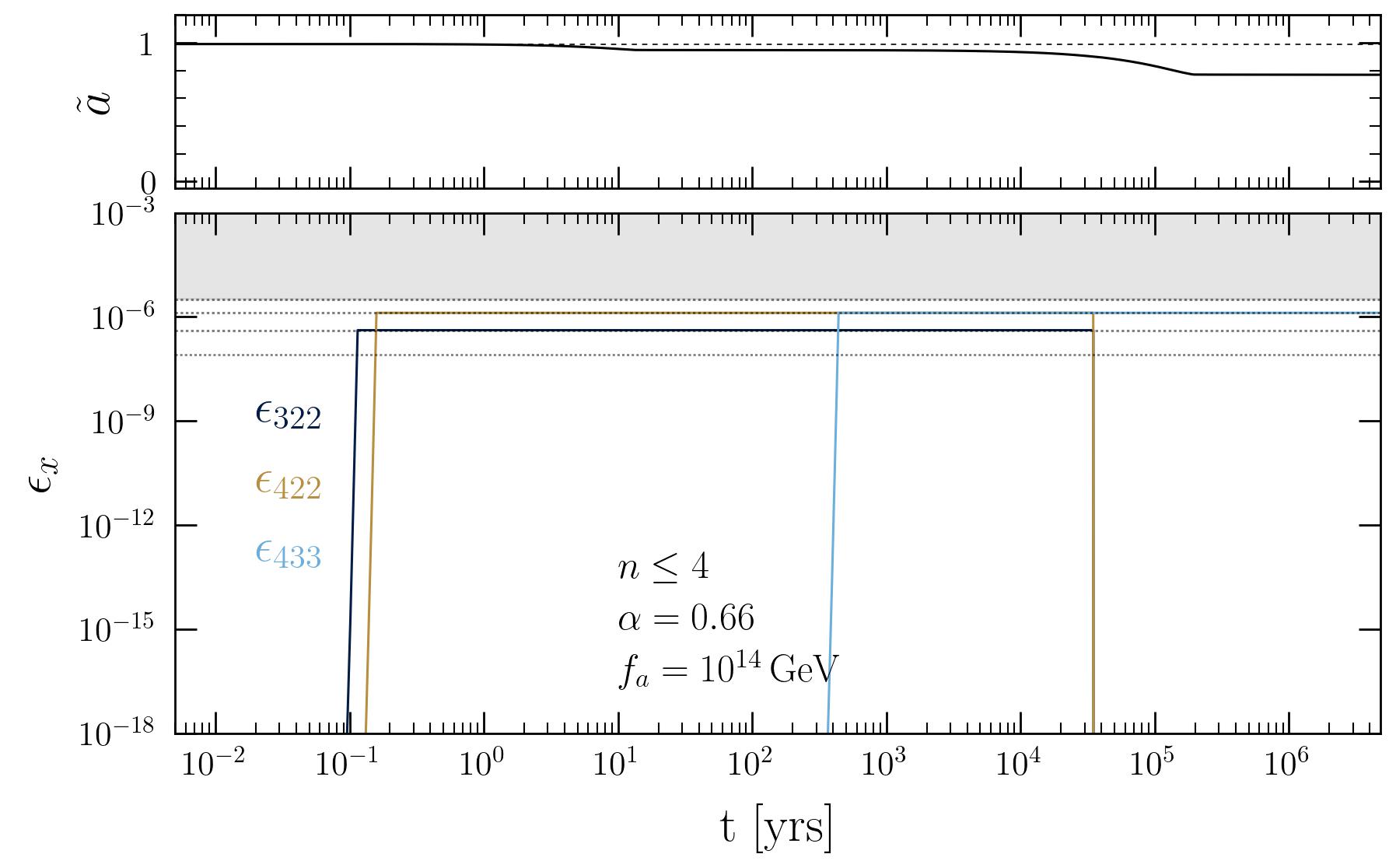}
      \includegraphics[width=0.49\textwidth, trim={0cm 0cm 0cm 0cm}, clip]{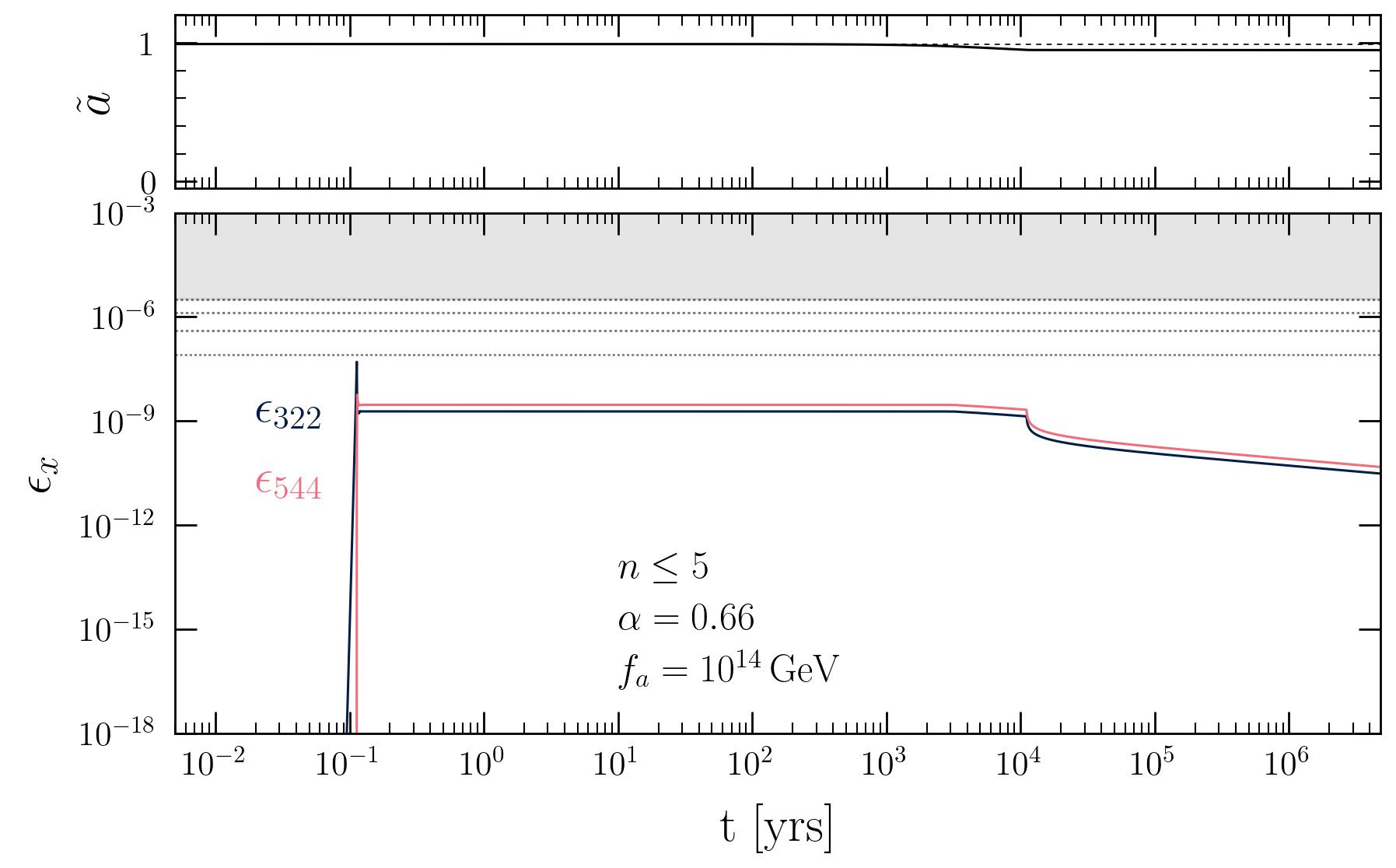}
     \caption{Same as Fig.~\ref{fig:evolve}, but including relativistic corrections to the scattering rates, and comparing the evolution between the $n \leq 4$ (left) and $n \leq 5$ (right) analysis for $\mu = 4 \times 10^{-12}$ eV ($\alpha \simeq 0.66$) and $f_a = 10^{14}$GeV. \label{fig:f14_n4_n5} }
\end{figure*}


Thus far, we have illustrated the evolution of the various quasi-bound states using the hydrogen-like scattering rates. In order to move beyond this approximation, we adopt the Green's function approach outlined in Sec.~\ref{subsec:si}, and re-analyze the growth of the system. An example of the evolution of the occupation numbers is shown in Fig.~\ref{fig:f12_lowmass} for $\alpha = 0.17$, both with and without relativistic corrections, and including $n\leq 4$ states. Here, one can see that the relativistic corrections serve to $(i)$ modify the equilibrium occupation numbers of the $\levtwo$ and $\levthree$ states, $(ii)$ allow for the growth of the $\levfourthree$ state, and $(iii)$ although it is not seen, enhance the spin down rate (which can be inferred from the elevated $\levtwo$ occupation number).

The left panel of Fig.~\ref{fig:f14_n4_n5} shows the evolution of the $n \leq 4$ states at a higher axion mass, corresponding to $\mu = 4 \times 10^{-12}$ eV (also including relativistic corrections). Here, one can see that the $\levthree$ (dark blue), $\levfourtwo$ (yellow), and $\levfourthree$ (light blue) states grow unimpeded to the bosenova threshold. This occurs purely because one has removed any efficient `energy sink' which would impede the growth of these states (at lower $\alpha$, scattering processes with, or into, the $\levtwo$ serve this role); energy sinks for these states do exist, but for large $\alpha$ they only arise at $n \geq 5$. Thus, in order to ensure one does not overstate the strength of the spin down limits, one should ensure that any state capable of extracting spin (on the relevant timescale of interest) has the leading order energy sink present in the analysis. For the $m=2$ states, this leading order contribution will arise at $\left| 544 \right>$, while for the $m=3$ states this will arise at  $\left| 766 \right>$. Reaching $n=7$ in a self-consistent and contained manner is non-trivial, and thus we will continue to expand the analysis to $n=5$, but do not attempt to analyze higher $n$ states -- this approach should allow us to establish the location of the $m=1$ and $m=2$ spin down regions. Future work will focus on extending this analysis to yet higher states, thereby understanding the spin down induced by yet heavier axions.



\subsubsection{Evolution at $n \leq 5$}

In order to understand the stability of the evolution at $n \leq 4$, we extend our analysis to include the $n=5$ states. With this in mind, the relevant states include: 
\begin{itemize}
    \item $\left. |511 \right>$: This state has a sufficiently short superradiance growth time to be relevant for the evolution of the system.
    We include this level in the evolution, although we note now that our numerical solutions do not show this state reaching sizable occupation numbers.
    \item $\left. |522 \right>$: This state has a sufficiently short superradiance growth time to be relevant for the evolution of the system.
    We include this level in the evolution.
    \item $\left. |532 \right>$: The superradiance timescale of this level is typically far too large to allow for the necessary $\sim 180$ e-folds of growth, except for a narrow range of axion masses  near $\alpha \sim 0.6$. The growth rate cannot be enhanced via mixing (as selection rules do not permit this for the relevant states), nor can it efficiently enhance or suppress the growth rate of other $n=5$ states. Given that the bosenova threshold naturally implies $\tau_{\rm bh} \gg 180 \times \tau_{\rm sr}$ for even reasonably large $f_a$, this state will not play a relevant role in the evolution of the system. Therefore, we do not include this level.
    \item $\left. |533 \right>$: The superradiance growth rate can be sizable, and this level can efficiently mix with other states. We include this level.
    \item $\left. |542 \right>$: The superradiance timescale is too large to be relevant, and growth via mixing would require large occupation numbers in the $m=1$ states (which should not be occupied at large $\alpha$). We therefore neglect this level.
    \item $\left. |543 \right>$: The  superradiance growth timescale is barely large enough to grow a few e-folds at $\alpha \sim 0.9$, but certainly not large enough to extract energy or alter the evolution of the system.  Efficient growth via mixing would require $m=1$ states, which we have argued should not be present at large $\alpha$. 
    \item $\left. |544 \right>$: This state can grow via superradiance, and the growth rate can be enhanced (\eg via the $\levthree$ state). We include this level.
\end{itemize}
In summary, we introduce four new states at $n=5$, include: $\left. |511 \right>$, $\left. |522 \right>$, $\left. |533 \right>$, and $\left. |544 \right>$.

For the sake of simplicity, we do not explicitly write the modifications to the evolution of the occupations numbers $\epsilon_x$ or mass or spin of the black hole (although the generalization from the equations outlined above is straightforward to infer). In Table~\ref{tab:rates_n5} we outline a subset of the relevant scattering rates involving the $n=5$ processes -- in practice, all scattering permutations between active states are included in the analysis even though we only show the analytic expressions for a subset of these. As mentioned above, we show relativistic corrections to a subset of the scattering rates in Fig.~\ref{fig:rates_n5}.

\begin{figure*}
    \includegraphics[width=0.49\textwidth]{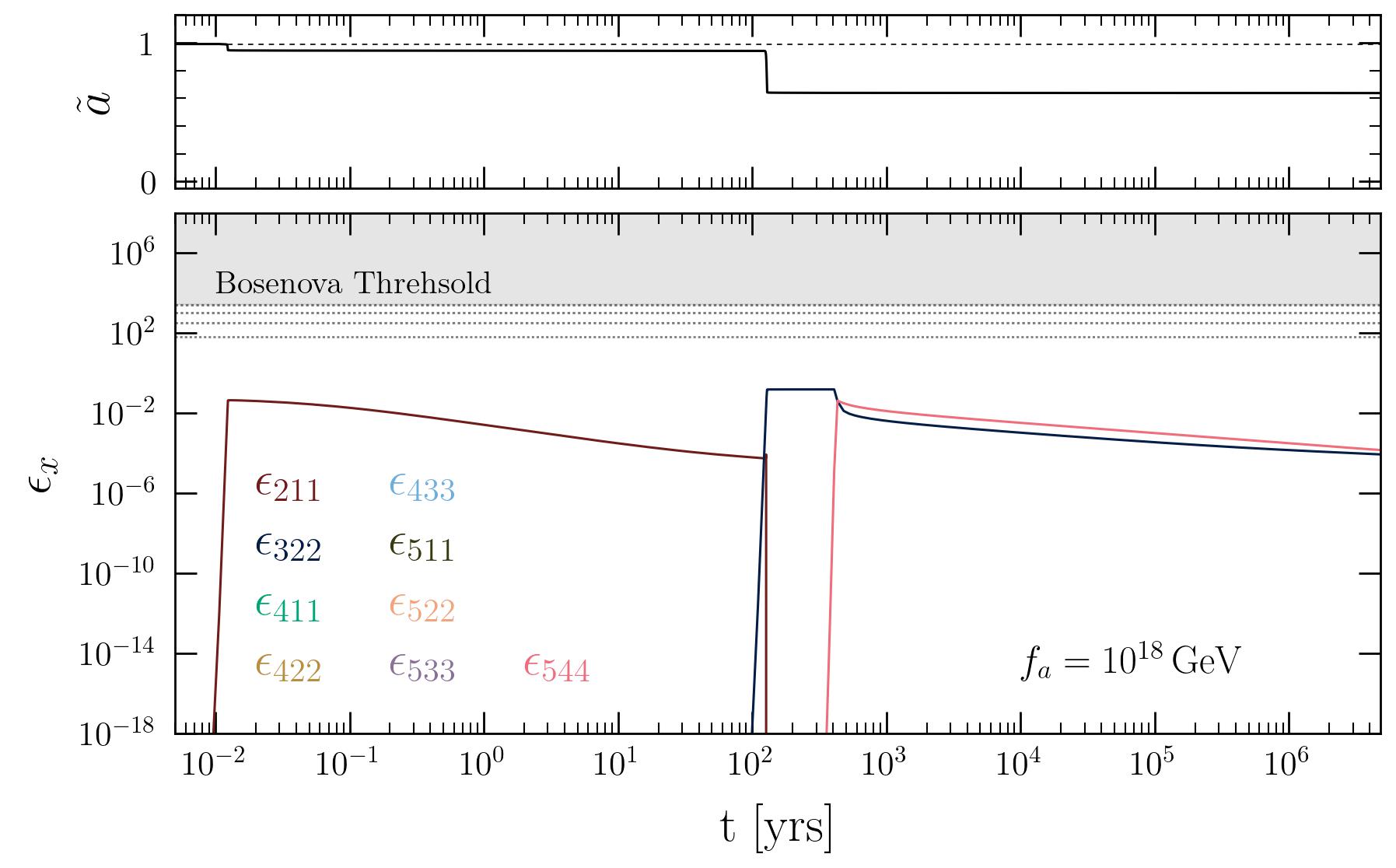}
    \includegraphics[width=0.49\textwidth]{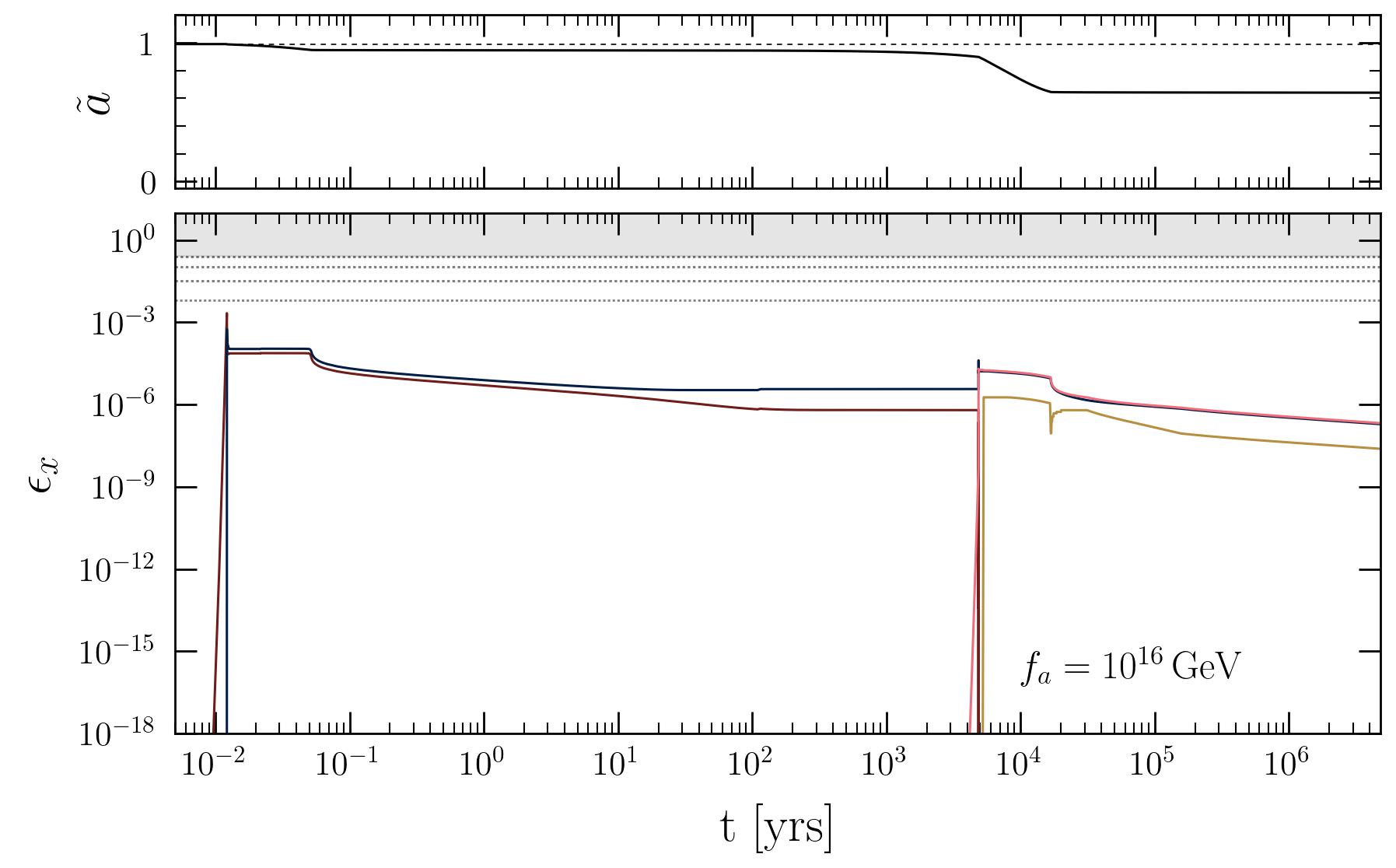}
    \includegraphics[width=0.49\textwidth]{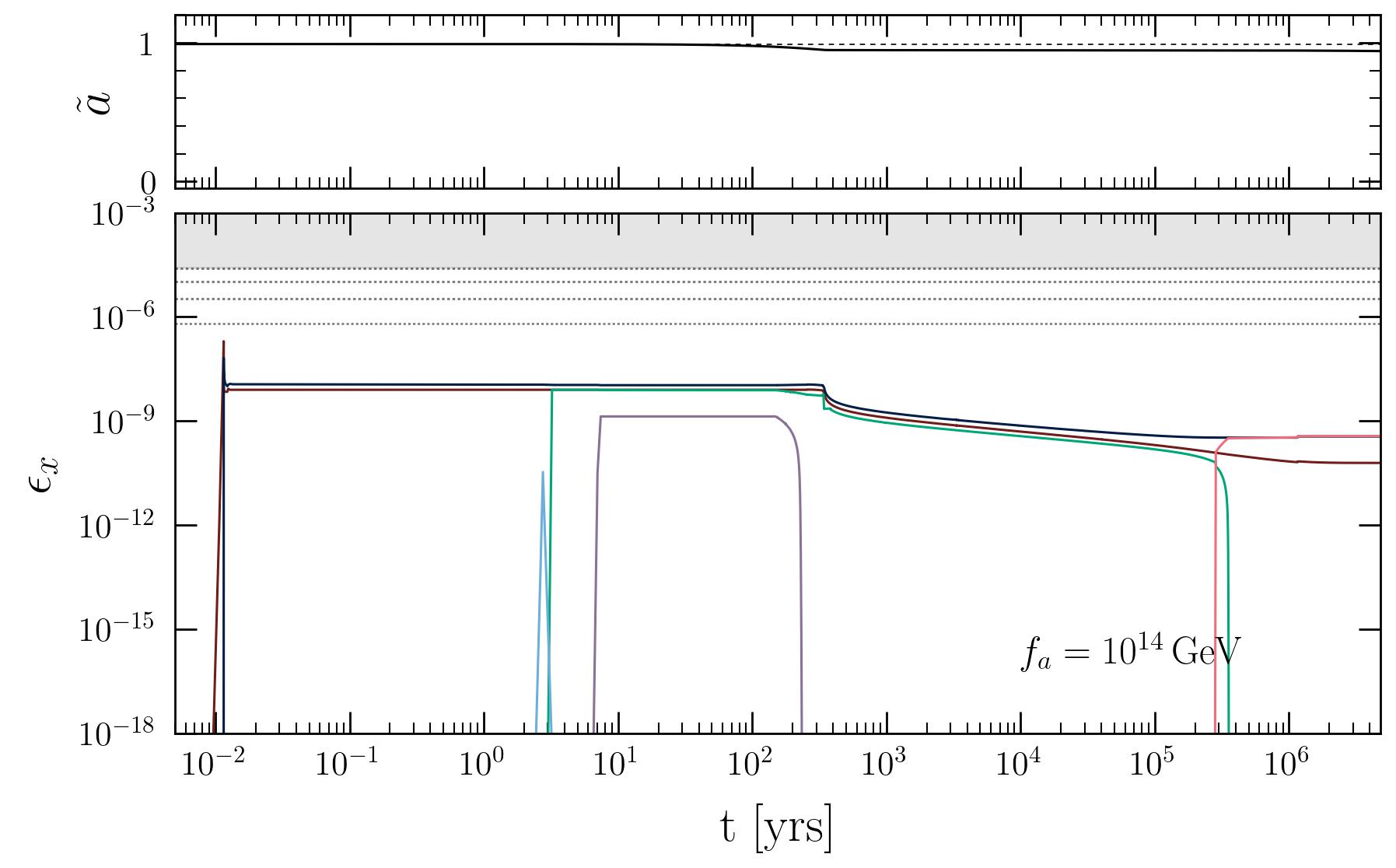}
    \includegraphics[width=0.49\textwidth]{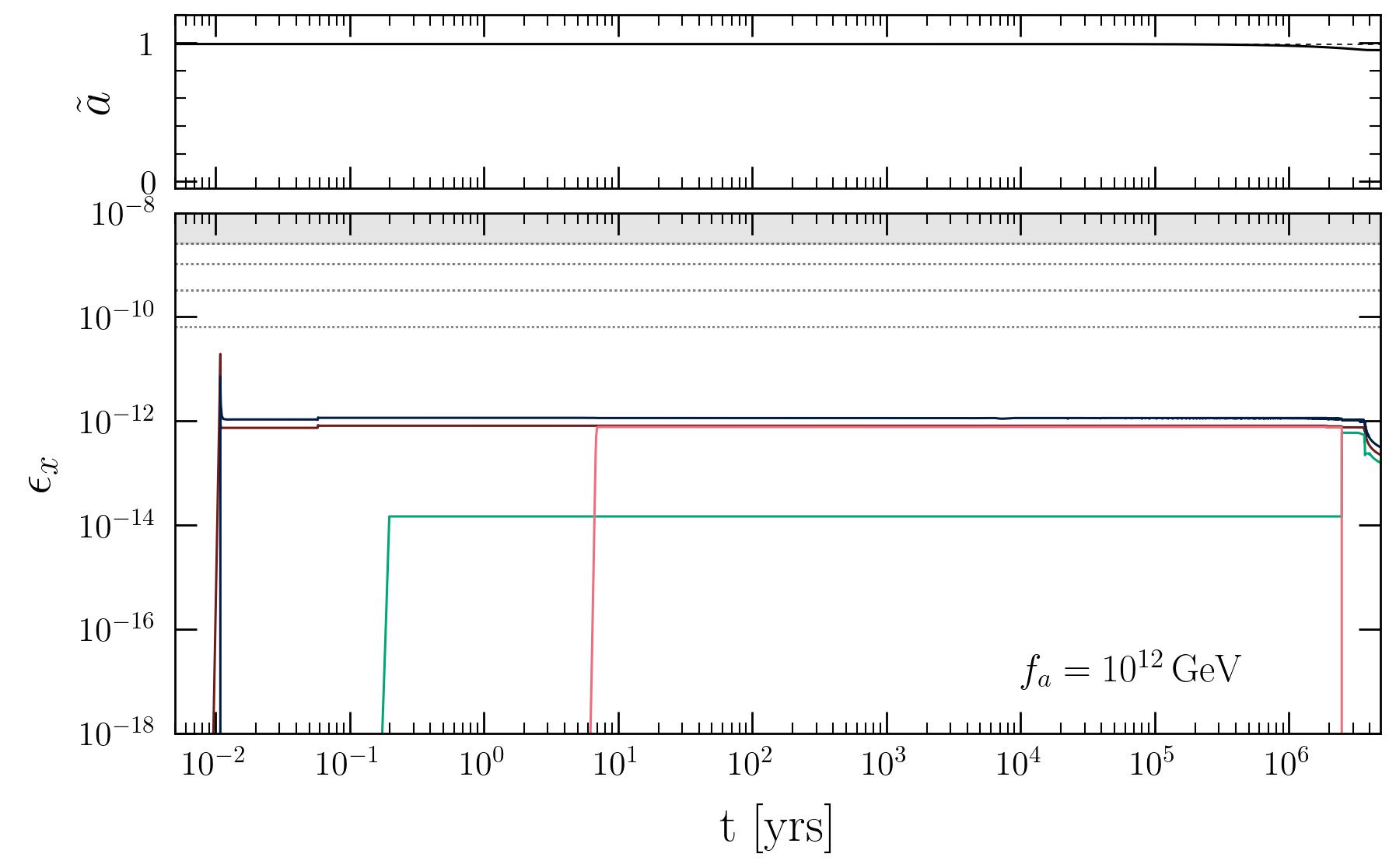}
    \caption{Same as Fig.~\ref{fig:evolve} ($\alpha \simeq 0.33$), but evolving all relevant states at $n \leq 5$, and including relativistic corrections to all scattering processes. }\label{fig:evolve_5_REL}
\end{figure*}

In Fig.~\ref{fig:f14_n4_n5}, we compare the evolution of the $n\leq 4$ and $n\leq 5$ systems for a fixed value of $\alpha$ and  $f_a$, including  relativistic corrections in both systems. Here, one can see that the inclusion of the $\left| 544 \right>$ state serves to quench the growth of the $\levthree$, and prevent the growth of the $\levfourtwo$ and $\levfourthree$ states (which were previously grown via superradiance). This state prevents the appearance of a bosenova, and delays and suppresses spin down.

In Fig.~\ref{fig:evolve_5_REL}, we show the evolution of the same state as Fig.~\ref{fig:evolve} (namely, $\alpha \sim 0.33$), but including all states up to $n \leq 5$ (including relativistic corrections). By comparing these figures one can see that the behavior qualitatively quite different, leading to a more complex evolution for values of $f_a \lesssim 10^{16}$ GeV.

\begin{figure*}
    \includegraphics[width=0.47\textwidth]{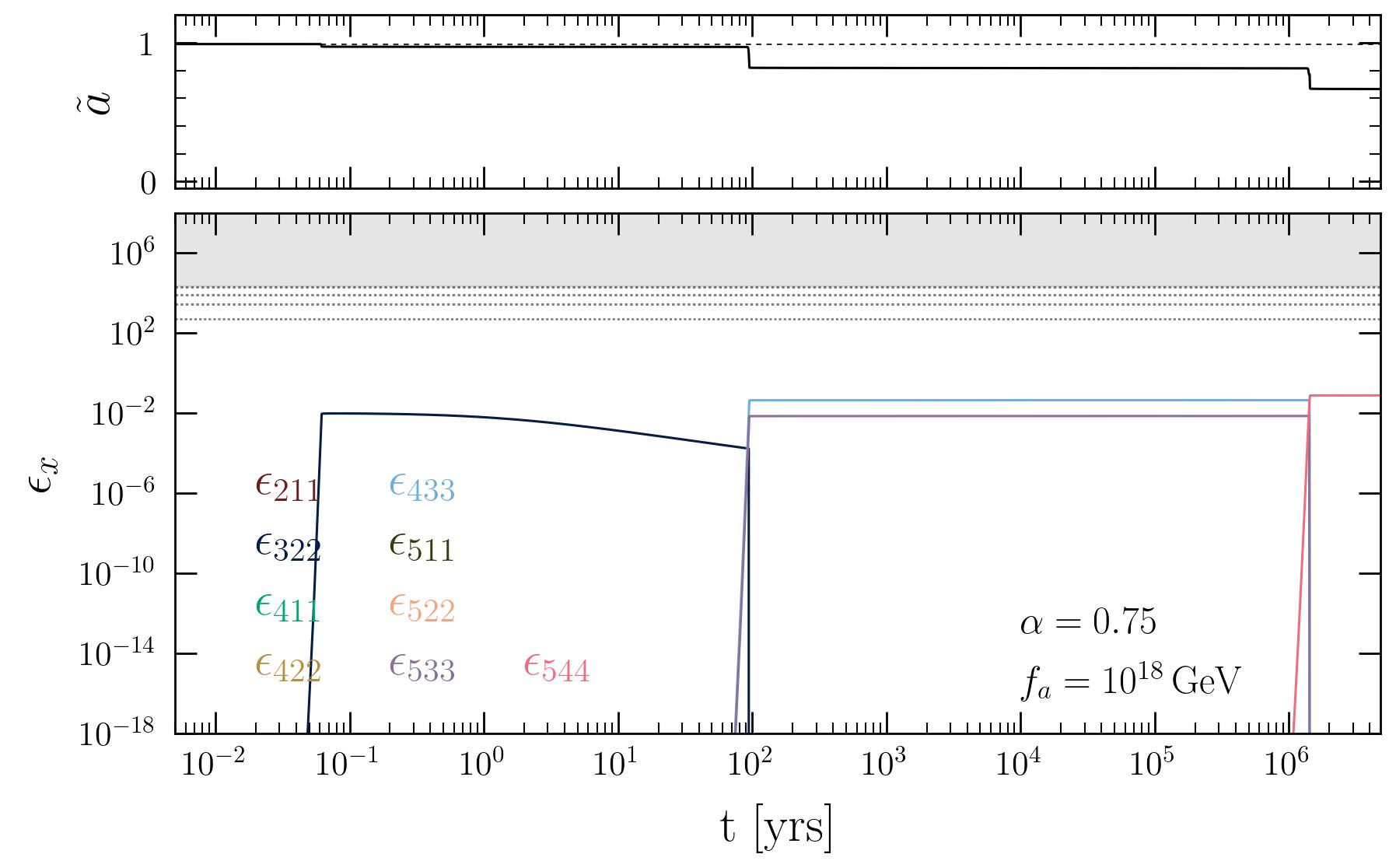}
    \includegraphics[width=0.47\textwidth]{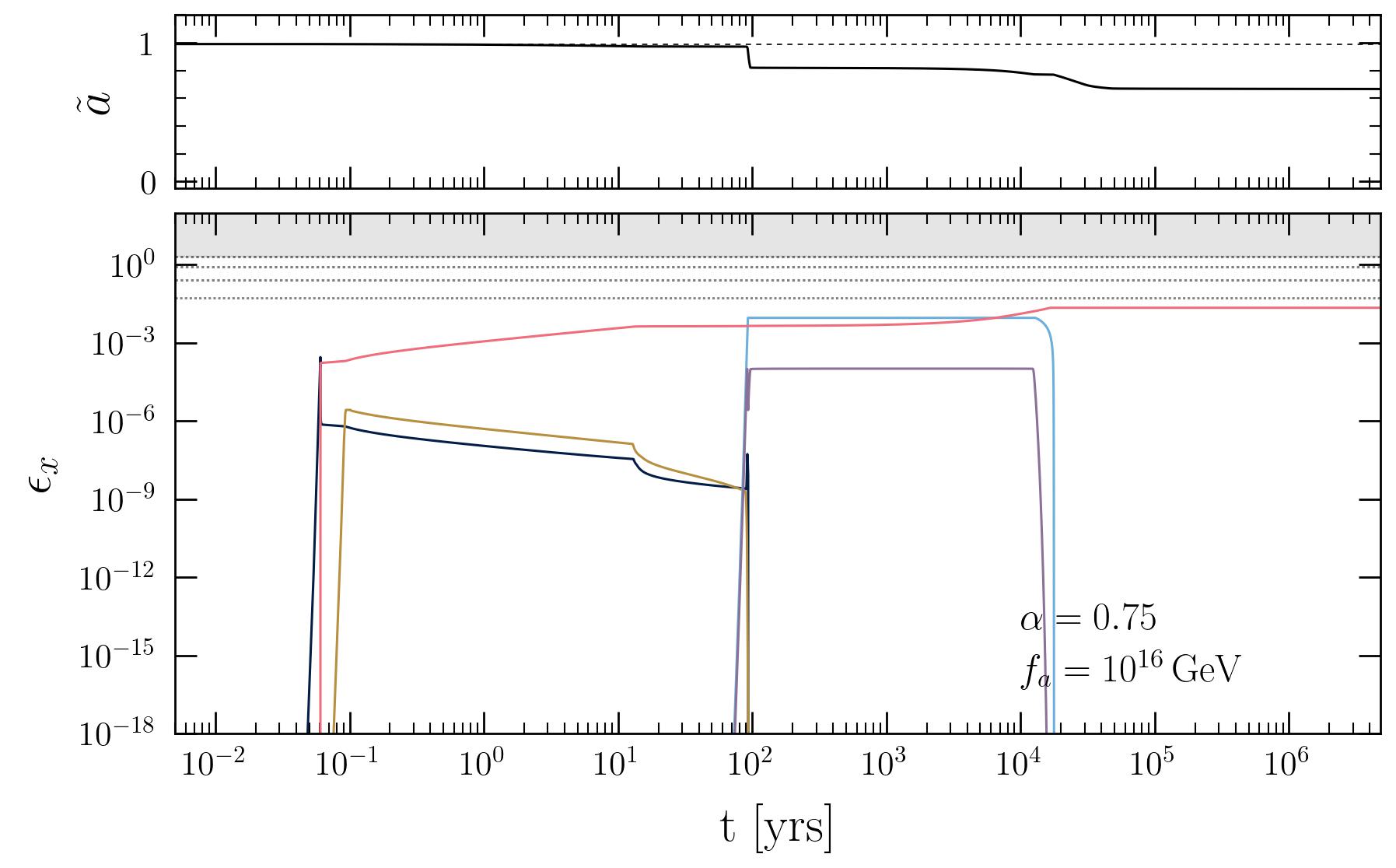}
    \includegraphics[width=0.47\textwidth]{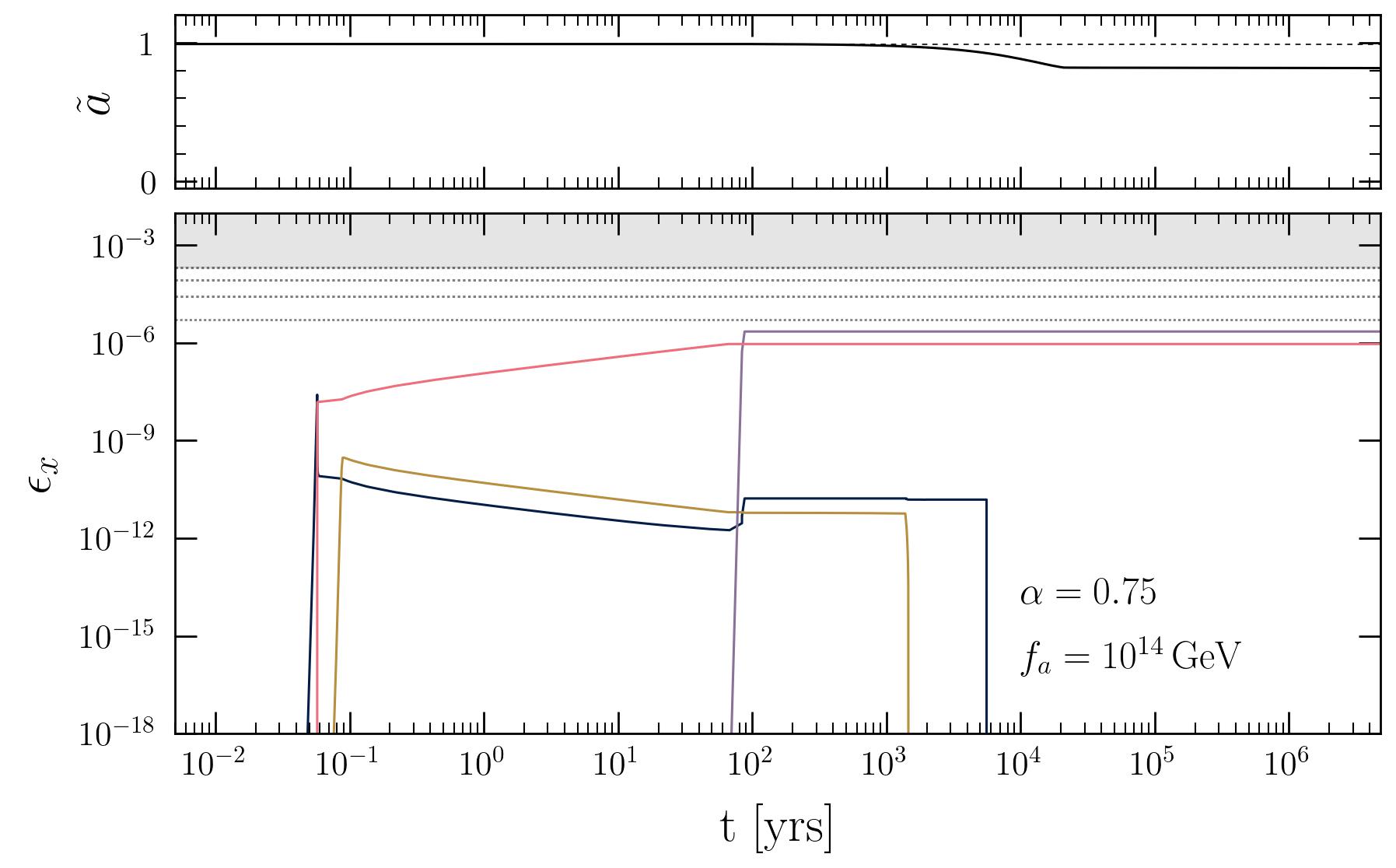}
     \includegraphics[width=0.47\textwidth]{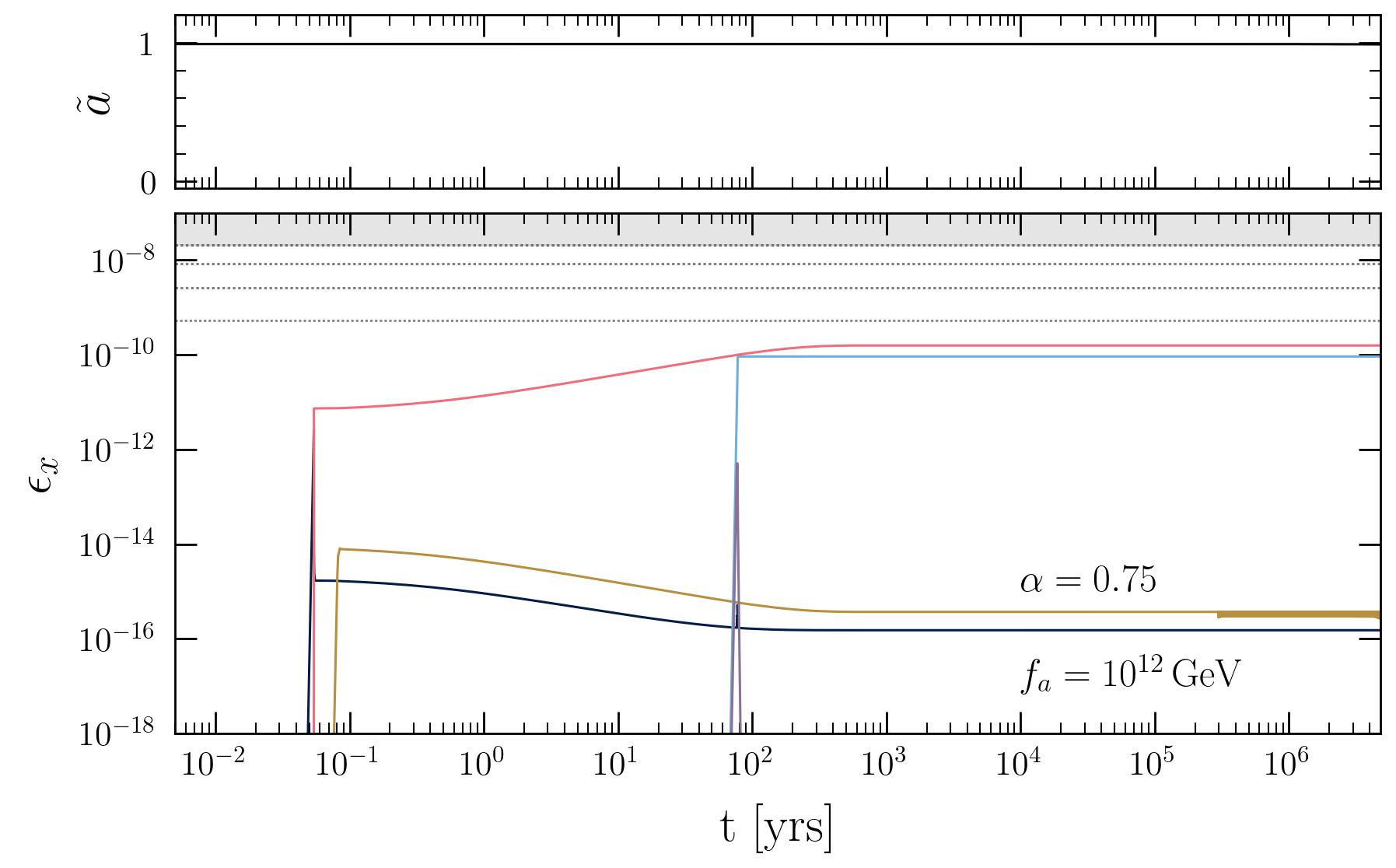}
    \caption{Same as Fig.~\ref{fig:evolve_5_REL}, but for $\alpha = 0.75$.  }
    \label{fig:evolution_highmass}
\end{figure*}

In Figs.~\ref{fig:evolution_highmass} and ~\ref{fig:evolution_highmass2} we show the evolution of the full system in the high mass regime, specifically at $\alpha = 0.75$ and $\alpha = 0.83$. Here, one sees that for values of $f_a \lesssim 10^{16}$ GeV, the $\left| 544 \right>$ state is grown via the $\levthree$ state, suppressing the characteristic occupation numbers of the $m=2$ states, and therefore  the $m=2$ spin down.  At smaller values of $f_a$, the suppressed $\levthree$ occupation numbers fully prevent spin down on all relevant timescales.  For the highest axion mass shown (see Fig.~\ref{fig:evolution_highmass2}), the occupation numbers of the $\levthree$ level are actually not produced via superradiance; the $\levthree$ is superradiant, but has a timescale much longer than the Salpeter timescale, and the large occupation numbers seen at small $f_a$ are instead realized via the scattering of higher order states (meaning that spin down is not arising at $m=2$, but rather at $m \geq 3$).

\subsubsection{Prospects for Going Beyond $n = 5$}

As mentioned above, truncating states at $n \leq 5$ prevents one from confidently inferring the evolution of spin down at $m \geq 3$ -- this is simply a consequence of the fact that one of the most efficient dissipation channels will proceed via $\left| n 33 \right> \times \left| n 33 \right> \rightarrow \left| n^\prime 6 6 \right> \times {\rm BH}$, with $n^\prime \geq 7$. If such states grow, they can establish equilibria with suppressed occupation numbers in a manner analogous to the $\levtwo-\levthree$ (or the $\levthree-\left|544\right>$) system. While this poses a challenge in understanding the evolution of $m \geq 3$, the dominant energy dissipation channels for the $m\leq 2$ states are included in the $n \leq 5$ system; as such, it is natural to ask whether this spin down at $m=2$ is stable with respect to the inclusion of the  $n \geq 6$ states. 

One way to approach this problem was outlined in~\cite{Baryakhtar:2020gao}, where they studied the stability of the two-level system to higher order states in the hydrogen-like limit. In this case, the two-level system admits a well-defined quasi-stable equilibria with occupations numbers $\epsilon_{322}^{\rm eq}, \epsilon_{211}^{\rm eq}$ that can be computed for any  value of $\alpha$ and $f_a$. Starting from these equilibrium occupation numbers, one can then analyze the size of terms contributing to the growth and dissipation of higher order states -- it turns out that for many states, dissipation in the low$-\alpha$ limit is favored, allowing one to argue that the two-level system is stable for $\alpha \lesssim 0.22$ (although, as mentioned above, relativistic corrections push this value slightly lower). 

While this should probably be thought of as the preferred `first' approach, it becomes dramatically more complicated at larger $\alpha$; this is because $(i)$ there exists not a single quasi-stable equilibrium configuration, but many such configurations, including configurations with more than two states (implying one needs to perform the equilibrium stability analysis across many subsets of parameter space, with each analysis being more complicated due to the larger permutations of scattering processes),  $(ii)$ the number of terms contributing to the growth or dissipation of higher order states increases rapidly with $n$, $(iii)$ relativistic corrections quickly break the simple scaling relations, complicating the clean analytic comparisons that are used in the hydrogen-like limit (this was already seen above, where relativistic corrections disrupted the two-level system at $\alpha \sim 0.15$), and $(iv)$ unlike in the two-level low-$\alpha$ system, many states {\emph{do}} grow at large $\alpha$ and large $n$ (the low-$\alpha$ hydrogen-like limit was simple exactly because one could argue that states do not grow). To some degree, the complexity of applying such an approach can already be seen merely by analyzing the complexity of the behavior seen in Figures~\ref{fig:f12_lowmass}-\ref{fig:evolution_highmass2}.

\begin{figure*}
    \includegraphics[width=0.47\textwidth]{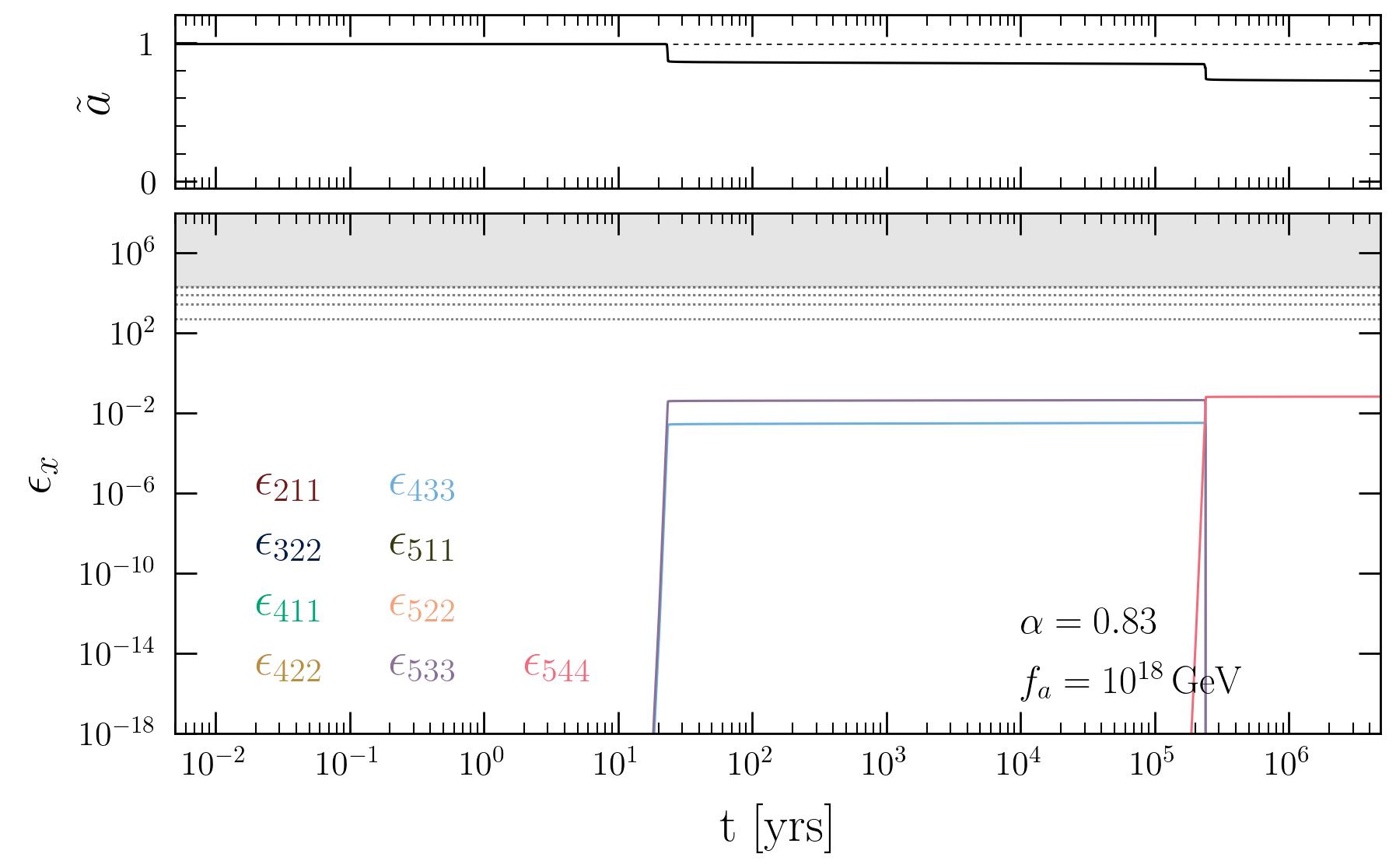}
    \includegraphics[width=0.47\textwidth]{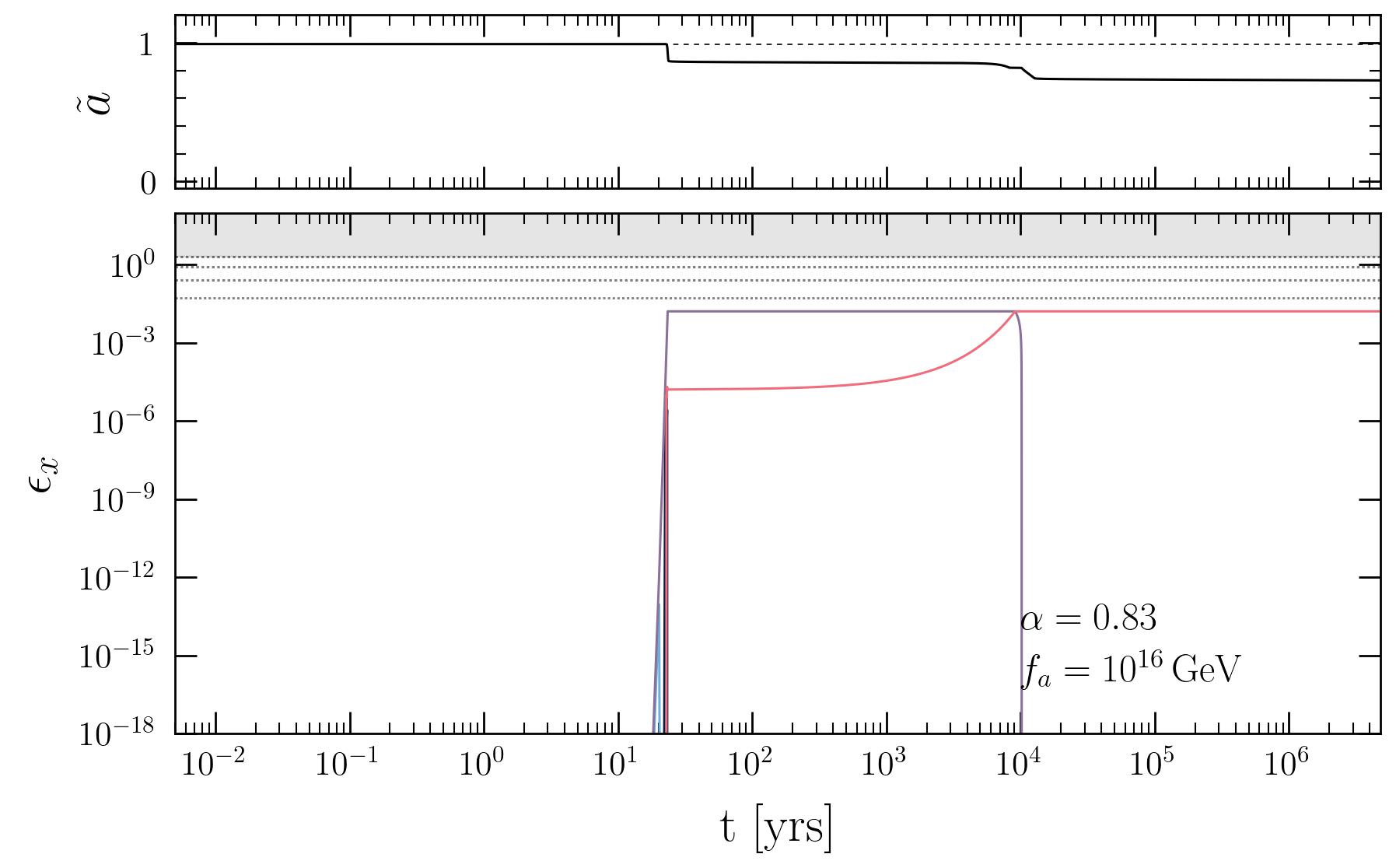}
    \includegraphics[width=0.47\textwidth]{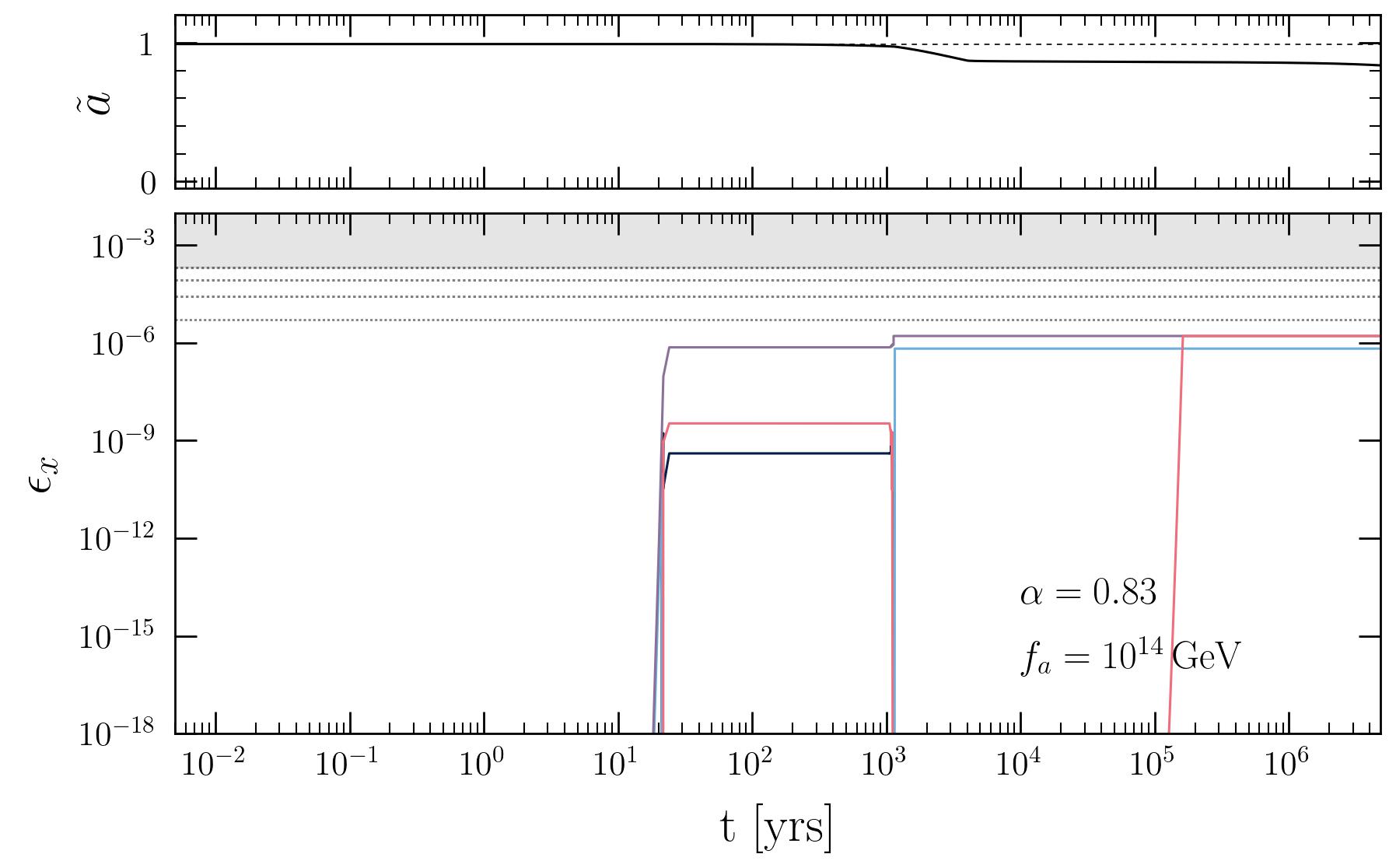}
     \includegraphics[width=0.47\textwidth]{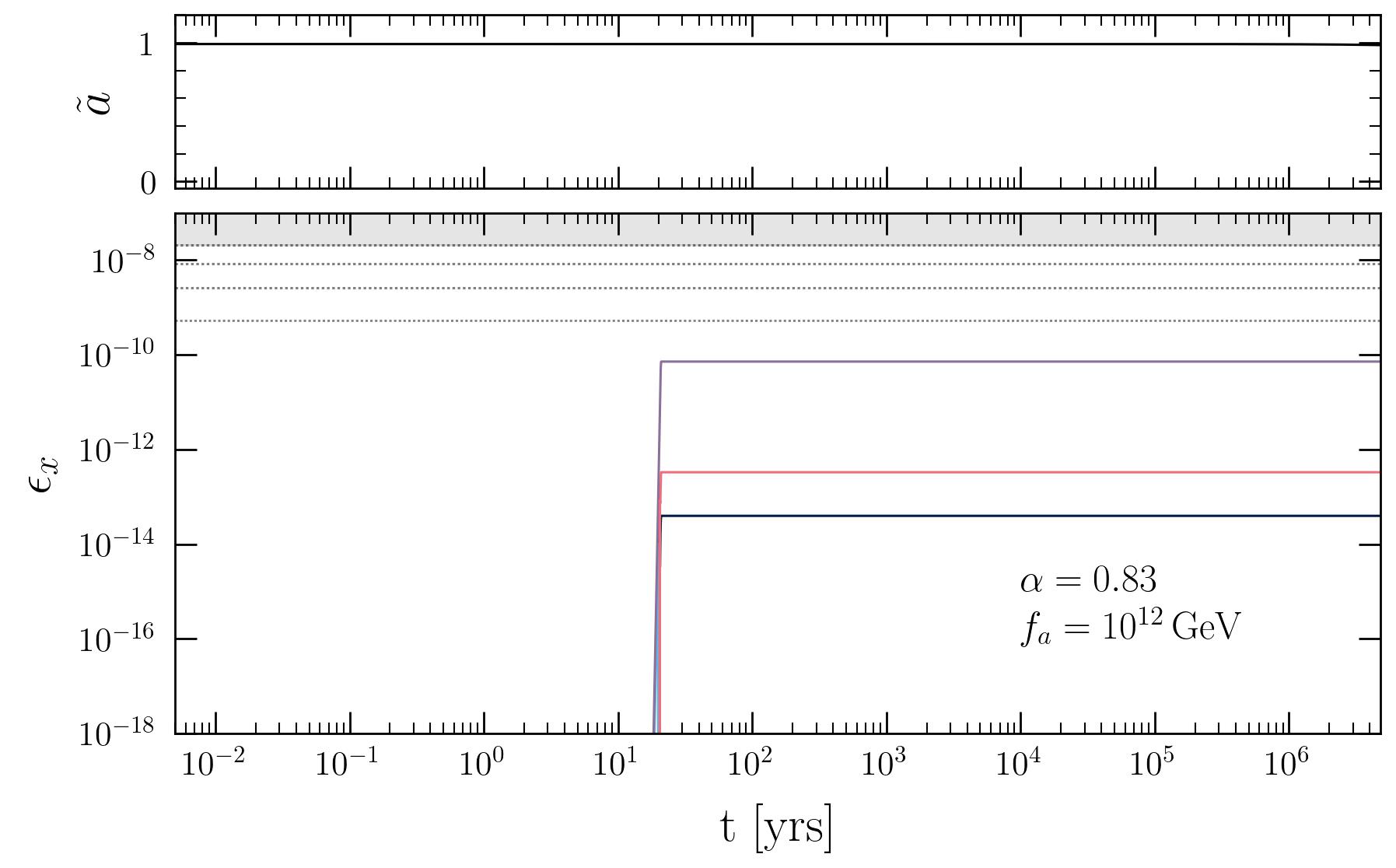}
    \caption{Same as Fig.~\ref{fig:evolve_5_REL}, but for $\alpha = 0.83$.  }
    \label{fig:evolution_highmass2}
\end{figure*}

As an illustrative example, we consider here how one might attempt to go about extending this approach beyond the two level system. For the sake of simplicity we focus on the range of $\alpha$ such that the  $m=1$ levels are no longer superradiant, but the $m=2$ levels are (e.g., this roughly corresponds to $ 0.4 \lesssim \alpha \lesssim 0.7$ for black holes with spin $\tilde{a} \sim 0.95$) -- this simplification allows one to neglect the $m=3$ levels as these cannot be grown via scattering processes, and the superradiant timescale is typically quite long. For this range of parameters, the $\levthree$ state is the fastest to grow, quickly driving the growth of the $\left| 544 \right>$ for even moderate values of $f_a$. These two states typically establish an equilibrium, with $\epsilon_{322} / \epsilon_{544} \sim \mathcal{O}(0.5)$ for the relevant range of $\alpha$. Once this equilibrium is established, one can show that the $\left| 522\right>$ state cannot grow -- this is because self-interactions cannot grow this state without the presence of an $m=1$ state, and because superradiance is inefficient relative to the $\levthree \times \left| 522\right> \rightarrow \left| 544\right> \times {\rm BH}$ and $ \left| 544\right> \times \left| 522\right> \rightarrow \levthree \times \infty$ scattering processes (assuming equilibrium distributions for $\epsilon_{322}$ and $\epsilon_{544}$)\footnote{We caution the reader, however, that the ratio $\gamma^{\rm SR}_{522} / (\left[ \gamma_{322 \times 522}^{544 \times {\rm BH}} + \gamma_{544 \times 522}^{322 \times \infty} \right]\epsilon_{322}^{\rm eq} \epsilon_{522}^{\rm eq}) < 1$ is extremely close to 1 in the large mass limit, and thus reasonable perturbations to the equilibrium occupation numbers may alter this statement.}. The $\levfourtwo$ state can grow via superradiance, but its growth is quenched by the $\levfourtwo \times \levfourtwo \rightarrow \left| 544\right> \times {\rm BH}$ process when its occupation number is still small (typically saturating near $ \epsilon_{322} $), and therefore it does not disrupt the $\levthree-\left| 544\right>$ equilibrium. Bearing this in mind, the question that remains is whether levels at $n \geq 6$, which have not been included in the numerical analysis presented here, could invalidate the inferred $m=2$ spin down timescale.

Starting from a three-state equilibrium, one can see that the leading states (leading here implying the smallest $n$ state) which could potentially be produced via self-interactions include: $\left| 644 \right>$ (from $\levthree \times \levthree \rightarrow \left| 644 \right> \times {\rm BH}$, or from $\levthree \times \levfourtwo \rightarrow \left| 644 \right> \times {\rm BH}$), $\left| 766 \right>$ (from $\levthree \times \left| 544\right> \rightarrow \left| 766 \right> \times {\rm BH}$, or from $\levfourtwo \times \left| 544\right> \rightarrow \left| 766 \right> \times {\rm BH}$), and $\left| 988 \right>$ (from $ \left| 544\right> \times \left| 544\right> \rightarrow \left| 988 \right> \times {\rm BH}$). For each of these states, one can compute all scattering permutations with the three pre-existing states; using the $\alpha$ and $f_a$-dependent occupation numbers of each state, one can then determine if the growth rate of any of these states is positive (and if so, for which values of $\alpha$). We find that neither the  $\left| 644 \right>$ state nor the $\left| 766 \right>$ state  grow for any value of $\alpha$\footnote{Let us emphasize that the conclusions drawn using this approach crucially depend on the inclusion of relativistic corrections everywhere, as the distinction between growth and decay often boils down to $\mathcal{O}(1)$ numbers appearing in the growth/decay rates.
} where the $m=2$ state is superradiant, while the $\left| 988 \right>$ state instead grows for all relevant values of $\alpha$. Despite the appearance of the $\left| 988 \right>$ state, the energy being extracted via this scattering process is sufficiently low so as to have a minimal effect on the equilibrium value of $\epsilon_{322}$ (which, in the end, is responsible for the $m=2$ spin down).

Despite the fact that the  $\left| 988 \right>$ level did not significantly disrupt the $m=2$ spin down, one cannot stop here. First, it remains possible that this new level pumps energy into other states, which in turn disrupt the $\levthree-\left| 544\right>$ equilibrium, thereby delaying $m=2$ spin down. Therefore, one must once again compute all scattering permutations, including the newly populated $\left| 988 \right>$ state, with this process stopping only when all scattering permutations do not populate new levels. Second, the story is again complicated by the fact that stopping at `lowest-order' in $n$ is not guaranteed to be sufficient. In fact,  the $\levthree \times \levthree \rightarrow \left| 544\right> \times {\rm BH}$ is {\emph{as efficient as}} the $\levthree \times \levthree \rightarrow \left| 644\right> \times {\rm BH}$ process in the $\alpha \rightarrow 0$ limit (actually, this process is nearly degenerate at $n=5,6,7,8,9,$ and $10$, only beginning to fall at $n \geq 11$)! Now it turns out that the presence of these higher order states (at least at leading order) alone do not significantly alter the $m=2$ spin down. This is because the equilibrium occupation number receives corrections which scale as $\epsilon_{322}^{\rm eq} \sim \sqrt{\frac{\gamma_{n44 \times n44}^{322 \times \infty}}{\gamma_{544 \times 544}^{322 \times \infty}}}$, and $\gamma_{n44 \times n44}^{322 \times \infty}$ falls off more quickly with $n$.   Nevertheless, the occupation numbers of such states are not negligible, and should be carefully considered if one is to argue for well-controlled equilibrium configurations. For reference, the ratio of the higher-$n$ scattering rates for the $\levthree \times \levthree \rightarrow \left| n44\right> \times {\rm BH}$ and $\left| n44\right> \times \left| n44\right> \rightarrow \left| 322\right> \times \infty$ processes are shown in Figs.~\ref{fig:n44}.

Rather than exhausting all permutations, one can instead try to intelligently guess which levels could potentially reduce the equilibrium occupation number of the $\levthree$ state (since this would be the only way in which constraints derived from the $n \leq 5$ levels would overestimate sensitivity). The lowest lying state with $m \geq 2$ which has a resonant scattering (in the hydrogen-like limit)  with the $\levthree$ level is $\left| 522\right>$. This state can be directly populated via $\left| 988 \right> \times \left| 988 \right> \rightarrow \left| 522 \right> \times {\rm BH}$. We had previously argued that the $\left| 522 \right>$ state does not grow -- we can now ask whether the appearance of the $\left| 988 \right>$ state changes this conclusion, and if so, how large can the occupation number of the $\left| 522 \right>$ state become. We find that the $\left| 522 \right>$  level remains unpopulated, even when including the additional channel. Moreover,  no such resonant processes are present at $n=6$ or $n=7$. This is again suggestive of the fact that the $\levthree$ state is difficult to deplete, but  numerical validation is likely necessary to confirm this conclusion.

An alternative approach is  to focus not on the exact stability of the occupation numbers themselves, but rather  on the characteristic stability of the spin-down process. The general idea here is to recognize that energy is injected into the bound states with low-$m$, and the role of self-interactions is to re-distribute some of that energy to higher order states (with energy re-distribution being most efficient when high-occupation states have strongly overlapping radial wavefunctions). This naturally introduces a hierarchy of scales, such that large$-n$ states should play an increasingly unimportant role in modifying the evolution of low$-n$ states. This statement does not imply that large $n$ states are not populated, and of course does not hold on a level-by-level basis (as is seen e.g. by the fact that the $\levthree$ state is primarily modified by the $\left| 544\right>$ state), but  rather should hold as a general trend. This is, however, reflected in the example provided above, and is also reflected in the hydrogen-limit scattering rates of Tables~\ref{tab:rates}-\ref{tab:rates_n5} (for the sake of simplicity we do not list all possible scattering rates, but rather a subset; all possible scattering permutations are nevertheless included in the evolution of the superradiant system).

\begin{figure*}
    \includegraphics[width=\textwidth]{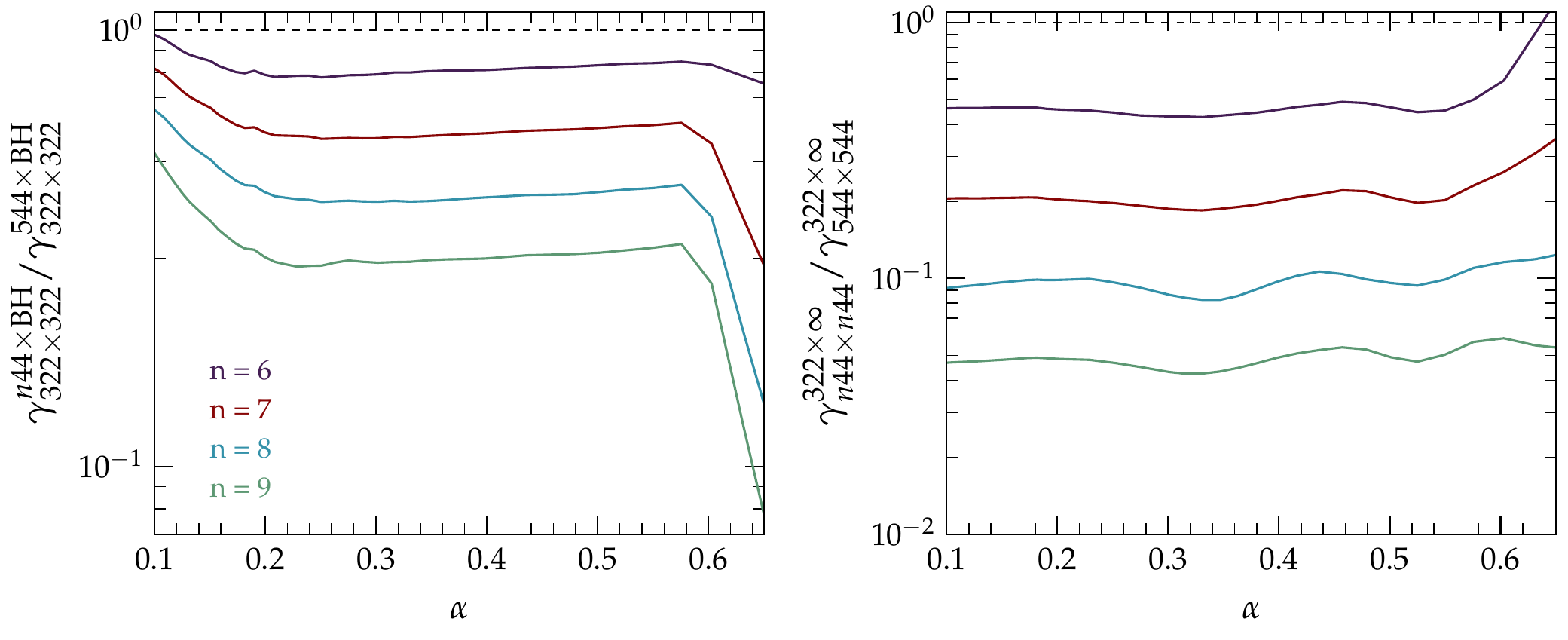}
    \caption{\label{fig:n44} Ratio of the $\levthree \times \levthree \rightarrow \left| n44\right> \times {\rm BH}$ (left) and $\left| n44\right> \times \left| n44\right> \rightarrow \levthree \times \infty$ (right) to the equivalent $n=5$ processes as a function of $\alpha$. Higher $n$ processes are efficiently produced from the $\levthree$ state, but do not lead to significant modifications of the $\levthree$ equilibrium occupation number (see text for discussion).  }
\end{figure*}

This introduces the possibility of a brute force approach, where one attempts to study of the stability of low-$m$ spin down (rather than stability of the occupation numbers themselves) for a particular system by extending the analysis level by level. The conclusion derived in Ref.~\cite{Baryakhtar:2020gao} that the two-level system is a complete description at $\alpha \lesssim 0.1$ would also be easily seen using this approach, as no higher order levels would be populated at small $\alpha$, regardless of how many levels were introduced into the analysis; this approach, the other hand, evades the issue of having to search for tens, hundreds, or even thousands of different quasi-stable equilibria and scattering permutations which may be present during the evolution of a single system (should such equilibria even exist). Needless to say, caution must be taken when applying this approach, since one must extend to sufficiently large $n$ to ensure that stability is achieved. In this work we truncate at $n=5$, which we argue is likely sufficient for the $m=2$ spin down of typical high spin solar mass black holes, but intend to extend to higher $n$ in future work.

In order to demonstrate this approach, we study the spin down stability of a black hole consistent with the fiducial parameters of Cygnus X-1. Specifically, we look at the final spin of the black hole for a range of values of $\alpha$ and $f_a$, performing the $n\leq 3$, $n\leq 4$, and $n \leq 5$ analysis. We perform two tests -- one in which only the $m \leq 2$ states can extract spin from the black hole\footnote{The analysis with only $m \leq 2$ spin down is technically not fully self-consistent, since other quasi-bound states grow and can in turn alter the mass of the black hole itself. In order to avoid instabilities in this case, we fix the black hole mass to its initial value and do not evolve it in time. This should not significantly alter the conclusions.}, and one in which all states can extract spin. The result of this analysis is shown in Fig.~\ref{fig:a_final}, with the left panel showing the special $m \leq 2$ case, and the right panel showing the general result. 

From Fig.~\ref{fig:a_final}, one can see that the $m\leq2$ spin down analysis (left panel) is stable across the $n\leq3, 4, $ and 5 analyses for nearly all values of $\alpha$ shown, up to $f_a \gtrsim 10^{14}$ GeV; including the higher order spin down (right panel), one can see that for $f_a \lesssim 10^{14}$ GeV the role of higher $m$ spin down is merely to enhance the amount of spin extracted (in other words, $m=1,2$ extract spin maximally, but so do $m=3$ and/or $m=4$). For smaller values of $f_a$, the inclusion of $n=5$ states tends to suppress spin down -- this is true for both analyses (left and right panels, with a small exception arising at large $\alpha$ and $f_a = 10^{13}$ GeV of the right panel), and arises merely because the $\left| 544 \right>$ state quenches the growth of the $\levthree$ state. What is interesting to note, however, is that as $f_a$ is decreased, the results of the $m \leq 2$ analysis converge to those of the full analysis, with convergence first arising at smaller value of $\alpha$ and larger values of $f_a$, and pushing toward covering the full parameter space. This is strongly suggestive of the fact that the $n \leq 5$ analysis is relatively robust, so long as the $m \leq 2$ states are capable of extracting sufficient spin to be constrained by observational measurements (note that this  naturally applies an upper axion mass for which results can be trusted, since for $\alpha \gtrsim 0.7$ one does not expect $m \leq 2$ to be capable of sizably altering the black hole spin). Finally, let us note that we have already argued above that states at  $n=6$ will not grow via self-interactions for $\alpha \gtrsim 0.4$, and the $\left| 622\right>$ will only have a minimal effect on the $m=2$ spin down -- as such, we expect the $n \leq 6$ result to look nearly identical to that of the $n=5$. Similar conclusions hold at $n=7$ (the $\left| 722\right>$ state is less influential than the $\left| 622\right>$, the $\left| 744\right>$ state will extract energy from the $m=2$ state less efficiently than the $\left| 544\right>$ state, and the $\left| 766\right>$ state does not grow). 

The above conclusion only applies to the specific example at hand, and thus we choose to repeat the same exercise, but adopting a slightly smaller initial spin, $\tilde{a} = 0.97$, and lifetime set to the Salpeter timescale (roughly reflecting parameters consistent with the spin down analysis of GRS 1915+105, see following section) -- the result is shown in Fig.~\ref{fig:a_final_2nd}. As before, one can see for $f_a \gtrsim 10^{14}$ GeV and $\alpha \gtrsim 0.4$, the left panel shows that the $n\leq 3,4$ and 5 analysis are  convergent. As before, the full analysis (right panel) consistently shows that the higher order states can always increase spin extraction, but this is as expected since the enhanced spin extraction is coming on top of the $m=2$ spin down. For smaller values of $f_a$, we again see that the $n \leq 5$ analysis seems to be consistent between the left and right panels, except for large values of $\alpha \gtrsim 0.66$. As before, this is suggestive of the fact that the $m = 3,4$ states are further enhancing spin extraction. Since we do not include the relevant states which quench $m = 3,4$ spin extraction (which arise at $n \geq 6$), one can still perform a conservative analysis at $\alpha \lesssim 0.66$ with $n\leq 5$ so long as the final spin of the $m\leq 2$ state is sufficiently low to be incompatible with observations. This is the relevant regime for the highly spinning black holes studied in this work.

In effect, let us reiterate that Figs.~\ref{fig:a_final} and ~\ref{fig:a_final_2nd} demonstrate that for black holes which are highly spinning and have strong constraints on the spin itself (meaning that the $m \leq 2$ spin down is sufficient to induce an incompatibility with observations), an analysis using the $n \leq 5$ states should be sufficiently to conservatively derive limits at $\alpha \lesssim 0.66$. Let us further note that for shorter timescales, this conclusion is strengthened -- this is merely because the $m \geq 3$ states require longer timescales in order to extract spin, and thus play an increasingly sub-dominant role in modifying the spin of the black hole.

\begin{figure*}
    \includegraphics[width=0.47\textwidth]{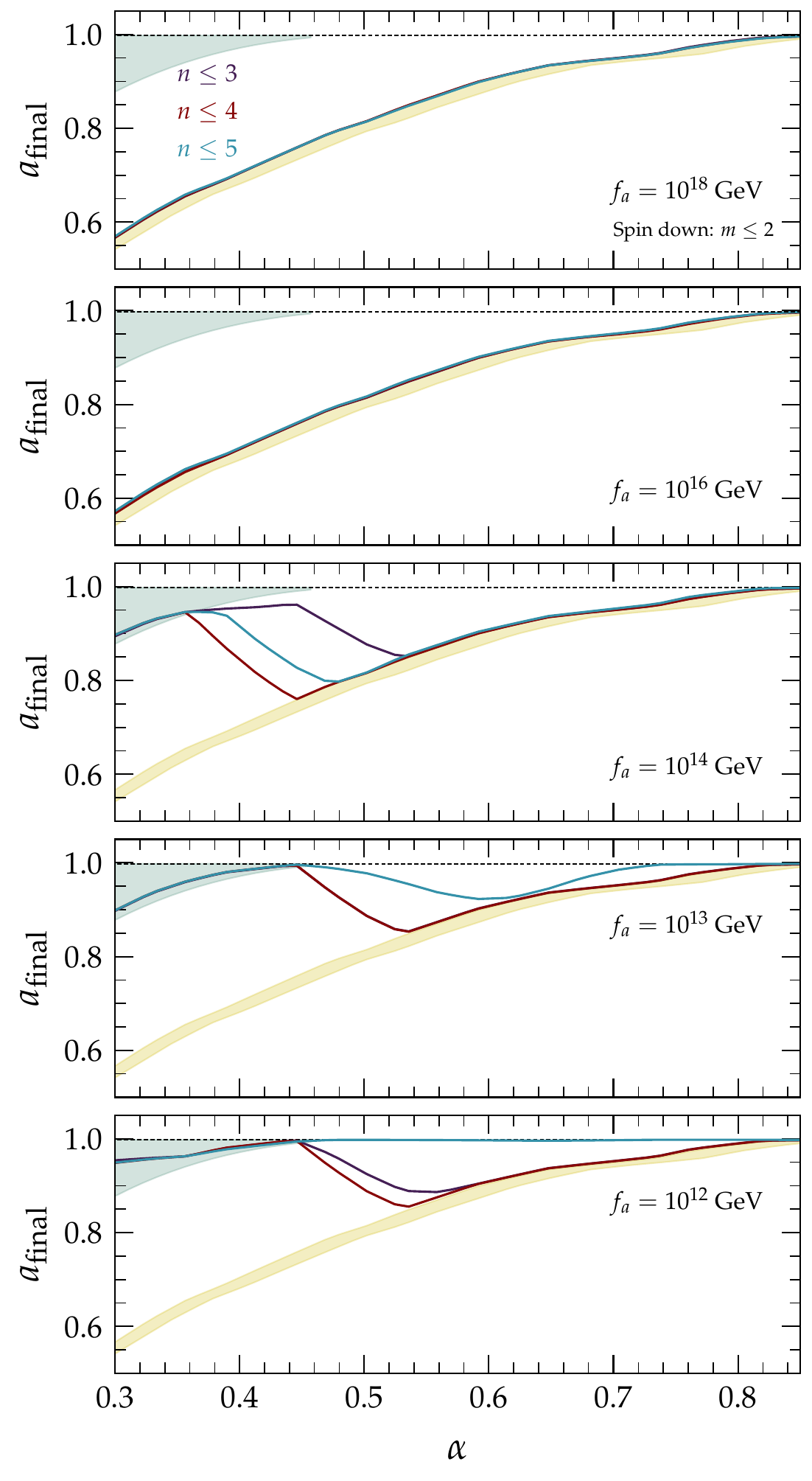}
    \includegraphics[width=0.47\textwidth]{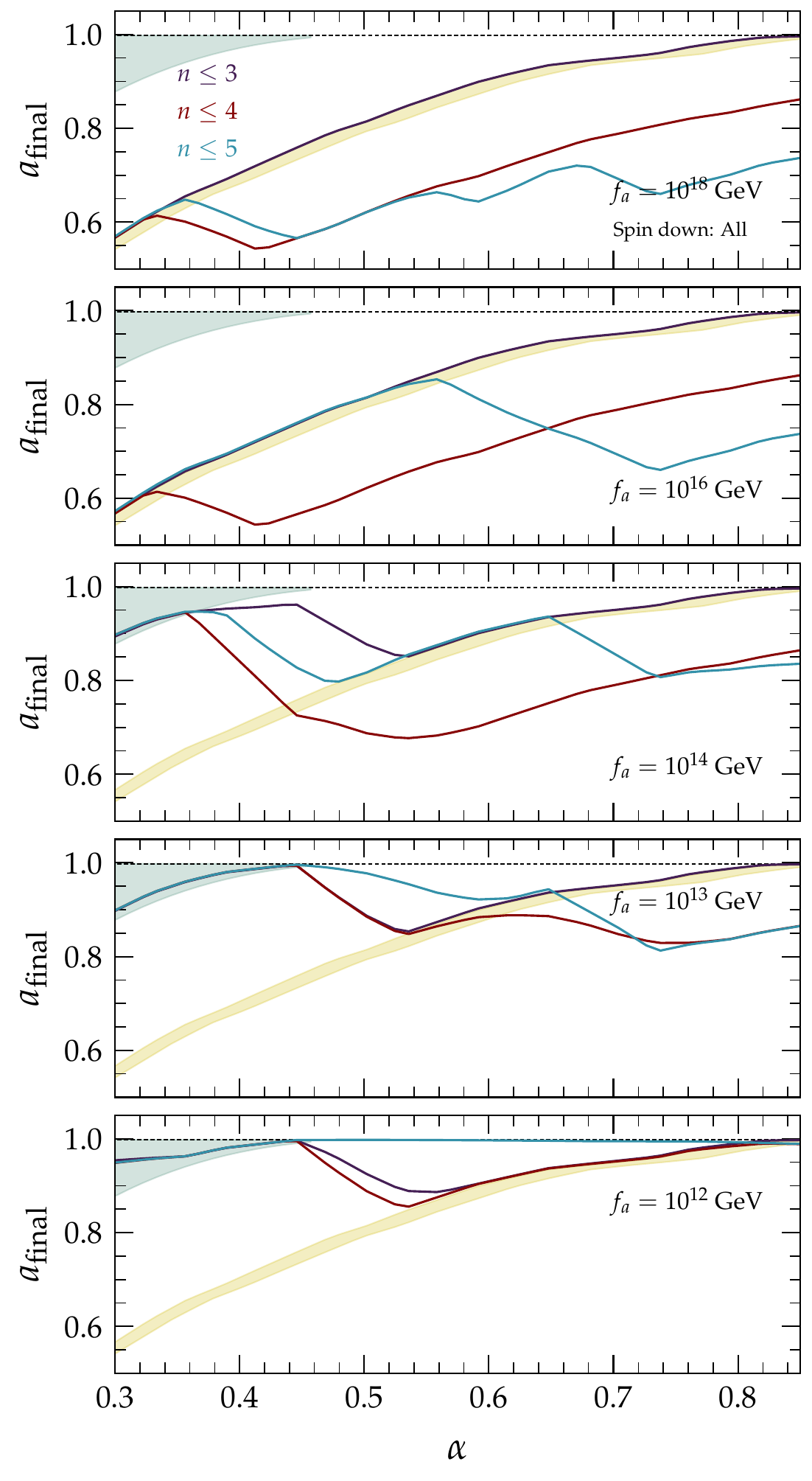}
    \caption{ Final spin at $\tau = 4\times 10^6$ years of a test black hole with the fiducial parameters of Cygnus X-1 as a function of $\alpha$, for fixed values of $f_a$ and for the $n \leq 3, n \leq 4,$ and $n \leq 5$ level systems. Left panel assumes that only the $m=1$ and $m=2$ states can extract rotational energy from the black hole, while right panel does not. The original black hole spin is shown with a vertical dashed line, and the gold shaded band approximately highlights where the $\levthree$ state will have fully spun down (with the band edges being computed by equating $\omega_{322} = 2 \Omega$, assuming a final black hole mass five percent smaller than the initial value). The green shaded regions denote the approximate $m=1$ spin down. }
    \label{fig:a_final}
\end{figure*}

\begin{figure*}
    \includegraphics[width=0.47\textwidth]{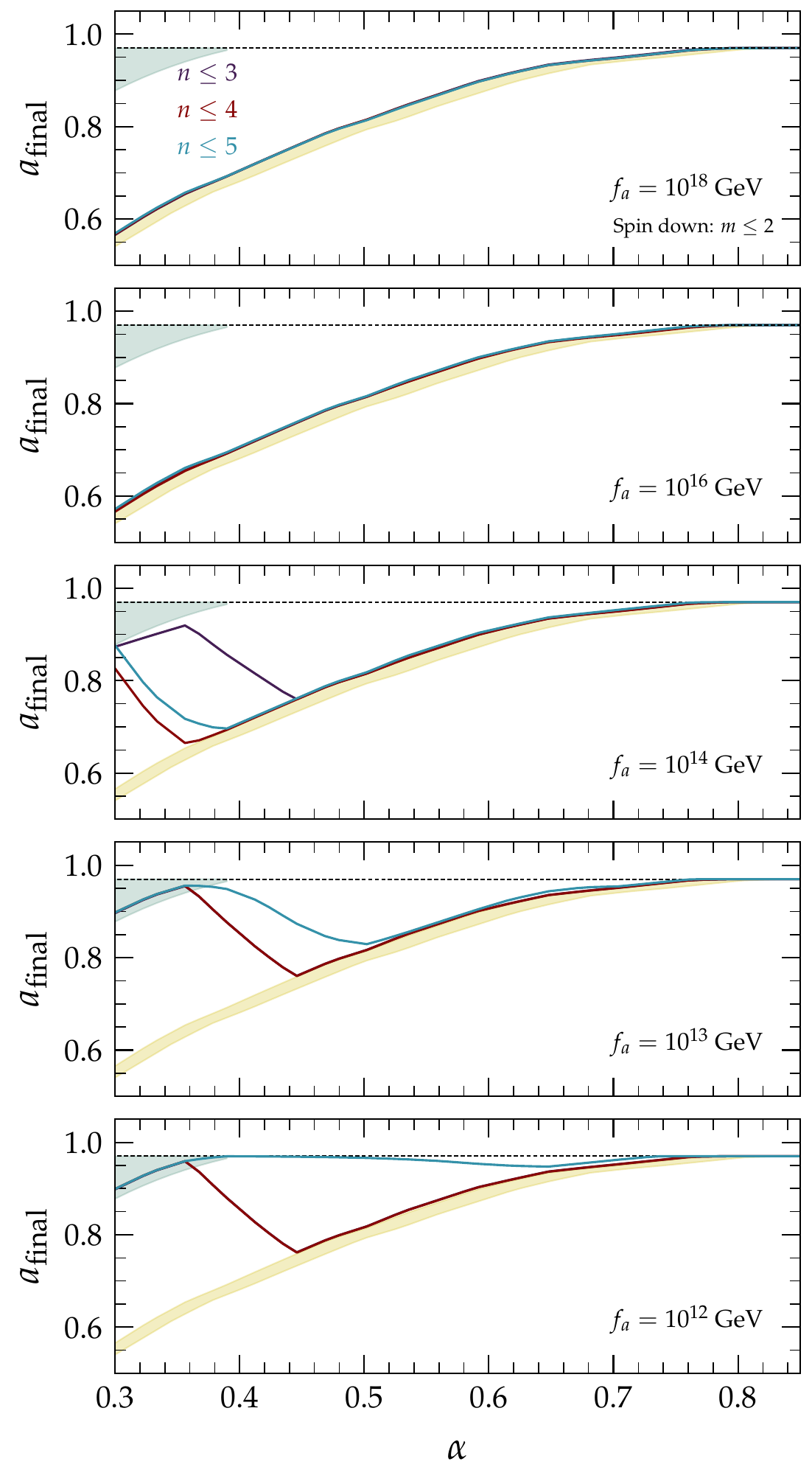}
    \includegraphics[width=0.47\textwidth]{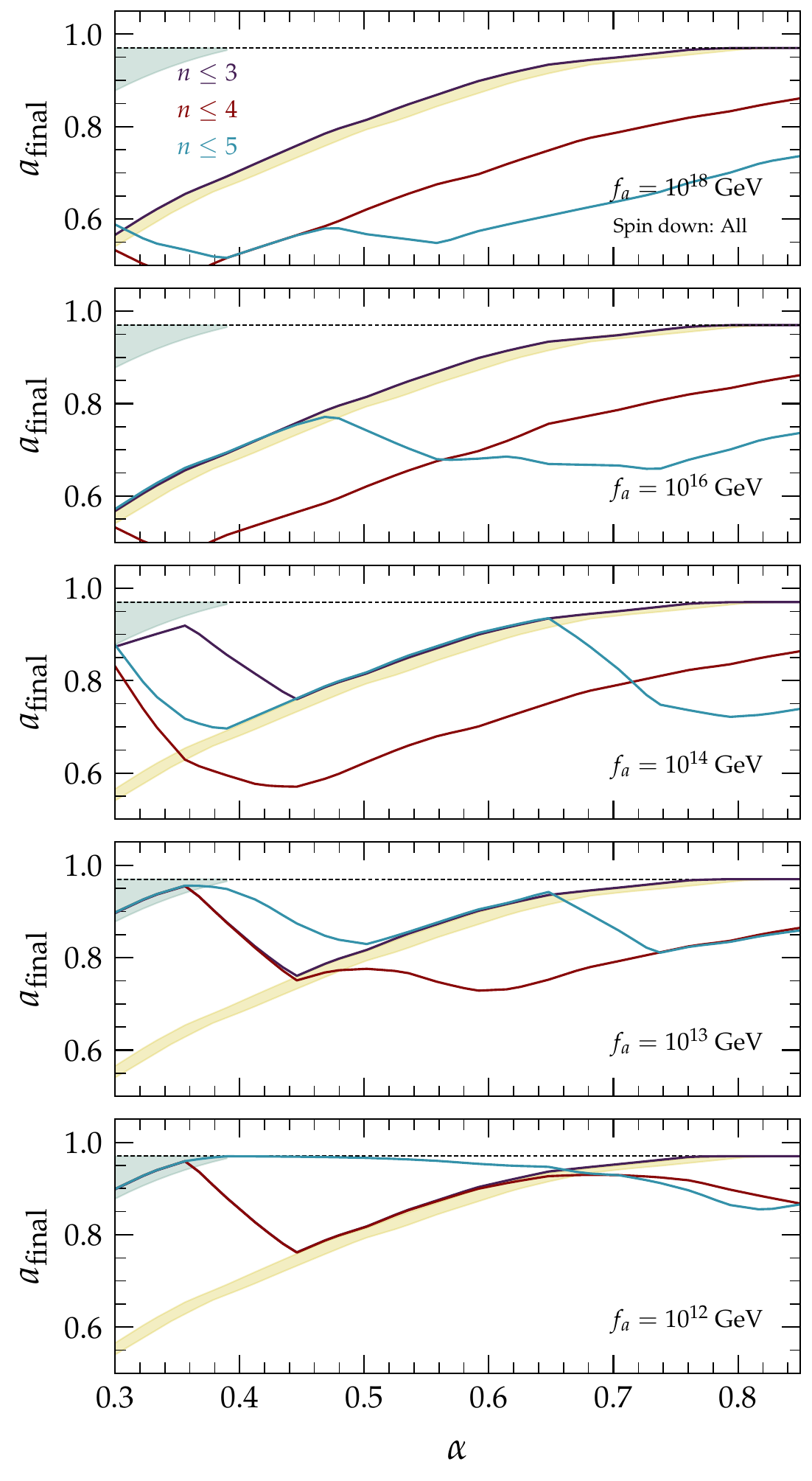}
    \caption{Same as Fig.~\ref{fig:a_final}, but for a spin $\tilde{a} = 0.97$ and a lifetime set to $\tau_{\rm Sal} = 5 \times 10^{7}$ years (this defines the so-called Salpeter timescale, which corresponds to the minimum timescale required to alter the black hole mass and spin -- see discussion in Sec.~\ref{sec:analysis}). This setup is intended to instead be reflective of the case of black hole GRS 1915+105. The green shaded regions denote the approximate $m=1$ spin down.   }
    \label{fig:a_final_2nd}
\end{figure*}

\section{Statistical Analysis and Re-visitation of Limits}\label{sec:analysis}
The above sections have served to demonstrate the formalism and impact of self-interactions on the superradiant evolution of scalar fields around rapidly rotating black holes. We now turn our attention to the question of how one should derive spin down constraints on observations of the black hole population, focusing specifically at observations of solar mass scale black holes.

From a statistical perspective, the question one would like to ask is whether an axion of fixed mass and decay constant $(m_a, f_a)$ is consistent with the observed set of black hole mass and spin distributions. One of the difficulties in addressing this question inherently stems from the fact that we observe black holes {\emph{today}}, and we are effectively ignorant of their properties and evolution at much earlier times. For sufficiently young black holes with lifetimes $\tau_{bh} \ll \tau_{\rm Sal}$, where the Salpeter timescale $\tau_{\rm Sal} \sim 4.5 \times 10^7$years is the minimum time required for a black hole to significantly alter its mass or spin (which assumes Eddington-limited accretion with radiative efficiency $\eta \simeq 0.1$)~\cite{Brito:2014wla}, one may be tempted to assume that spin at birth $\tilde{a}_{i}$ is equal to the spin observed today $\tilde{a}_0$. Up to caveats about extended periods of super-Eddington accretion (which are not expected in x-ray binary systems over these long time periods, although GRS 1915+105 has reached rates close to its Eddington limit during its current 30 year outburst, and GRO J1655-40 may have briefly entered a super-Eddington phase in 2005 \cite{Neilsen16}), this is certainly likely to be true in the conventional scenario\footnote{Dynamically it is exceedingly unlikely that any x-ray binary system was previously a part of a triple (BH, BH, star) system which has since undergone a merger, as all three objects would have to be extremely densely packed to leave the final star + BH system able to accrete, a setup which is inherently chaotic and liable to the ejecting of one/more bodies.}; however, in the presence of light bosons, this assumption may no longer hold. In other words, we cannot be confident whether a black hole is {\emph{currently}} undergoing spin-down -- while such a distinction is irrelevant for non-interacting bosons (since spin-down occurs exponentially fast),  the same is not guaranteed to be true when self-interactions are included in the evolution. As a result, it is natural to perform statistical inference by including the initial spin and mass of each black hole as model parameters $\{M_i, \tilde{a}_i\}$, and marginalizing over them (or profiling them out, depending on whether one performs a Bayesian or frequentist analysis) when deriving constraints on $(m_a, f_a)$. Since it is difficult to make strong statements about the properties of black holes on timescales $t \gtrsim \tau_{\rm Sal}$\footnote{In general, one could attempt to extend analyses of older black holes to times $\gg \tau_{\rm Sal}$ via a self-consistent model which includes the evolutionary history of black holes in the presence of accretion, merges, and superradiance. This is beyond the scope of this work, but would be of great interest as it could allow for a significant improvement in sensitivity (see \eg~\cite{Brito:2014wla} for initial attempts to perform such an analysis).}, we assume that the `initial' mass and spin distributions of black holes with ages $\tau_{bh} \geq \tau_{\rm Sal}$ are defined at a time 
 $t \sim t_0 -  \tau_{\rm Sal}$ (rather than $t \sim t_0 - \tau_{bh}$, with $t = t_0$ being today).

\begin{table*}
    \begin{tabular}{|c|c|c|c|c|c|}
         \hline  
         Name & Mass $[M_\odot]$ & $\tilde{a}$ & $\tau_{bh}$ [yrs] & Bin. Period [days] & $M_{\rm comp}$ [$M_\odot$] \\ \hline \hline
         Cyg X-1 [Fid.]& 21.2 $\pm \,  2.2$~\cite{miller2021cygnus} & $> 0.9985$~\cite{gou2014confirmation,zhao2021re} & $4.8 - 7.6 \times 10^6$ ~\cite{wong2012understanding}\footnote{It is worth noting that various groups have quoted a `conservative' age of Cygnus X-1, $\tau \sim \mathcal{O}(0.05)$ Myr, much lower than the value quoted here; this claim comes from Ref.~\cite{russell2007jet}, which infers the characteristic age of a shock wave which seems to have been produced by the jet of Cygnus X-1. This observation merely puts a lower limit on the age of the black hole (or more precisely, the age over which Cygnus X-1 was rapidly rotating). In fact, it was speculated in ~\cite{russell2007jet} that the shock wave formed only after Cygnus X-1 moved into a dense HII region, implying the age of the black hole is likely to be fully consistent with the inferred value quoted here~\cite{wong2012understanding}.} & 5.599829~\cite{gou2011extreme} & 40 ~\cite{zhao2021re}\\ \hline Cyg X-1 [Cnsrv.]& 21.2 $\pm \,  2.2$~\cite{miller2021cygnus} & $ 0.92^{+0.05}_{-0.07}$~\cite{Zdziarski24} & $4.8 - 7.6 \times 10^6$ ~\cite{wong2012understanding} & 5.599829~\cite{gou2011extreme} & 40 ~\cite{zhao2021re}\\ \hline 
           GRS 1915+105 & $12.4^{+2.0}_{-1.0}$ ~\cite{reid2014parallax} & $> 0.95$ ~\cite{reid2014parallax} & $3 - 5 \times 10^9$~\cite{dhawan2007kinematics} & 33.85 ~\cite{steeghs2013not}& 0.47 $\pm 0.27$ ~\cite{steeghs2013not}\\ \hline
        GRO J1655-40   & $6.3 \pm 0.5$ ~\cite{mcclintock2015black} & $0.72^{+0.16}_{-0.24}$ ~\cite{Baryakhtar:2017ngi} & $3.4 - 10 \times 10^8$ ~\cite{willems2005understanding} & 2.622~\cite{willems2005understanding} &  2.3-4 ~\cite{willems2005understanding} 
        \\ \hline 
          LMC X-1 &  $10.91 \pm 1.4$ ~\cite{mcclintock2015black} & $0.92^{+0.06}_{-0.18}$~\cite{gou2009determination} & $5 - 6 \times 10^6$~\cite{orosz2009new} & 3.9093~\cite{ruhlen2011nature} & 31.79$\pm 3.48$~\cite{ruhlen2011nature} \\ \hline 
          M33 X-7 & 15.65 $\pm 1.45$  ~\cite{mcclintock2015black} & $0.84 \pm 0.1$~\cite{Baryakhtar:2017ngi} & $2 - 3 \times 10^6$ ~\cite{orosz200715}& 3.4530~\cite{pietsch2006m33} & $\gtrsim$20~\cite{pietsch2006m33} \\ \hline 
    \end{tabular}
    \caption{ Black hole data used in this work. Mass and spin errors are quoted at $1\sigma$ and $2\sigma$ (except when an upper limit is listed, in which case we adopt the $3\sigma$ threshold), respectively. In the case of Cygnus X-1, we include a fiducial model (`Fid.') based on conventional spin-inference modeling, and a conservative model  (`Cnsrv') which was obtained when relaxing the priors on radiative transfer in the disc atmosphere~\cite{Zdziarski24} (see Sec.~\ref{sec:discussion} for more details).  A more extensive table of measurements of highly spinning solar-mass-scale black holes can be found, for example, in~\cite{draghis2023systematic,draghis2023systematic} (note, however, that not all black holes have mass or age estimates).   }
    \label{tab:bhs}
\end{table*}

In this work we perform a Bayesian analysis\footnote{Since the spins of various black holes are measured extremely precisely, our analysis will be largely data driven, and thus a frequentist approach is expected to yield similar results.} using the solar-mass-scale black holes listed in Table~\ref{tab:bhs}. For each black hole in the sample, we adopt the following priors on the initial mass and spin
\begin{eqnarray}
    p(\tilde{a}_i) &=& \begin{cases} \frac{1}{2(\tilde{a}_{\rm max} - \tilde{a}_c)} \hspace{.8cm} & { \tilde{a}_i > \tilde{a}_c} \\[10pt]
    \mathcal{N}(\tilde{a}_i, \tilde{a}_c, \sigma_{\tilde{a}}^-) \hspace{.8cm} & { \tilde{a}_i \leq \tilde{a}_c} \label{eq:a_prior}
    \end{cases} 
    \\[10pt] 
    p(M_i) &=& \begin{cases} \mathcal{N}(M_i, M_c, \sigma_M^+) \hspace{.5cm} & { M_i > M_c} \\[10pt]
    \mathcal{N}(M_i, M_c, \sigma_M^-) \hspace{.5cm} & { M_i \leq M_c} \label{eq:mass_prior}
    \end{cases}
\end{eqnarray}
where $M_c$ and $\tilde{a}_c$ are the central values of the inferred mass and spin, and $\mathcal{N}(x, x_c, \sigma_x)$ is  a Normal distribution evaluated at $x$, and with central value and standard deviation $x_c$ and $\sigma_x$, respectively.  When only a lower limit on the spin is observed, we adopt a flat prior over the values of spin. The maximum spin is set to $\tilde{a}_{\rm max} = 0.998$.

In Eqns.\ref{eq:a_prior}-\ref{eq:mass_prior}, we have assumed the initial mass and spin distribution to be uncorrelated, which in general need not be the case -- we note, however, that~\cite{Hoof:2024quk} has recently shown that using fully correlated spin-mass measurements does not significantly alter the superradiant sensitivity (at least for the examples provided). We take $\sigma_{M}^\pm$ to be the $1-\sigma$ upper/lower uncertainty in the inferred mass in Table~\ref{tab:bhs} (note we focus on timescales $t \lesssim \tau_{\rm Sal}$, where the mass of the black hole is nearly constant -- consequently, the mass prior in Eq.~\ref{eq:mass_prior} effectively serves to replace the need for a mass-dependent term in the likelihood itself), and $\sigma_{\tilde{a}}^-$ to be the lower $1-\sigma$ uncertainty on the measured spin distribution. Our fiducial analysis adopts log-flat priors on both the axion mass and the axion decay constant, which span the intervals $m_a \in [10^{-14} \, {\rm eV}, 10^{-10} \, {\rm eV}]$ and $f_a \in [10^{9} \, {\rm GeV}, 10^{18} \, {\rm GeV}]$. 

Let us elaborate briefly on the choice of priors in Eqns.\ref{eq:a_prior}-\ref{eq:mass_prior}. In the case of the black hole spin, we are effectively ignorant of the value of $\tilde{a}_i$, with observations only constraining the value of the spin {\rm today}. Therefore, a logical choice would be to merely adopt flat priors over the interval $\tilde{a} \in [0, \tilde{a}_{\rm max}]$. This would be a good choice, except in our analysis we are not including any mechanism that can spin up the black hole, only a mechanism capable of spinning down the black hole. Therefore, any of the samples we draw with $\tilde{a}_i \ll \tilde{a}_c - 2 \times \sigma_{\tilde{a}}^-$ are guaranteed to be strongly disfavored by the data, regardless of the presence of an axion. This implies our sampling procedure is inefficient (potentially highly inefficient). The obvious way to avoid this issue is to adopt a prior which preferentially does not sample this region of parameter space, while leaving the ``acceptable'' region of parameter space unaltered. The logical choice which avoids introducing arbitrary hard boundaries in the prior is the one of Eq.~\ref{eq:a_prior}. In the case of the black hole mass, the story is slightly different. Here, we have not included any mechanism capable of significantly altering the mass of the black hole itself (superradiance can change the mass, but does so only by an amount which is much less than the uncertainty in the measurement). A natural choice of priors in this case would be a log-flat distribution in masses that is sufficiently broad to be well beyond the mass uncertainty of any object. In this case, one would incorporate the actual mass measurement directly into the likelihood itself (described below). The problem, as before, is that one ends up with a highly inefficient sampling procedure where most of the sampled points will inherently be rejected purely because of the poor overlap between the prior and the likelihood. An easy fix is to fold the mass measurement directly into the prior, and remove it from the likelihood in order to avoid double counting. In this sense, one is directly sampling from the allowed mass posterior in a way that is compatible with the observation -- since there is no strong evolution in the mass, the final mass is guaranteed to be consistent with the observations (with the statistical sampling procedure already reflecting the relative posterior weighting). Before continuing, let us emphasize that these choices are valid here only because we are working on sufficiently short timescales where we can neglect mass and spin evolution -- a generalized analysis including these effects would need to modify both the priors and the likelihood.

For fixed axion and initial black hole parameters, we evolve Eqs.~\ref{eq:dE_1}-\ref{eq:dE_1_end} (or the equivalent versions which include higher order superradiant states) from $t = t_0 - \tau_{\rm max}$ to $t_0$, where $\tau_{\rm max} = {\rm Min}[\tau_{bh}, \tau_{\rm Sal}]$ (for the sake of being conservative, we always adopt a value of $\tau_{bh}$ consistent with the lower limit in Table~\ref{tab:bhs}); this yields a prediction of the mass and spin today, $(M_0, \tilde{a}_0)$. The likelihood is then given by 
\begin{eqnarray}
    \mathcal{L} = \prod_i  \begin{cases}\mathcal{N}(\tilde{a}_0, \tilde{a}_c, \sigma_{\tilde{a}}^+)  \hspace{.5cm} & { \tilde{a}_0 > \tilde{a}_c} \\[10pt]
    \mathcal{N}(\tilde{a}_0, \tilde{a}_c, \sigma_{\tilde{a}}^-) \hspace{.5cm} & { \tilde{a}_0 \leq \tilde{a}_c}
    \end{cases}  \, ,
\end{eqnarray}
where the central values $\tilde{a}_c$ and upper/lower uncertainties are those quoted in Table~\ref{tab:bhs}, and the product runs over all black holes in the sample. In practice, we perform an independent analysis for each black hole -- this is done for computational simplicity, but a joint analysis may give marginally stronger limits. For black holes whose spin inference leads only to lower limits on $\tilde{a}$, we instead take the likelihood to be an asymmetric Gaussian centered  on the two-sigma limit, and with an excessively large tail in high-spin regime (truncated at $\tilde{a} = 0.998$, thereby mimicking a flat distribution), and a lower standard deviation of $10^{-3}$ -- this choice effectively serves to ensure high spin black holes are not penalized, while those below the limit are immediately excluded (and does so without imposing any hard thresholds in the analysis).  In general, one could adopt a procedure that directly uses the inferred distributions themselves.

In Table~\ref{tab:bhs} we have also included information on the binary systems to which these black holes belong (namely, the binary period and the companion mass). Various groups have investigated the role of binary companions in perturbing the growth and evolution of superradiant systems (see \eg~\cite{Baumann:2018vus,Baryakhtar:2020gao,Baumann:2022pkl,Tong:2022bbl,Fan:2023jjj,Zhu:2024bqs}), where the general expectation is that the binary can induce an effective mixing between the superradiant levels, allowing axions in growing  superradiant modes to dissipate some of their energy. This effect is expected to be small so long as~\cite{Baryakhtar:2020gao}
\begin{eqnarray}
    \alpha \gtrsim 0.013 \left(\frac{M}{M_\odot} \, \frac{1 \, {\rm day}}{P_{\rm orb}} \right)^{2/9} \, \tilde{a}^{-1/9} \, \left(\frac{M}{ M_{comp} + M } \right)^{1/9}
\end{eqnarray}
where $P_{\rm orb}$ is the binary period and $M_{comp}$ the companion mass. Binaries can also resonantly deplete superradiant levels when the binary periods are close to the energy splittings -- this can be computed for each system by determining if $P_{\rm orb} \times (\omega_{nlm} - \omega_{n^\prime l^\prime m^\prime}) \sim 1$~\cite{Baumann:2018vus,Baryakhtar:2020gao} (see also~\cite{Takahashi:2024fyq} for a recent study of the interplay of binaries with axion self interactions).  The systems shown in Table~\ref{tab:bhs} are safely away from both of these thresholds for the parameter space of interest\footnote{Technically, the $\levfourthree$ and $\levfourtwo$ levels (and some of the $n=5$) levels are sufficiently close to potentially allow for a resonance in Cyg X-1 at very low axion masses (near $\alpha \sim 0.03$), however these states only play a relevant role in the evolution of the superradiant system for larger values of $\alpha$, and thus such a resonance would not alter the spin down of the black hole.}, and thus we do not need to impose any additional constraints on the parameter space of interest.

\begin{figure}
    \centering
    \includegraphics[width=0.45\textwidth]{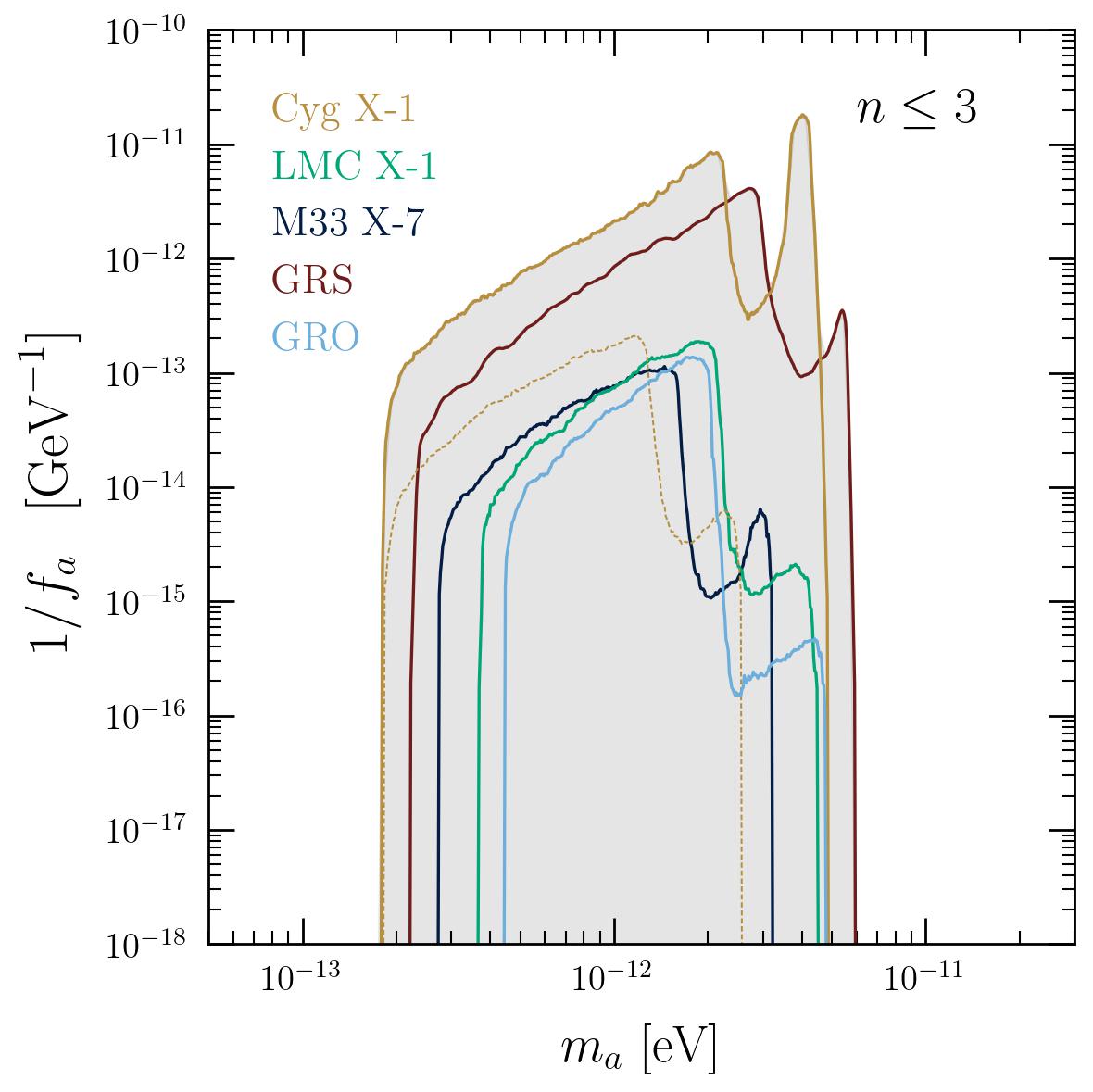}
    \caption{95$\%$ upper limits on $f_a$ derived using the $n\leq 3$ states, and including relativistic corrections to the scattering rates, for each of the black holes listed in Table~\ref{tab:bhs} (results for the conservative spin measurement of Cyg X-1 is shown using a dotted line). The region excluded by the combination of these constraints in shown in grey.}
    \label{fig:limits_n3}
\end{figure}

\begin{figure*}
    \centering
    \includegraphics[width=0.45\textwidth]{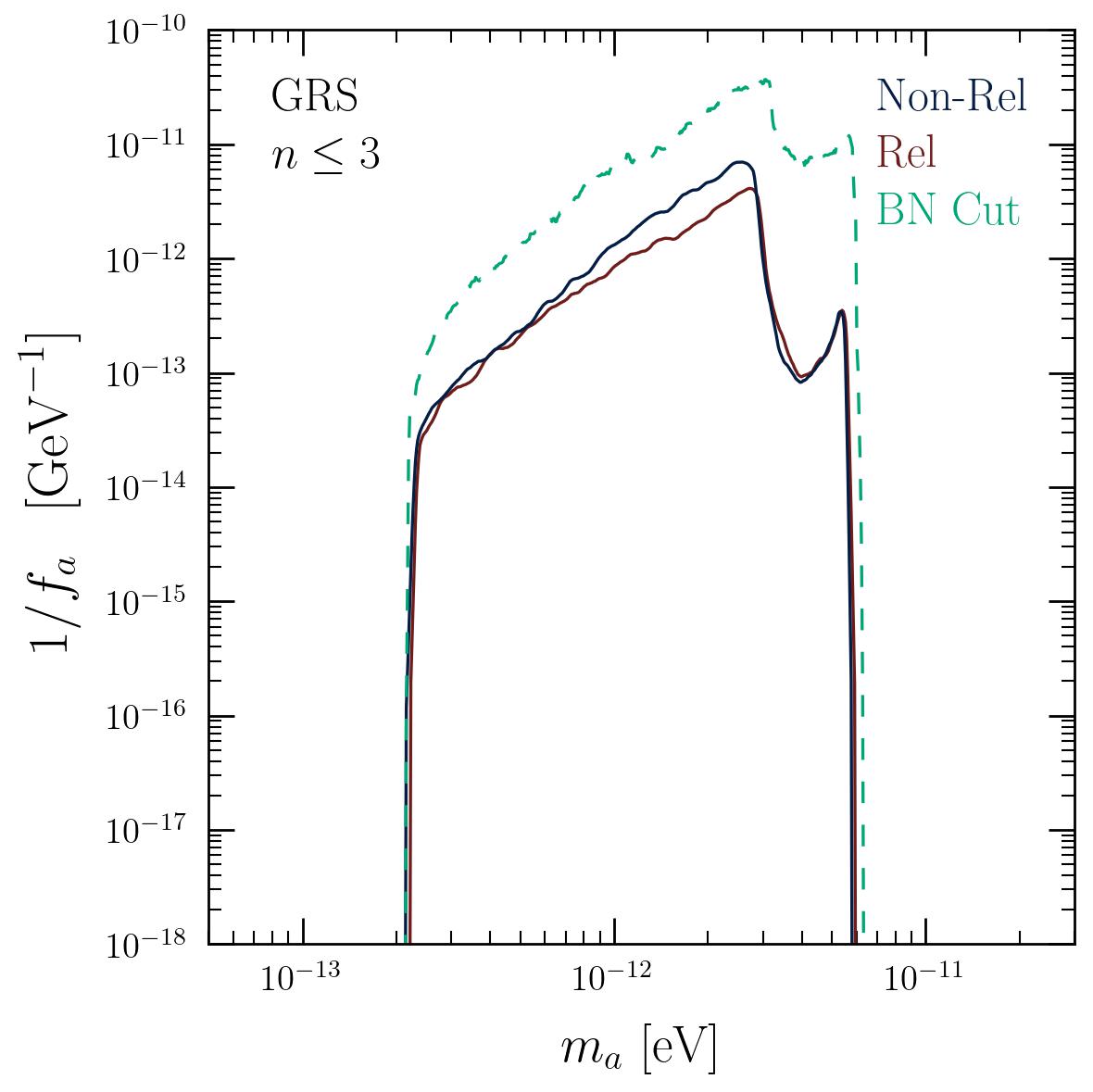}
    \includegraphics[width=0.45\textwidth]{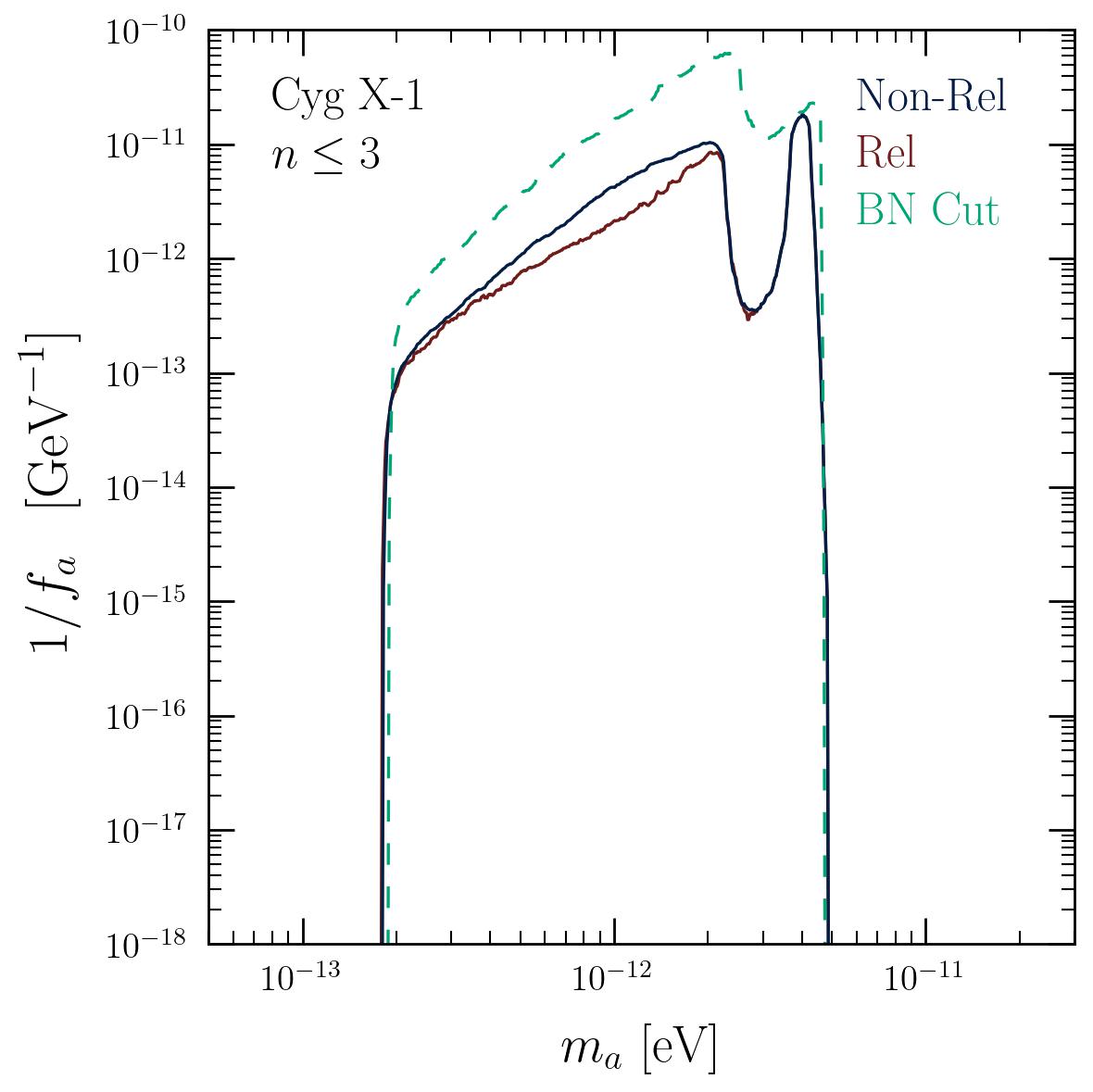}
    \caption{95$\%$ upper limits on $f_a$ derived using the $n\leq 3$ states for GRS 1915+105 (left) and Cyg X-1 (right). Results are shown using the non-relativistic scattering rates (blue), relativistic scattering rates (red), and by neglecting scattering and only imposing a maximum upper limit on the occupation number set by the bosenova threshold (cyan, labeled `BN Cut').}
    \label{fig:N3_NR_Rel}
\end{figure*}

We begin by running an independent MCMC for each black hole listed in Table~\ref{tab:bhs} using the two-level system (\ie only including the $n\leq 3$ states). The results are presented in  Fig.~\ref{fig:limits_n3}. All calculations include relativistic corrections to the scattering rates. One should be careful in interpreting the $m=2$ spin down region (corresponding to the strengthening in constraints at large masses), as there is no scattering process included in this model which prevents the growth of the $\levthree$ state (any suppression of the limit is instead being driven by the statistical uncertainties in the black hole properties, which instead blur the boundaries of the $m=1,2$ spin down regimes). In the case of Cygnus X-1, we plot both the fiducial (solid) and the conservative result (dotted). The union of all constraints is highlighted in gray. As expected, we see that constraints are dominated by the high-precision spin measurements of Cygnus X-1 and GRS 1915+105, although the conservative Cygnus X-1 limit suggests that the strength of the fiducial bound should be treated with caution. We return to the motivation for including the `conservative' model below.

\begin{figure*}
    \centering
    \includegraphics[width=0.49\linewidth]{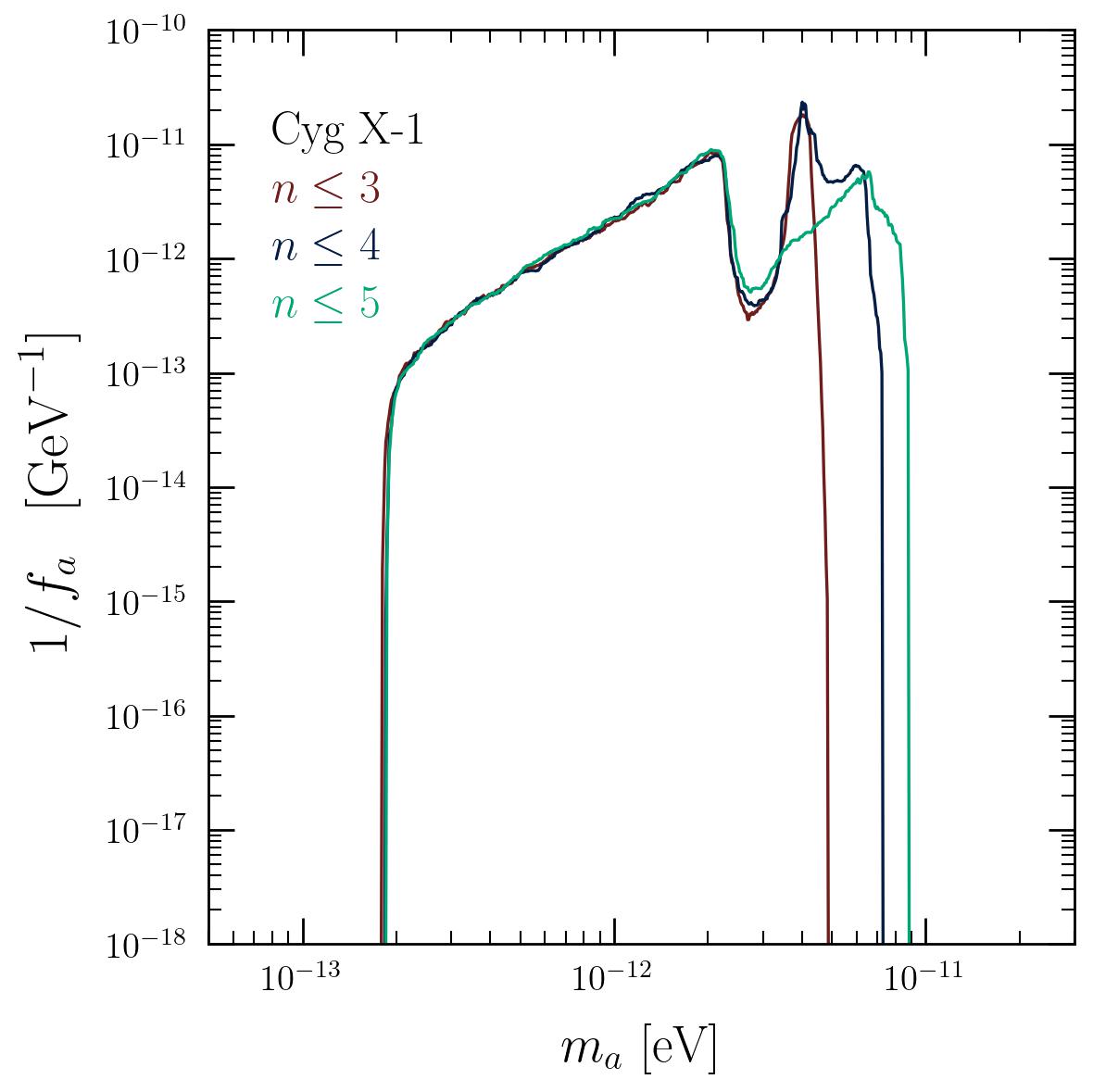}
    \includegraphics[width=0.49\linewidth]{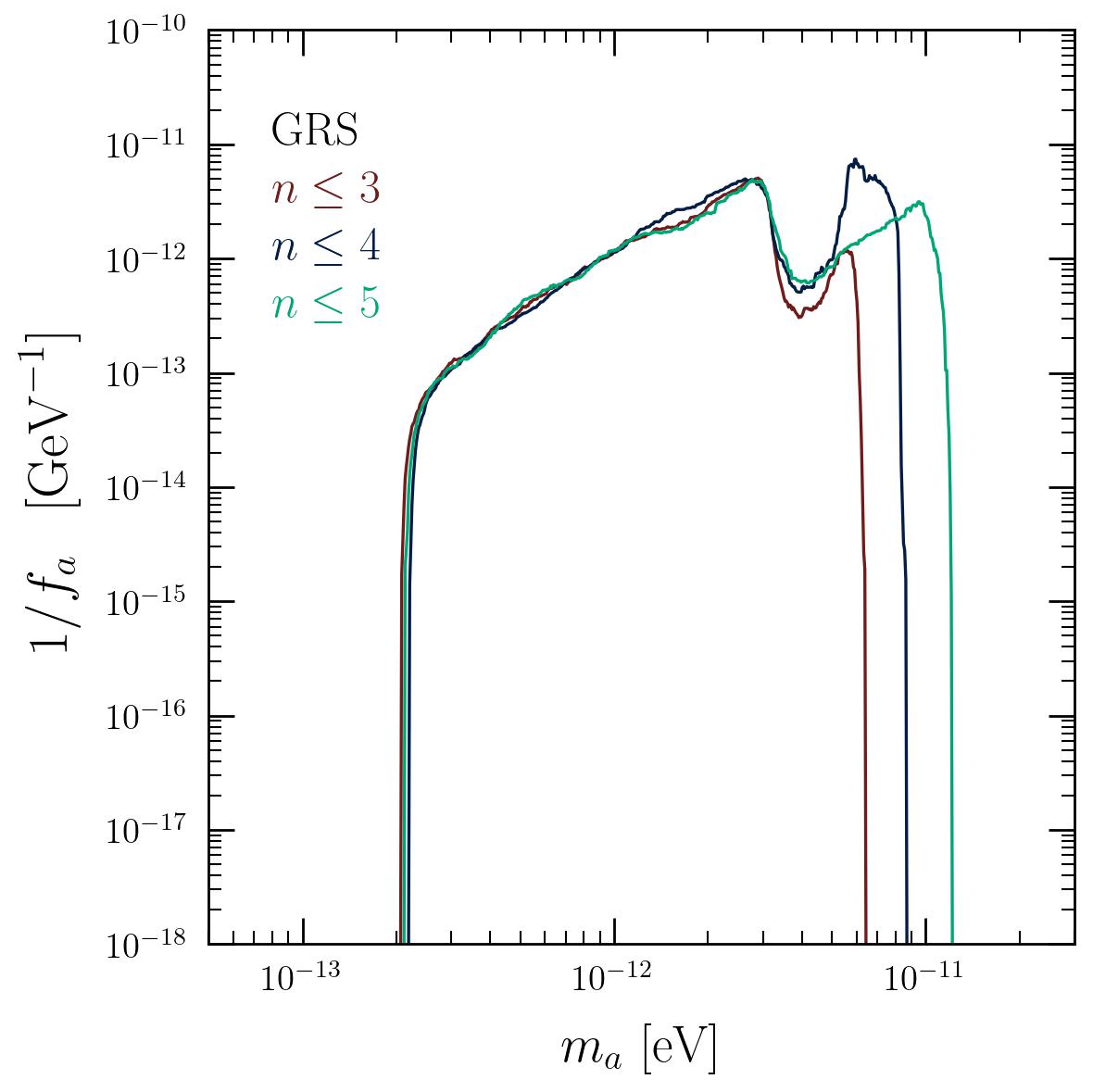}
    \caption{\label{fig:cyg345}  Comparison of the 95$\%$ CI limits derived using the fiducial properties of Cygnus X-1 (left) and GRS 1915+105 (right), and evolving the $n\leq 3$, $n\leq 4$, and $n\leq 5$ states. Relativistic corrections are included everywhere. }
    
\end{figure*}

In Fig.~\ref{fig:N3_NR_Rel} we plot the constraints derived on GRS 1915+105 (left) and Cygnus X-1 (right) for the $n \leq 3$ states, including (red) and neglecting (blue) relativistic corrections to the scattering rates.  We also include an analysis in which the only role of self-interactions is to impose an upper limit on the occupation number -- this limit is referred to as the bosenova cut (`BN Cut'), and is shown in cyan.  For both black holes, we see that relativistic corrections serve to slightly suppress the constraint coming from the $m=1$ level (this occurs because the enhanced spin extraction rate of $m=2$ slightly modifies the equilibrium distributions).  The impact of the  relativistic corrections on the $m=1$ spin down is sub-dominant to the role of self-interaction induced level mixing, as can be seen by the fact that the `BN Cut' limit extends to as much as an order of magnitude smaller $f_a$ in both cases. Note that the absence of the $\levtwo$ state at large $\alpha$ implies the $\levthree$ state can grow unimpeded to the bosenova threshold. For Cygnus X-1, one finds all constraints tend toward degeneracy at large masses, while for GRS 1915+105 the larger uncertainty in black hole mass and spin are sufficiently great to prohibit the $m=2$ spin down region from obtaining full sensitivity.

 In the left panel of Fig.~\ref{fig:cyg345} we plot the limits derived using Cygnus X-1, applying the analysis with $n \leq 3$, $n\leq 4$, and $n\leq 5$ states (with relativistic corrections included everywhere). Here, we find that the inclusion of the $n = 4$ and $n = 5$ states have a minimal impact on the $m=1$ spin down region. The $m=2$ spin down limit, instead, is strongly affected by the inclusion of the $n= 5$ states, since this in turn is predominantly responsible for suppressing the growth of the $\levthree$ state. The extension of the $n \leq 4, 5$ models to masses above the $m=2$ region should be discarded, as we have not included the relevant levels for suppressing the $m=3$ spin down (which arises at $n = 7$). The net suppression observed in the $m=2$ limit amounts to a factor of nearly $10$ in $f_a$ at its peak.

One can perform the same comparison with GRS 1915+105 -- the result is shown in the right panel of Fig.~\ref{fig:cyg345}. As before, one sees that the results converge in the $m=1$ spin down region, however for the $m=2$ spin down region only the $n \leq 4$ analysis sees enhanced sensitivity (being driven not by the $m=2$ level but rather the $m=3$) -- instead, the $n \leq 3,5$ analyses display similar sensitivities in this regime.

In the previous figures, the uncertainty in the black hole mass tends to reduce the range of axion masses which can be constrained, and induces a smearing of what would typically be sharp features at the boundaries of the various spin down levels. In order to illustrate the effect of the mass uncertainty of the black hole on the derived limits, we re-run the $n\leq 5$ analysis of GRS 1915+105 assuming the black hole mass is known precisely, and compare it to our fiducial limit from GRS 1915+105 -- the result is shown in Fig.~\ref{fig:grs_massres}. The net effect is rather minimal, generating a modest enhancement near the peak spin down of the $m=1,2$ spin down regions, a slight suppression of the region where the $m=1$ level turns off (before, the mass uncertainty smeared this sensitivity), and an extension of the constraints to marginally higher and lower masses. In this sense, precision mass measurements are not enough to sizably extend constraints to heavier axions -- this needs to be accomplished by studying the coupled spin down at high$-n$.

\begin{figure}
    \centering
    \includegraphics[width=0.9\linewidth]{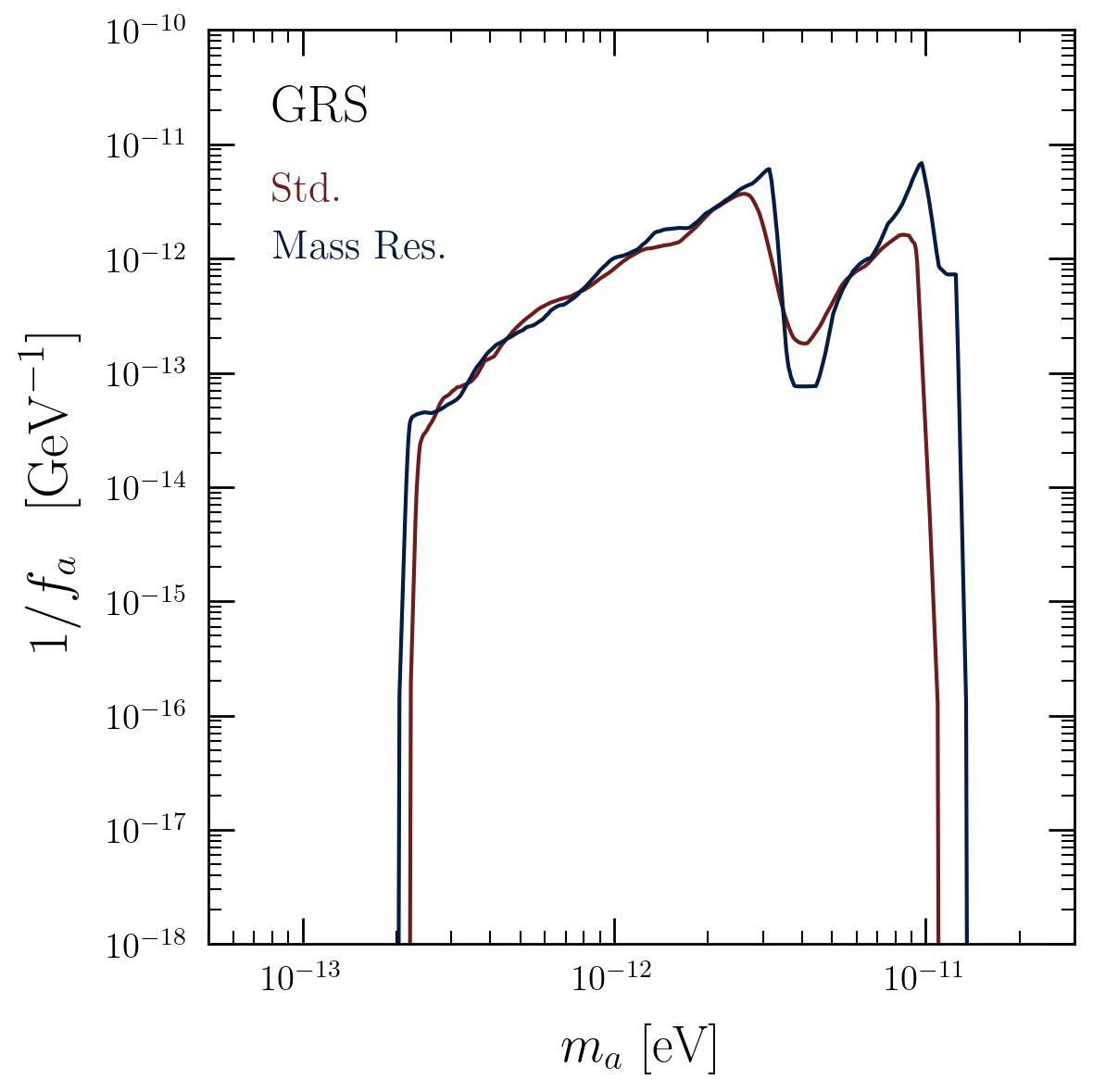}
    \caption{ Comparison between the 95$\%$ CI obtained for the fiducial parameters of GRS 1915+105 (labeled `Std.', meaning Standard), and the limit obtained assuming no uncertainty on the mass inference of the black hole (labeled `Mass Res.', meaning Mass Resolution analysis). Both analyses are performed for $n \leq 5$ and including relativistic corrections.  }
    \label{fig:grs_massres}
\end{figure}

Superradiance limits on solar-mass-scale black holes have previously been derived in Ref.~\cite{Baryakhtar:2020gao}, Refs.~\cite{Stott:2018opm,Stott:2020gjj}, and Ref.~\cite{Hoof:2024quk}. The approaches of these groups are known to differ significantly, as are their derived limits. We briefly discuss the differences in these approaches below, and provide a comparison with the most relevant of these below.

We begin with a discussion of the limits derived in Refs.~\cite{Stott:2018opm,Stott:2020gjj} (which have also been applied to the analyses of~\cite{Mehta:2020kwu,Mehta:2021pwf}). The primary differences between the analysis of Refs.~\cite{Stott:2018opm,Stott:2020gjj} and that provided here are:
\begin{itemize}
    \item Refs.~\cite{Stott:2018opm,Stott:2020gjj} do not include self-interaction induced level mixing. Rather, the role of self-interactions is only to establish a maximal occupation number at which each level ceases to grow (\ie they impose a bosenova threshold\footnote{The bosenova threshold differs slightly from Eq.~\ref{eq:bnova} by a small pre-factor, however this only induces a minor shift in the derived sensitivity.}). As discussed above, self interactions quench the growth at lower occupation numbers, reducing spin down and significantly suppressing the derived sensitivity.
    \item Refs.~\cite{Stott:2018opm,Stott:2020gjj} do not time evolve the superradiant system, but rather set constraints by comparing the growth rate, the black hole lifetime, and the relevant thresholds on the occupation numbers. In principle, such an approach can be used to obtain rough ideas of where constraints should lie, but only for high-precision measurements of highly spinning black holes (in other cases, the superradiant timescale can evolve strongly in time that only allows for a minimal change, with respect to the measurement uncertainties, in the black hole spin). We believe that this is one of the primary reasons Refs.~\cite{Stott:2018opm,Stott:2020gjj} appear (enormously) to obtain constraints using black holes which have spins perfectly consistent with $\tilde{a} \sim 0$, We note that this error has also propagated into the works of \cite{Mehta:2020kwu,Mehta:2021pwf}. 
    \item There are slight differences between the data sets used in Refs.~\cite{Stott:2018opm,Stott:2020gjj} and the dataset adopted here, see e.g. Table 1 of~\cite{Mehta:2021pwf}, but these do not contribute appreciably to the differences in the limits. 
    \item Refs.~\cite{Stott:2018opm,Stott:2020gjj} perform an analysis up to $n = 6$, and work in the non-relativistic limit. In the absence of self-interaction induced level mixing, working at large $n$ leads to an over estimation of the constraints at large $\alpha$. 
    \item Finally, there is an important distinction in the statistical approach. In this work, limits are obtained here by looking at the 2$\sigma$ marginal posterior in $(f_a,m_a)$. Instead, in Refs.~\cite{Stott:2018opm,Stott:2020gjj}, limits are set by computing the weighted fraction of the black hole posterior which overlaps with the region of parameter space where one would expect spin down to occur. Focusing momentarily only on the technique (rather than the application to superradiance) and putting asides the issues mentioned above, one could imagine a pathological scenario in which an axion is excluded at $90\%$ despite the best fit point (and the region of parameter space around it) being fully consistent with the presence of an axion. In this case, various model comparison methods would yield little or no preference for a model either with, or without, the axion. Furthermore, a higher-precision and fully consistent measurement (e.g. with the same inferred central value of the mass and spin) would in fact {\emph{not}} exclude the presence of such an axion. This behavior suggests that the type I error could deviate (potentially strongly) from the quoted exclusion confidence level, which is not a desirable feature of a limit setting procedure. While we believe this issue is subdominant to the issues discussed above (and likely not to have introduced any such pathological problems in previous analyses), it is nevertheless worth highlighting. 
\end{itemize}
Owing to the large differences in the analysis, we do not directly compare with the results of these works.

Recently, Ref.~\cite{Hoof:2024quk} (which includes a subset of the authors in Ref.~\cite{Stott:2018opm,Mehta:2020kwu,Mehta:2021pwf}) performed an analysis on one solar mass black hole (M33 X-7) using the posterior distributions from the inference of the black hole mass and spin itself. As before, Ref.~\cite{Hoof:2024quk} does not evolve the states, but rather defines a characteristic spin-down timescale which is set by assuming  the $\levtwo-\levthree$ states are fixed to their equilibrium values (as derived in Ref.~\cite{Baryakhtar:2020gao}). The authors also look at the impact of using correlated mass-spin uncertainties directly from the inferred posteriors, find a small, but not significant difference in these results. As can be seen in the right panel of Fig.~\ref{fig:compare}, the approach of Ref.~\cite{Hoof:2024quk} typically yields similar results, except near the lower- and upper-mass thresholds where our analysis shows no (or a large reduction) in sensitivity. While the origin of this discrepancy is not immediately clear, one can compute the $m=1$ spin down timescale for a mass in this region in the limit $f_a \rightarrow \infty$ (i.e. the limit of a non-interacting boson), taking black hole parameters in the central $1-\sigma$ contour (see Fig. 2 of~\cite{Hoof:2024quk}), and show that this timescale is longer than the age of the black hole itself -- this point cannot be physically excluded, suggesting an error in the analysis of~\cite{Hoof:2024quk}.

The limits obtained in Ref.~\cite{Baryakhtar:2020gao} were derived by solving the coupled $\levtwo-\levthree$ level system for each black hole in Table~\ref{tab:bhs} (note that the properties of Cyg X-1 have recently been revised, with older estimates yielding a mass of $M \sim 14.8 \pm 1.0 \, M_\odot$, and a spin roughly consistent with the fiducial model listed in Table~\ref{tab:bhs}). Their approach is similar to the two-level analysis ($n \leq 3$) provided here, with the notable difference being the approach to limit setting. In this work, we perform a Markov Chain Monte Carlo (MCMC) over the parameters $\xi \equiv (f_a, m_a, M_i, a_i)$, marginalize over the initial mass and spin, and derive a limit on each $f_a$ by binning the MC samples in $m_a$. This can be contrasted with the approach of Ref.~\cite{Baryakhtar:2020gao}, which evolves the $\levtwo$ and $\levthree$ states, and obtains a limit by equating the age of the black hole with the dynamical timescale of spin depletion (evaluated at the quasi-equilibrium occupation numbers)~\cite{MashaEmail}. In addition, Ref.~\cite{Baryakhtar:2020gao} truncates the $m=2$ spin-down for values of $f_a \lesssim 10^{17}$ GeV, as they identify the $\left. |544 \right>$ level as becoming important as this threshold, but do not include this level in the evolution. The approach of Ref.~\cite{Baryakhtar:2020gao} errors on the side of being conservative. For high-precision spin measurements, this error can become significant; this can be seen in Fig.~\ref{fig:compare}  by comparing the limits derived in Ref.~\cite{Baryakhtar:2020gao} using Cygnus X-1 and M33 X-7 with the limits obtained here. In the case of a high precision measurement like Cygnus X-1, the approach of Ref.~\cite{Baryakhtar:2020gao} tends to underestimate sensitivity by roughly one order of magnitude, while this reduces to a factor of $\sim 2$ for M33 X-7. Finally, notice that that the extension of these results to the $m=2$ spin-down allows one to extend the constraints at larger masses by a factor of $\sim 2$.

\section{Discussion of Spin Measurements}\label{sec:discussion}
A wide variety of approaches to inferring the spin of astrophysical black holes from astronomical observations have been developed over the past few decades. Broadly speaking, the general approach involves the generation of some form of model template (be that a gravitational waveform or an x-ray spectrum produced by an accretion flow) which are explicitly a (often very sensitive) function of black hole spin. These templates are then matched to observations through a process of likelihood maximisation, which typically leads to tight posteriors on black hole spins (e.g., Table \ref{tab:bhs}).

\begin{figure*}
    \includegraphics[width=0.48\textwidth]{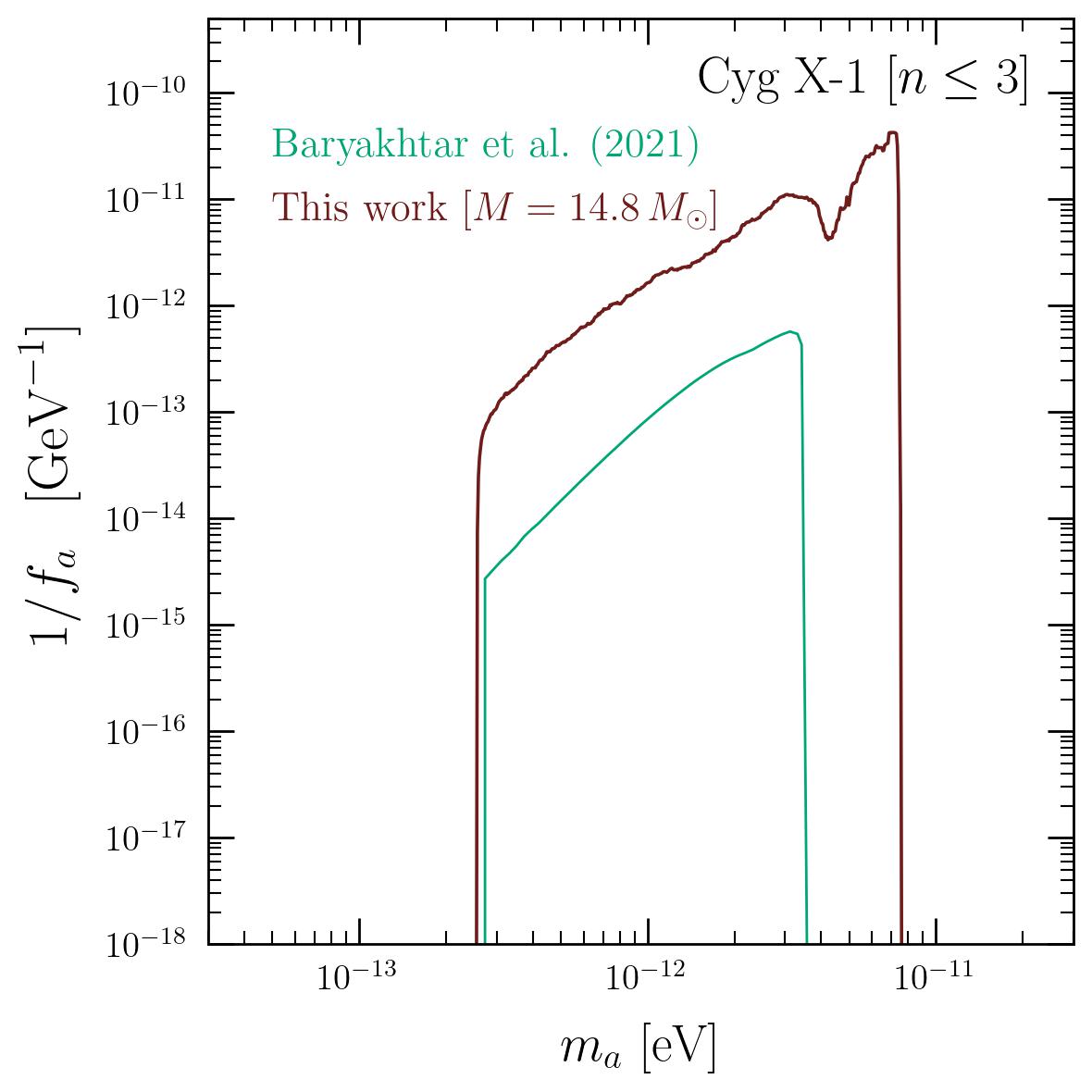}
    \includegraphics[width=0.48\textwidth]{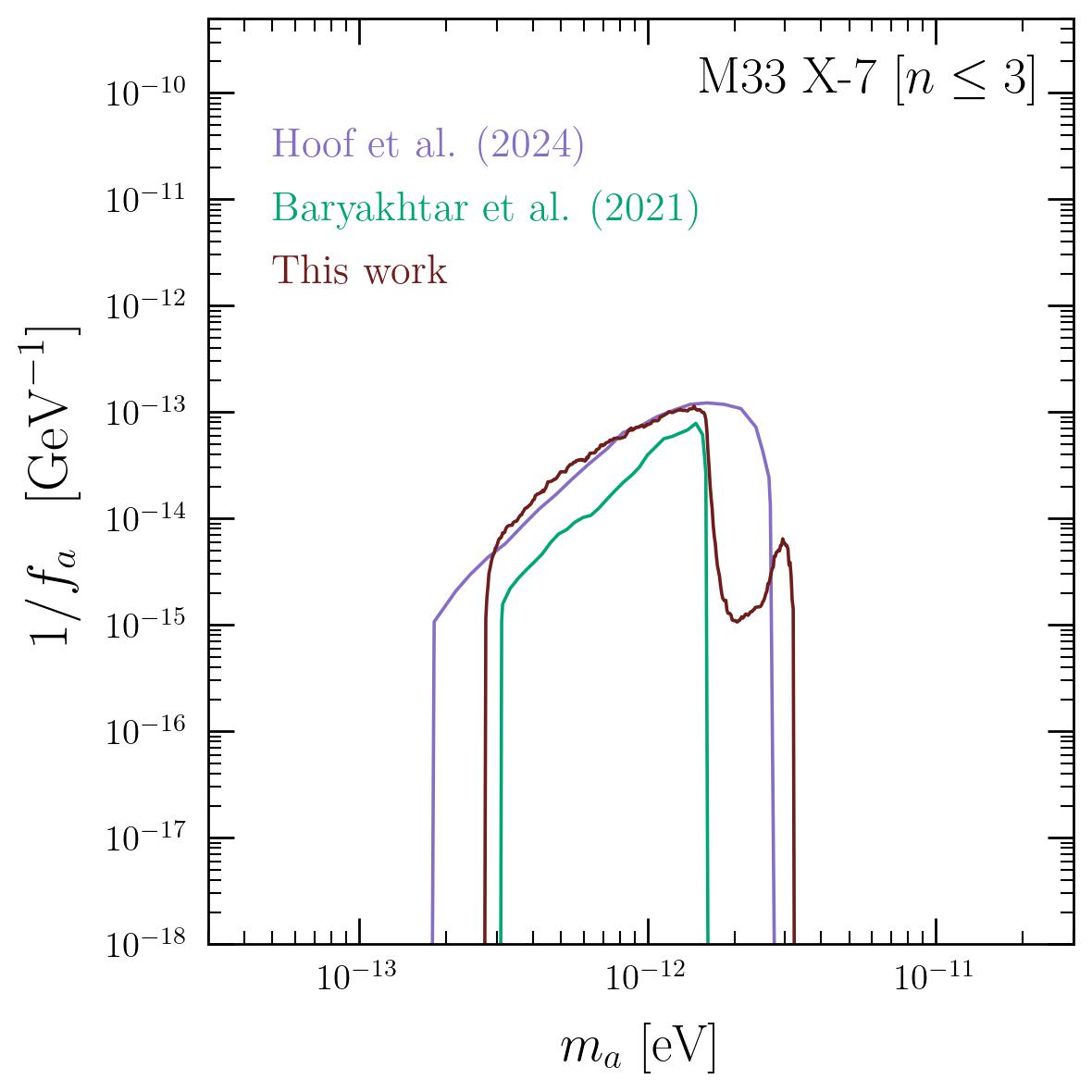}
    \caption{{\emph{Left:}} Comparison between the $m=1$ spin-down limit derived in Ref.~\cite{Baryakhtar:2020gao} using Cygnus X-1 (green; see text), and the equivalent comparison obtained using the procedure defined here (where we have run an analysis for $n \leq 3$, and used a black hole mass consistent with the outdated value; see text). Note that the $m=2$ spin down region arising at large masses should be ignored, as this is not included in the limit of Ref.~\cite{Baryakhtar:2020gao}. Differences in the $m=1$ spin down are a result of the improved statistical analysis (see text). {\emph{Right:}} Comparison between the upper limit derived using M33 X-7 and the result obtained using the procedure defined here (for $n \leq 3$), the upper limit of Ref.~\cite{Hoof:2024quk}, and the $m=1$ spin down limit derived in Ref.~\cite{Baryakhtar:2020gao}. Differences with Ref.~\cite{Baryakhtar:2020gao} are driven by the statistical analysis, while differences with Ref.~\cite{Hoof:2024quk} arise primarily from the limit setting procedure, and a hypothesized error in the analysis (with sub-dominant differences also appearing due to the fact that mass and spin measurements here are assumed to be uncorrelated), which we argue over-estimates the sensitivity at high and low axion  masses -- see text for detailed discussion.   }
    \label{fig:compare}
\end{figure*}

Black hole spins can be estimated from the properties of the gravitational waveforms detected by LIGO \cite{Abbott16}. While measuring any one individual black hole spin with this approach is difficult, the population of binary mergers is now large enough that some structure is clear.  The spins of merging binary black holes  inferred from gravitational wave analysis are low. The most recent population study of \cite{Abbott23} presents the results from 70 binary black hole mergers detected as part of the Gravitational Wave Transient Catalog 3 (GWTC-3)~\cite{KAGRA:2021vkt}. GWTC-3 combines observations from the first three observing runs (O1, O2, O3) of the Advanced LIGO and Advanced Virgo gravitational-wave observatories. The distribution of the individual spins peaks at $\tilde a \simeq 0.13^{+0.12}_{-0.11}$, and half of the individual black holes  in GWTC-3 are found to have $\tilde a  \lesssim 0.25$. There is no strong evidence of any rapidly rotating black holes in the GWTC-3 sample \cite{Abbott23}. 

On the contrary, techniques that make use of electromagnetic observations of astrophysical black hole systems often find large black hole spin parameters (see \eg  Table \ref{tab:bhs})\footnote{The discrepancy between electromagnetic and gravitational wave inferences may strike the reader as odd, and it has been suggested that this discrepancy may well relate to different evolutionary pathways of both populations \cite{Fishbach22}, although this is a controversial statement \cite{Belczynski24}}. So far these have typically been restricted to x-ray observations (\eg see the following reviews \cite{Reynolds13,McClintock14}), although optical techniques are being developed \cite{Mummery20, Mummery:2023meb}. The accretion flows which  form in black hole binary systems can reach, for typical values of the free parameters of the theory,  temperatures of order a few keV in their innermost regions. X-ray photons are therefore naturally produced in abundance, and in principle carry characteristic signatures of the highly relativistic regions of spacetime from which they originated. The general approach therefore is to search for (and find) these signatures in the observed spectral energy distributions (SEDs) of accreting black hole systems. 

As it is x-ray observations of accreting systems which typically provide the highest black hole spin constraints, which are of most practical use for superradiant studies, we spend some time here discussing in more detail these methods and their potential systematics. 

There are two ways in which x-ray observations of accreting black holes can be leveraged to produce spin constraints. Both methods utilise the fact that the accretion flows in these systems penetrate deep into the spacetime of the black hole, but differ in that they probe different ways in which this accretion flow can source  x-ray photons. The so-called ``continuum fitting'' technique \citep[see e.g.,][for a detailed review]{McClintock14} is perhaps the simplest to qualitatively understand. As a disc fluid element spirals in towards the black hole, its gravitational potential energy is liberated by turbulent dissipation in the flow (the process by which accretion is itself mediated \cite{BalbusHawley91}). This liberated energy ultimately escapes the disc as a hot thermal photon field, and is observed as a broad continuum SED. The constraints of mass, angular momentum and energy conservation within the accretion fluid can be manipulated into a simple one-dimensional temperature profile of the disc $T(\tilde r, \tilde a)$ \cite{NovikovThorne73, PageThorne74}, which depends sensitively on black hole spin in the hottest innermost regions. The reason for this sensitive dependence on spin can be traced to the efficiency of the accretion process itself, as the fraction of the rest mass energy of each fluid element which is ultimately radiated away during accretion is \cite{PageThorne74}
\begin{equation}\label{eq:acceff}
    \eta = 1 - \left(1 - {2 \over 3\tilde r_I}\right)^{1/2}, 
\end{equation}
where $\tilde r_I$ is the (dimensionless) radius of the innermost stable circular orbit, which ranges from $6$ for a Schwarzschild black hole to $1$ for a maximally rotating black hole \cite{Bardeen72}. The efficiency of accretion therefore varies from $\sim 0.06$ to $\sim 0.42$ as the spin is increased, making the discs around more rapidly rotating black holes significantly hotter, and their SEDs therefore peak at higher photon energies. It is this principle which is ultimately utilised in the continuum fitting approach. 

For the continuum fitting approach to work, a clear detection of the emergent disc photon field must be available. This, however, is often not the case. Many x-ray binary systems (and in particular AGN) are observed in (and often to transition between) various so-called accretion ``states'', one of which (referred to as the ``hard'' state) is dominated by a non-thermal x-ray component, understood to result from the Comptonisation of thermal disc photons by a ``corona'' of relativistic electrons.  The physical origin and geometry of this corona is still hotly debated in the community, but its emission often dominates over that of the bare disc itself, preventing the use of continuum fitting techniques. 

Fortunately, some of this coronal emission is directed onto the accretion flow itself where it is absorbed and remitted (a process referred to as ``reflection'', although this is a misnomer), opening up further opportunities for black hole spin constraints. The so-called ``reflection spectrum'' observed from black holes in this state contains asymmetric and broadened iron emission lines, owing to the presence of iron atoms in the flow which are not fully ionised and can therefore undergo atomic transitions.  The iron K$\alpha$ fluorescence line at $\sim 6.4$ keV is a prominent feature of the reflection spectrum, which also includes emission lines and absorption edges from all astrophysically abundant elements and a broad Compton scattering feature at $\sim 20-30$ keV known as the `Compton hump' \citep{Matt1991,Ross2005,Garcia2013}. The iron line is a particularly powerful diagnostic because it is narrow in the emission rest frame, whereas the observed line profile is heavily distorted by a combination of the relativistic orbital motion of the emitting material (and associated Doppler shifting of the lines), and the gravitational energy-shifting of the emitted photons over their trajectory to the observer \citep{Fabian1989,Laor1991,Dauser2010}. 

This reflected component reveals itself as an excess of flux above the coronal power-law component in the energy range surrounding $\sim 6.4$ keV. This excess is assumed to be the relativistically broadened rest frame iron K$\alpha$ line profile.   The line profile is sensitive to the location of the inner edge of the accretion disc, as this sets the fastest orbital frequency and the largest gravitational redshift. As such, it promises to be a powerful probe of the dimensionless spin parameter of the Kerr metric, $\tilde a$, under the assumption that the disc inner edge is set by the innermost stable circular orbit, which is a function of spin. The spin parameter also influences the line profile via the spin-dependence of orbital frequency, gravitational redshift, and photon trajectories, although these are more subtle effects. Now that both techniques have been introduced, we proceed to a discussion of some potential sources of  systematic errors which have been suggested in the community.  

Firstly, as can be seen in Eq.~\ref{eq:acceff} and in the above discussion, the presence of an innermost stable circular orbit (ISCO) plays a key role in both x-ray fitting techniques.  The reason for this is that the ISCO is assumed to represent the inner edge of the accretion flow, with the disc temperature (in the case of continuum fitting \cite{Li05}) and disc density (in the case of reflection \cite{Ingram2019}) assumed to be zero within this region. In the case of reflection modelling the rationale behind this is relatively simple, although an oversimplification. Upon crossing the ISCO each fluid element is rapidly accelerated and therefore the disc density must drop by a compensating factor to conserve the radial mass flux. However, the disc density of course does not drop truly to zero (see e.g., \cite{MummeryBalbus2023}) and recent calculations suggest that the intra-ISCO region may contribute non-negligibly to the overall reflection spectrum \cite{Dong23} (see also \cite{Reynolds97, Wilkins20}). The concern here is that the uncertainty in the location of the inner edge of the disc occurs in the same spacetime regime at which the largest signatures of the black hole spin are imprinted on the data (i.e., close to the event horizon). This may be particularly of concern for stellar mass black hole systems, which have large densities in the run up to the ISCO (when compared to AGN discs which are not thought to suffer from this contamination \cite{Wilkins20}, although see \cite{Dong23}), and therefore may suffer most from any intra-ISCO contamination. It is not yet clear if the intra-ISCO region acts as a degeneracy with black hole spin measurements, but further calculations such as those performed in \cite{Dong23} will likely illuminate this issue in the near future.

 The effect of the plunging region on continuum fitting analyses has been more widely studied, typically by numerical approaches which can self-consistently compute the thermodynamic quantities of accretion flows as a function of radius. Numerical general relativistic magnetohydrodynamic simulations \citep[e.g.,][]{Noble10, Penna10, Zhu12, Schnittman16, Lancova19, Wielgus22} generically find non-zero temperatures at and within the ISCO, with non-zero associated thermal emission. This emission will inevitably modify the SED of an observed black hole disc, and may induce systematic errors in inferred black hole properties. The degree to which this region biases spin measurements is unclear, with recent work \cite{Lancova23} suggesting that this bias could be substantial (although this is contention with earlier work which found minor \cite{Zhu12}, or only moderate \cite{Noble11} degeneracies). More recent high quality observations of stellar mass black hole binaries \cite{Fabian20, Lazar21} are not able to be satisfactorily fit by existing models (which neglect this intra-ISCO region), and it has been suggested that emission in the plunging region may be the source of this model-data discrepancy. This suspicion has recently been confirmed \cite{Mummery24PlungeA, Mummery24PlungeB}, and it remains to be seen if this has a systematic impact on continuum fitting spin measurements. 

Another potential source of systematic bias in continuum fitting spin measurements was recently proposed and discussed in detail by  \cite{Salvesen21}. The physical origin of this potential bias is as follows. It has long been well known \cite{Shapiro83} that the photons leaving the disc atmosphere will not have completely thermalised their energy. On their path through the disc atmosphere, photons can either be absorbed and re-emitted (thus totally thermalising their energy), or they can undergo elastic scattering. Elastic scattering however, by definition, does not change the energy of the photon, and so if this process dominates in the disc atmosphere, photons will be observed to have hotter temperatures associated with the altitudes closer to the disc midplane, not the discs upper surface. As an alternative way in which the energy of the photons emitted from the disc surface can be increased is to increase the black hole spin (and liberate more gravitational energy), a rather natural degeneracy between the physics of the radiative transfer in the disc atmosphere and the black hole spin can be set up.  Models for the effects of radiative transfer on continuum emission are well tested in the community (e.g., \cite{DavisHubeny06, Davis06, Davis19}), but \cite{Salvesen21} demonstrated that reasonably small deviations from these models can produce moderate changes in inferred black hole spin constraints.

Another potential concern for all modelling approaches (continuum and reflection), is that there may be additional physical components which are not included in conventional models but which produce flux in the key spectral regions used for spin measurements. The concern being that high inferred spin values may be an artifact of conventional models compensating for the lack of these spectral components. For instance, the thermal spectra of AGN are known to be poorly described by a simple bare accretion disc like that developed in classical models \cite{NovikovThorne73, PageThorne74}, which generally do not produce sufficient flux in the x-ray bands to satisfactorily describe observations. Additional components (a popular model is a second, ``warm'', corona) are then required to match the data. This is of concern particularly for reflection constraints of black hole spins, as this approach necessitates the subtraction of a continuum component. It  is only in the excess above the continuum that the iron reflection components become visible, and if this subtraction involves systematic errors in the continuum, then this may propagate into spin biases.

A recent case study of some of these potential systematic effects was performed by \cite{Zdziarski24}, who re-analysed Cygnus X-1 data which is usually found to favour high spins (as utilised in this work, Table \ref{tab:bhs}). The authors reproduced the high spin values inferred by conventional models $\tilde a > 0.9985$, but showed that a reduced value value $\tilde{a} = 0.92^{+0.07}_{-0.05}$ was found if the priors on radiative transfer in the disc atmosphere were relaxed (note that the inclusion of the `conservative' Cyg X-1 model in the proceeding section was motivated by this analysis). Allowing for an additional ``warm'' coronal component on the other hand completely changed the spin posterior, leading to a best fit $\tilde{a} = 0.04^{+0.26}_{-0.04}$ (note that the existence of warm coronal components are rather controversial, and thus this model should be interpreted as a heuristic, and somewhat provocative,  example of the potential effects of additional spectral components). The authors in \cite{Zdziarski24} did not consider the effects of the plunging region in their analysis. Naturally, modifications of spin measurements at the warm-corona level would have substantial implications for superradiance calculations. 

We conclude this section by stressing that the purpose of this somewhat pessimistic overview was to bring to light some of the discussions currently taking place in the x-ray astronomy community, which represent an important caveat to axion superradiance studies. Many of the discussions of potential systematics presented here are themselves controversial, and the physical reality of (e.g.,) ``warm coronae'' are by no means settled or universally accepted. In a statistical sense, x-ray spin measurements are repeatable, produce good fits to the data, and are based upon physically sound models (which may or may not be complete). The study of black hole accretion is an active field of research, and it is overwhelmingly likely that spin measurements will be refined and improved upon in the coming years, particularly as future observatories come online \cite{Madsen24}. 

\section{A Comment on Supermassive Black Holes}\label{sec:smbh}

In this work we have chosen to focus on solar mass scale black holes for a number of different reasons. First, the superradiance growth rate of a given mode scales proportionally to the black hole mass, implying the characteristic time required to extract a non-negligible fraction of the black hole spin will, in most cases, be significantly longer than the Salpeter timescale. While supermassive black holes are much older than solar-mass-scale black holes (and are unlikely to have been accreting near the Eddington limit for all of that time), the need for long evolutionary timescales makes deriving robust limits extremely difficult. The second issue that arises for massive black holes stems from the fact that these objects are subject to larger environmental effects. The presence of dense accretion disks, ambient stellar objects, binary companions, etc. can complicate the superradiant evolution (see \eg~\cite{Du:2022trq}) -- and this is particularly true when superradiant growth is slow. Finally, spin measurements are also subject to large uncertainties. The slow growth and long timescales associated to supermassive black holes tend to imply that self-interactions are likely to play a prominent role in suppressing spin extraction (suggesting the limits derived \eg in~\cite{Unal:2020jiy} may need to be revised). We intend to address this quantitatively in future work.  

\vspace{.5cm}
\section{Conclusions}\label{sec:conc}

In this work we have revisited superradiance constraints on axions using the inferred spins of solar-mass-scale black holes. The motivation for focusing on light black holes is three-fold: (1) these objects are relatively young, meaning their typical age is short relative to the timescale required for accretion to alter their properties, (2) the local environments around these objects are not expected to alter the short-term evolution of superradiance itself, and (3) the superradiant growth timescale is shorter for light black holes, meaning for a fixed black hole, there is a larger range of axion masses for which pertubative control at small $\alpha$ is obtainable. Extending such an analysis to the case to higher mass black holes will be the focus of future work.

In this study we have gone beyond the two-level system and solved for the evolution of axion superradiance including all relevant states at $n \leq 5$, and computing relativistic corrections to all scattering processes using a Green's function approach.  We have confirmed that higher-order level mixing plays a negligible role in the $m=1$ spin down at $\alpha \lesssim 0.15$ (a threshold slightly marginally lower than previously argued~\cite{Baryakhtar:2020gao}, but in agreement with the numerical result of~\cite{Omiya:2022mwv}), but we find these levels can significantly alter the evolution at higher masses. In particular, we have argued that the coupled evolution of the $n\leq5$ levels are likely sufficient to capture the estimated $m=2$ spin down constraints on typical high spin solar mass black holes, which allows one to extend constraints to larger values of the axion mass (for a fixed mass black hole). We have also introduced a statistical formalism that allows for a more rigorous and meaningful interpretation of observed black hole spins; comparing with previous results in the literature, we find that prior limits had significantly underestimated the sensitivity to the axion decay constant (by roughly an order of magnitude). The main result of this work is highlighted in Fig.~\ref{fig:main}, which shows the combined spin-down constraints on GRS 1915+105 and the conservative interpretation of Cygnus X-1 \footnote{All code used to produce the results shown here can be found at~\cite{WitteGit}}.

The role of the $n=4$ and $n=5$ states highlighted in this work encourages further studies for mixing at $n > 5$; such levels will be important for confirming stability of spin-down constraints at $m =2$, establishing spin down constraints at $m > 2$, and for studies of gravitational waves (where the signal produced from these objects is strongly sensitive  to characteristic occupation number of each state, and the mass of the black hole itself), and may prove to be important for supermassive black holes, where the characteristic ages of black holes are notably larger than their solar-mass counterparts. Finally, we have concluded with an extensive discussion on the potential systematics associated with black hole spin inference performed using x-ray observations, arguing that care needs to be taken when interpreting the robustness of these results. 
\newline


\section{Acknowledgments}%
SJW would like to thank Masha Baryakhtar, Enrico Cannizzaro, Andrea Caputo, Marios Galanis, Doddy Marsh, Hidetoshi Omiya, Mario Reig, and Olivier Simon for their comments, discussions and feedback, as well as Ciaran O'Hare for his comments which initially inspired some aspects of this work. SJW acknowledges support from a Royal Society University Research Fellowship (URF-R1-231065). This article/publication is based upon work from COST Action COSMIC WISPers CA21106, supported by COST (European Cooperation in Science and Technology).  This work was supported by a Leverhulme Trust International Professorship grant [number LIP-202-014].


\bibliography{biblio}

\end{document}